\documentclass[aip,amsmath,amssymb,reprint]{revtex4-1}

\usepackage{graphicx}
\usepackage{dcolumn}
\usepackage{bm}

\usepackage[utf8]{inputenc}
\usepackage[T1]{fontenc}
\usepackage{mathptmx}
\usepackage{etoolbox}
\usepackage{nicefrac}
\usepackage{xcolor}
\usepackage{url}
\usepackage{float}
\usepackage{hyperref}

\usepackage{color}

\definecolor{blueryb}{rgb}{0.01, 0.28, 1.0}

\definecolor{greenryb}{rgb}{0.1, 1.0, 0.1}

\makeatletter
\def\@email#1#2{%
 \endgroup
 \patchcmd{\titleblock@produce}
  {\frontmatter@RRAPformat}
  {\frontmatter@RRAPformat{\produce@RRAP{*#1\href{mailto:#2}{#2}}}\frontmatter@RRAPformat}
  {}{}
}%
\makeatother
\begin{document}

\preprint{AIP/123-QED}

\title[Developments and Further Applications of EDDPs]{Developments and Further Applications of Ephemeral Data Derived Potentials}
\author{Pascal T. Salzbrenner*}
\email{pts28@cam.ac.uk}
\author{Se Hun Joo}
\affiliation{ 
Department of Materials Science \& Metallurgy, University of Cambridge, Cambridge, U.K.
}

\author{Lewis J. Conway}
\affiliation{ 
Department of Materials Science \& Metallurgy, University of Cambridge, Cambridge, U.K.
}
\affiliation{Advanced Institute for Materials Research, Tohoku University, Sendai, Japan}

\author{Peter I. C. Cooke}
\affiliation{ 
Department of Materials Science \& Metallurgy, University of Cambridge, Cambridge, U.K.
}

\author{Bonan Zhu}
\affiliation{Department of Chemistry, University College London, London, U.K.}

\author{Milosz P. Matraszek}
\affiliation{Trinity College, University of Cambridge, Cambridge, U.K.}

\author{William C. Witt}
\affiliation{ 
Department of Materials Science \& Metallurgy, University of Cambridge, Cambridge, U.K.
}
\author{Chris J. Pickard*}
\email{cjp20@cam.ac.uk}
\affiliation{ 
Department of Materials Science \& Metallurgy, University of Cambridge, Cambridge, U.K.
}
\affiliation{Advanced Institute for Materials Research, Tohoku University, Sendai, Japan}

\date{\today}

\begin{abstract}
Machine-learned interatomic potentials are fast becoming an indispensable tool in computational materials science. One approach is the ephemeral data-derived potential (EDDP), which was designed to accelerate atomistic structure prediction. The EDDP is simple and cost-efficient. It relies on training data generated in small unit cells and is fit using a lightweight neural network, leading to smooth interactions which exhibit the robust transferability essential for structure prediction. Here, we present a variety of applications of EDDPs, enabled by recent developments of the open-source EDDP software. New features include interfaces to phonon and molecular dynamics codes, as well as deployment of the ensemble deviation for estimating the confidence in EDDP predictions. Through case studies ranging from elemental carbon and lead to the binary scandium hydride and the ternary zinc cyanide, we demonstrate that EDDPs can be trained to cover wide ranges of pressures and stoichiometries, and used to evaluate phonons, phase diagrams, superionicity, and thermal expansion. These developments complement continued success in accelerated structure prediction.
\end{abstract}

\maketitle

\section{Introduction}
\label{sec:intro}

Machine learning methods have transformed computational materials science \cite{Machine-learning-materials-general-review, Machine-learning-roadmap}. They have been applied to predict material properties~\cite{Bhadeshia-NN-alloy-properties, Machine-learning-property-prediction-competition, Temperature-machine-learning, Machine-learning-phase-space-averages}, develop new density functionals~\cite{Machine-learning-XC-potentials, Burke-ML-Density-Functionals, ML-DFT-molecules}, and train highly accurate interatomic potentials on large quantities of data~\cite{Behler-Parrinello-NNPs, Gaussian-Approximation-Potentials, Gabor-MLIP-review}. These machine-learned potentials enable simulations over much larger length- and time-scales than those feasible with \textit{ab initio} methods such as density functional theory (DFT) \cite{DFT-Hohenberg-Kohn, DFT-Kohn-Sham}, at a comparable level of accuracy. Some successful applications of machine learned potentials include crack propagation in silicon~\cite{Gabor-Silicon-GAP}, the high-temperature high-pressure phase diagram of hydrogen~\cite{Bingqing-Chris-hydrogen-MLP}, demonstrating the influence of quantum effects on phases of water and ice ~\cite{Bingqing-water-MLP} and Ge-Te-based phase change materials~\cite{GeTe-MLP, Gabor-GeSbTe-MLP}, to name a handful of examples among many.

The development of machine-learned potentials is a highly active field and there are already many implementations to choose from. The primary differences between them arise from the choice of atomic descriptors and the fitting method~\cite{Behler-Parrinello-NNPs, Gaussian-Approximation-Potentials, SNAP-MLP, DeepMD-MLP, Chris-EDDPs, MTP, ACE-I, ACE-II, MACE, Kozinsky-Allegro-NNP-systematic, Gabor-Behler-MLP-perspective, Behler-MLP-Four-Generations-Review, Universal-graph-IAP, Unified-graph-NN}. Pioneering developments in machine learning the energy landscapes of extended systems were made by Behler and Parrinello~\cite{Behler-Parrinello-NNPs}, as well as Csányi and co-workers~\cite{Gaussian-Approximation-Potentials}. More recently, classes of local potentials that are systematically improvable by increasing the size of the basis, such as the moment tensor potentials (MTP)~\cite{MTP} and the atomic cluster expansion (ACE)~\cite{ACE-I, ACE-II}, have been developed. Message-passing neural networks take inspiration from chemical intuition~\cite{Atom-centred-message-passing-unified, MACE}. Another promising development is the construction of large-scale graph neural networks covering extensive swathes of the periodic table~\cite{Universal-graph-IAP, Unified-graph-NN}.

In addition to research into the development of appropriate atomic descriptors~\cite{Behler-descriptors, Gabor-SOAP, Atomic-density-representations, Gabor-descriptors}, substantial effort has been devoted to developing efficient schemes for the training database generation~\cite{Efficient-database-generation}. Traditionally, training data consist of experimental structures, supplemented by manually constructed defect and surface prototypes~\cite{Gabor-MLIP-review}. However, more recent methods have enabled automated construction of datasets without experimental input.

Molecular dynamics (MD) simulations can be used to train the potential on selected uncorrelated snapshots~\cite{Behler-Parrinello-NNPs, Shapeev-MLIP, Gabor-MLIP-review, MLP-construction-strategies}. This frequently includes an element of active learning~\cite{Active-Learning-Shapeev, Active-Learning-Car}, where the potential is updated when it encounters a badly described structure. It is possible to consciously bias the MD runs towards configurations badly described by the potential, requiring fewer steps to obtain sufficient sampling of the energy landscape \cite{Hyperactive-learning, Uncertainty-driven-active-learning}. Another promising method is to use techniques inspired by structure prediction, in particular random search, to create \textit{ab initio} datasets~\cite{Chris-EDDPs, USPEX-MLP, Maise-MLP-I, Maise-MLP-II, GAP-RSS-boron, GAP-RSS-phosphorus}.

Structure prediction is a prominent component of modern materials science; understanding materials' structures at the atomic level allows for a quantum mechanical prediction and description of their properties~\cite{Structure-prediction-review, Lewis-Chris-CSP-review}. Perhaps the most straightforward technique for first-principles structure prediction is \textit{Ab Initio} Random Structure Searching (AIRSS)~\cite{Chris-Silane, AIRSS}, which explores energy landscapes through the relaxation of many randomly generated structures to nearby local minima. Through sufficient samples and aided by constraints ensuring the initial random structures are chemically sensible, low-enthalpy arrangements of the constituent atoms are identified. Other approaches also exist, notably evolutionary~\cite{USPEX, Eva-Zurek-XtalOpt} and particle-swarm optimisation~\cite{Calypso}, as well as minima/basin hopping~\cite{Basin-hopping}.

The success of \textit{ab initio} structure prediction over recent decades~\cite{Chris-hydrogen-phase-III, Chris-HG-aluminium, Chris-helium-ammonia, Chris-Silane, H3S-theoretical-prediction, Chris-clathrate-hydrides, Hemley-LaH10-prediction} is largely due to the development of DFT-based methods which can accurately determine candidate structure energies. DFT's quantum mechanical foundation allows it to replicate the underlying smoothness of the energy landscape, resulting in robust behaviour for all configurations of atoms. This is essential for structure prediction. However, the need in DFT to account for the electronic structure incurs a significant computational cost. This cost can scale cubically with the number of atoms, while the size of the configuration space to be searched ultimately increases exponentially~\cite{AIRSS}. 

Machine learned potentials can ameliorate the cubic scaling problem~\cite{Maise-MLP-I, GAP-RSS-phosphorus, Au-cluster-NNP, Gabor-Carbon-GAP-searching, Evolutionary-algorithms-learn, Calypso-NNP, USPEX-MTP, Shapeev-MLP-searches, Kolmogorov-Au-cluster-searching, Surrogate-potential-Hammer, Surrogate-potential-Jacobsen} and a class of neural-network potentials, Ephemeral Data-Derived Potentials (EDDPs)~\cite{Chris-EDDPs}, were recently developed specifically for high-throughput structure prediction. To date, the approach has been validated with two published searches. Its first application was the prediction of a new high pressure phase of silane (SiH$_{4}$). Silane was one of the first systems to be studied with AIRSS\cite{Chris-Silane}, but the recent work using EDDPs has uncovered a previously unseen structure which contains twelve formula units (f.u.) in the primitive cell and becomes thermodynamically stable around $300$ GPa~\cite{Chris-EDDPs}. Recently, EDDPs were employed in the study of a ternary hydride system. Following the recent experimental claim of room temperature superconductivity at close-to-ambient pressures in the Lu-H-N system~\cite{Dasenbrock-Gammon2023}, ternary hydride systems have been the focus of several computational studies~\cite{Xie2023,Huo2023,Hilleke2023,Lewis-Chris-LuNH-search}. Of these studies, the only one to identify a thermodynamically stable ternary structure at ambient pressure leveraged EDDPs for the structure search, exploring more of the composition space and searching with larger unit cells~\cite{Lewis-Chris-LuNH-search, LuH-colour-theory}. 

The training scheme for the EDDPs is inspired by the AIRSS approach and closely integrated with the AIRSS software package. In AIRSS, sensible structures are generated by placing atoms randomly into a unit cell and then adjusting their positions to satisfy a set of chemically inspired constraints. The structures are then relaxed to their nearest local minima. Normally, this is done using the \textsc{CASTEP} DFT package~\cite{CASTEP}, but with the EDDPs, \texttt{repose} (see Sec.~\ref{sec:eddp_calculations}) fulfils this role. Through sufficient samples, low-enthalpy arrangements of the constituent atoms are identified.

This work is organised as follows. In Sec.~\ref{sec:main_features}, we summarise the main features of the EDDPs, notably their tightly-constrained, physically inspired descriptors. In Sec.~\ref{sec:how_to_train}, we explain our training scheme, which is based on a wide exploration of chemical space in small cells and uses a light-weight neural network of typically just five nodes in a single hidden layer. In Sec.~\ref{sec:practical_use}, we explain the software implementation of EDDPs, how they can be used to carry out simulations, and suggest some best practices for generating a high-quality potential. In Sec.~\ref{sec:case_studies}, we highlight several key features of the EDDPs through a series of case studies. The EDDPs are smooth and exhibit size transferability (for instance, a Pb potential trained on no more than six atoms can be used to successfully run MD simulations with thousands of atoms). EDDPs can be used to study a wide range of stoichiometries and pressures and can describe complex systems such as metal-organic frameworks (MOFs). The EDDPs, originally designed to accelerate structure search, can be used to calculate phonon dispersions, run MD, and accurately predict phase diagrams.

\section{Main Features of the EDDPs}
\label{sec:main_features}

\begin{figure*}[htpb]
\includegraphics[width=0.95\linewidth]{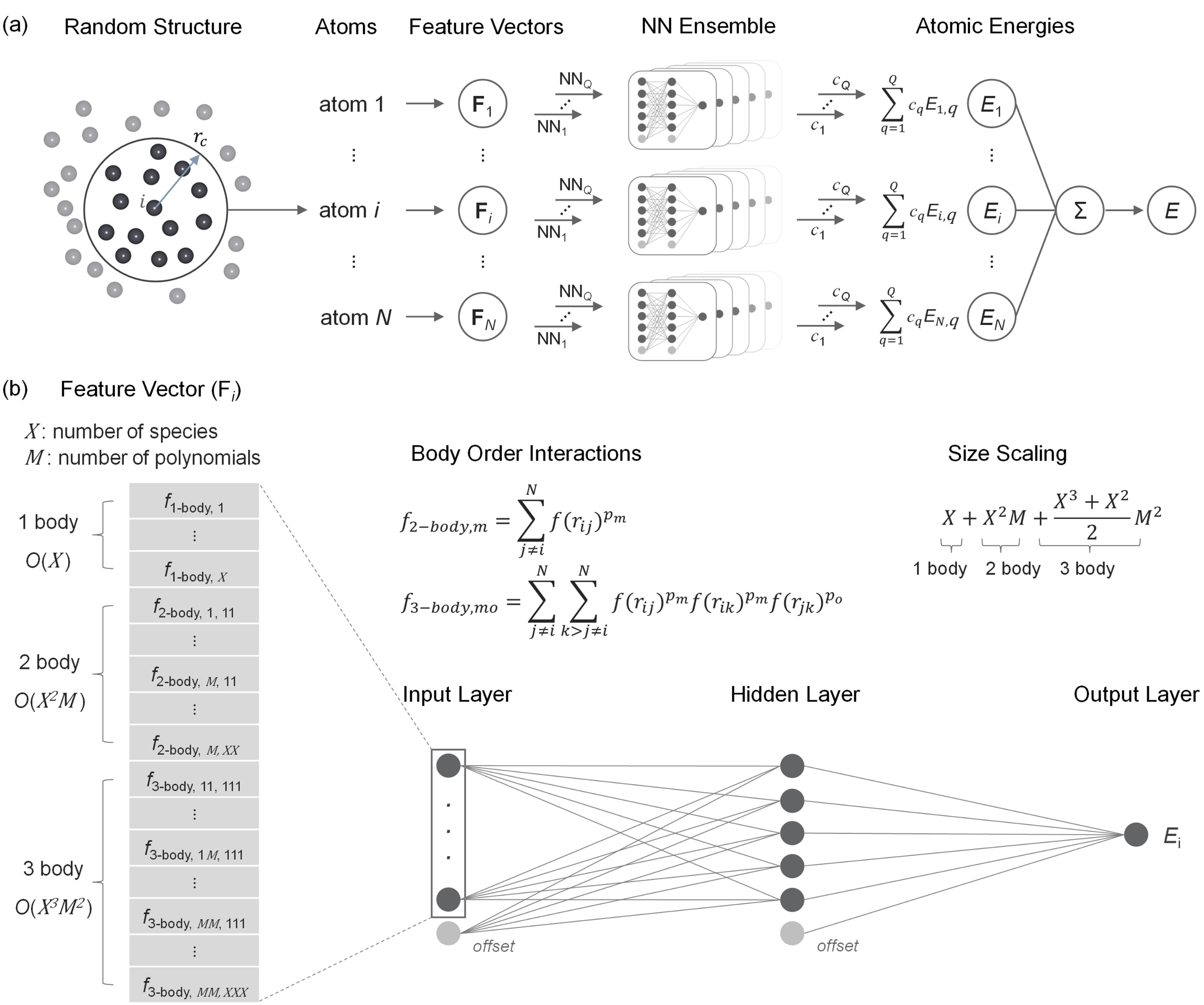}
\caption{\label{fig:eddp} (a) Key stages of generating a potential: building feature vectors from atomic environments, training of multiple individual potentials to generate an ensemble before obtaining a single potential from a weighted fit across this ensemble (Q is total size of ensemble, with each individual potential indexed by q). (b) Feature vectors are constructed by concatenating the basis functions for each species and each body-order interaction. The total size of this vector depends on the number of functions (M), the number of body-order interactions included in the potential and the number of species (X). These feature vectors are passed through a shallow neural network to yield an atomic energy. The atomic energies for all atoms are summed up to give the total energy.}
\end{figure*}

In this section we outline briefly the EDDP method; a more complete description is available in Ref.\setcitestyle{numbers}~\cite{Chris-EDDPs}.\setcitestyle{super} Like many interatomic potentials, EDDPs approximate the total potential energy by a sum of atomic contributions~\cite{Behler-Parrinello-NNPs}:

\begin{equation}
\label{eq:NN_atomic_energy}
E_{\mathrm{tot}} = \sum_i E_i = \sum_i E(\mathbf{F}_i),
\end{equation}
where \(\mathbf{F}_i\)  is a feature vector encoding a local atomic environment and \textit{i} runs over all atoms in the system. In principle, the function \(E(\mathbf{F}_i)\) may be determined in numerous ways, including linear, Gaussian process, or neural network regression. In the Subsection \ref{sec:eddp_feature_vectors}, we introduce the EDDP feature vector in a linear setting, after which we summarise the neural networks employed in practice (shown in Fig.~\ref{fig:eddp}(a)). 

\subsection{EDDP Feature Vectors}
\label{sec:eddp_feature_vectors}

The EDDP feature vector is rooted in the body-order expansions utilised by many classical interatomic potentials. Such expansions express the atomic energies in Eq. \ref{eq:NN_atomic_energy} in terms of interactions with neighbouring atoms:

\begin{equation}
\label{eq:total_atomic_energy}
E_i = E^{(1)}_i + E^{(2)}_i + E^{(3)}_i + ...
\end{equation}
with each term corresponding to an increasing body-order interaction ($E^{(1)}$ is one-body, etc.). Taking inspiration from the physically motivated Lennard-Jones~\cite{Lennard-Jones-I, Lennard-Jones-II} and extended Lennard-Jones~\cite{Extended-Lennard-Jones} potentials, these interactions can be modelled with a linear combination of functions composed of a radial function \textit{f(r)} raised to a power $p_{m}$.  Wang et al.~\cite{Cutoff-LJ} have previously proposed using radial functions that are naturally cut-off beyond a radius $r_c$. We use a function that follows this approach with the form:

\begin{equation}
\label{eq:eddp_basis_functions}
    f(r) = 
     \begin{cases}
     2(1-r/r_{c}) & r \leq r_{c}\\
     0 & r > r_{c}.
     \end{cases}
 \end{equation}
The two-body interaction expressed as a sum of these functions is:

\begin{equation}
\label{eq:twobody}
E^{(2)}_i = \sum^N_{j\ne i}\sum^M_{m}w^{(2)}_mf(r_{ij})^{p_m} = \mathbf{w}^T_{(2)}\mathbf{F}_i^{(2)},
\end{equation}
where the first summation is over the neighbours \textit{j} of the central atom \textit{i}, with distance $r_{ij}$ between them. The second summation is over the total number of functions \textit{M} with corresponding weights $w^{(2)}_m$. The analogous expression to Eq.~\ref{eq:twobody} for the three-body interactions is:

\begin{equation}
\label{eq:threebody}
E^{(3)}_i = \sum^N_{j\ne i}\sum^N_{k>j\ne i}\sum^M_{m}\sum^M_{o}w^{(3)}_{mo}f(r_{ij})^{p_m}f(r_{ik})^{p_m}f(r_{jk})^{q_o} = \mathbf{w}^T_{(3)}\mathbf{F}_i^{(3)}.
\end{equation}
The first two summations now run over neighbouring atoms, with the index \textit{k} corresponding to a third atom that contributes to the interaction. The summations in $m$ and $o$ run over functions with corresponding weights $w_{mo}$. The powers $p_{m}$ are distributed geometrically between a minimum and maximum exponent. By default, the minimum exponent is $2$, ensuring continuity of energies and forces at the cutoff radius~\cite{Chris-EDDPs}.

The right-hand sides of Eq.~\ref{eq:twobody} and Eq.~\ref{eq:threebody} show how the two- and three-body interactions can be expressed as a scalar product between a number of weights $\mathbf{w}^T$ and a vector $\mathbf{F_i}$. When the $\{\mathbf{F}_i^{(j)}\}$ for all body-order interactions are concatenated they constitute the feature vector (shown in figure \ref{fig:eddp} (b)), which defines the local environment of atom $i$: 

\begin{equation}
\label{eq:FeatureVec}
    \mathbf{F}_i = \mathbf{F}^{(1)}_i \oplus \mathbf{F}^{(2)}_i \oplus \mathbf{F}^{(3)}_i.
\end{equation}
The weights ($\mathbf{w}^T$) could be found from a linear fit. In practice these linear weights are not used in the EDDP scheme, rather a small neural network is employed with the feature vectors $\mathbf{F_i}$ as input.

In the case of multiple species and truncation at three-body terms, this concatenation gives a feature vector of length $X + X^2M + \frac{X^3 +X^2}{2}M^2$, where $X$ is the number of species and \textit{M} the number of polynomials. This scaling results from species-centred representations of each body order and the feature vectors are sparse in the case of many species. While the EDDPs include interactions up to three-body by default, two-body potentials can be sufficiently accurate to accelerate structure prediction in non-covalently bonded systems and considerably faster.

\subsection{Neural Network}

After constructing the feature vectors, a fit is carried out using a neural network with a single hidden layer typically containing no more than $5$ nodes. This small and light-weight neural network makes the cost of fitting the potential negligible compared to the cost of the first-principles calculations. An EDDP can be trained on a small number of central processing units (CPUs), such as might be contained in a laptop.

Since training neural networks involves non-convex optimisation with a stochastic initialisation, two subsequent fits on the same data will typically produce two different potentials. In the EDDP framework, this is exploited to construct an ensemble of several potentials to obtain a composite EDDP. By regularising the fit, ensembles are expected to outperform individual potentials~\cite{Committee-NNs, NN-ensembles}.

\section{How to Train Your EDDP}
\label{sec:how_to_train}

Several recent studies have demonstrated the success of small-cell training schemes~\cite{GAP-RSS-boron, GAP-RSS-phosphorus, Chris-EDDPs, Magnesium-small-cell-training, Hart-Zirconium-small-cell-training}. These methods cover a large region of chemical space by using a training set containing many small cells with a wide variety of structures. The computational cost associated with evaluating the DFT energies of datapoints in this way is low compared to methods which rely on MD trajectories of larger systems.

\begin{figure}[!bt]
\includegraphics[width=0.95\linewidth]{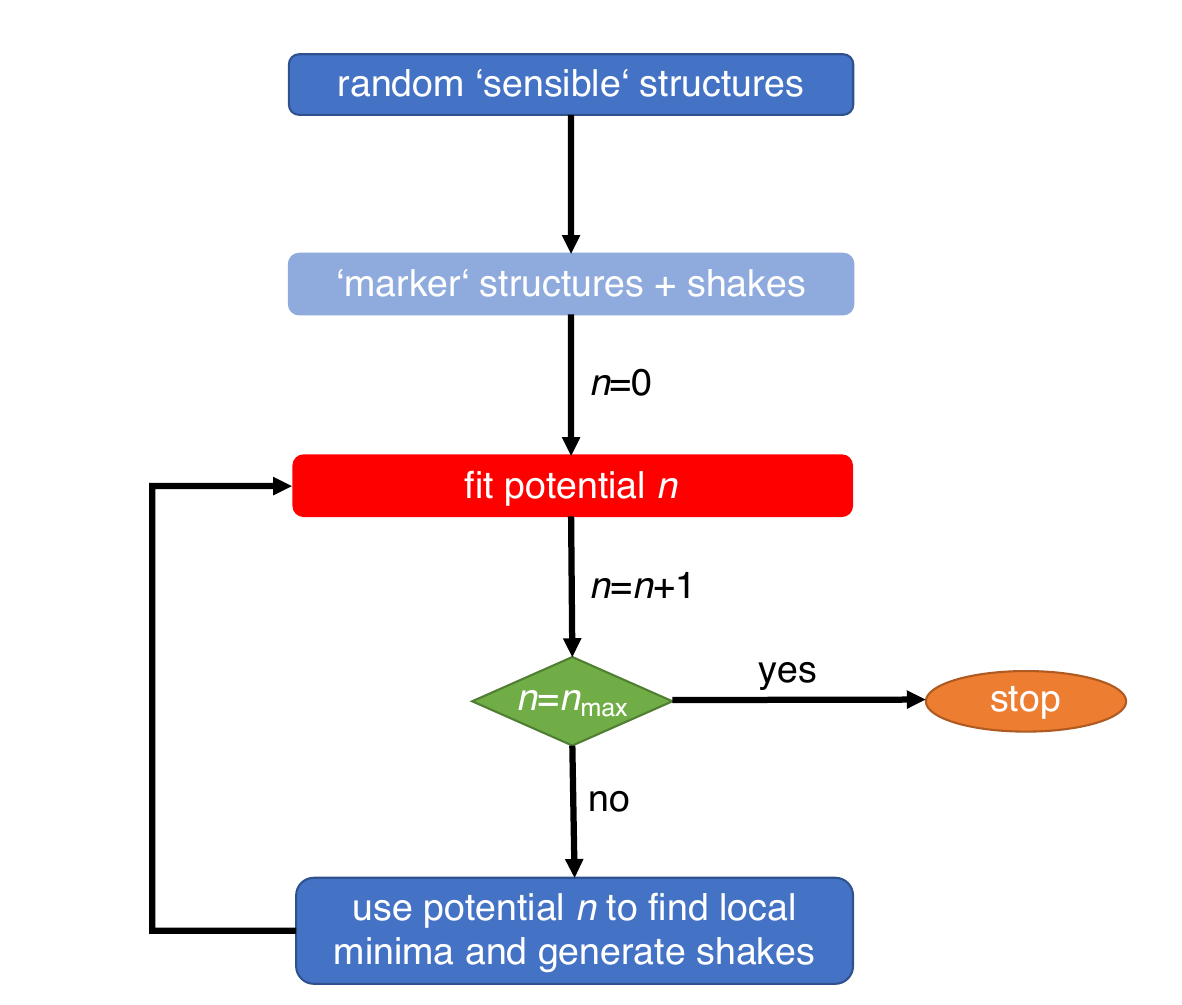}
\caption{\label{fig:eddp_training_flowchart} The iterative scheme used to generate a dataset of small-cell structures and train an EDDP on it. Blue rectangles indicate DFT singlepoint energy calculations, with light blue indicating the optional marker structure step. The red rectangle indicates the process of fitting the potential.}
\end{figure}

The scheme for training EDDPs uses a small-cell method designed to exploit random structure searching to provide diverse, and challenging, datasets. Figure \ref{fig:eddp_training_flowchart} summarises the iterative training. The training dataset is initially comprised of randomly generated structures. It is possible to specify a variety of constraints on how these structures are generated, such as minimum interatomic distances and a number of applied symmetry operations. The exact choices will depend on the task at hand, but the general principle is that this set of structures should be diverse, and include high-energy configurations in order to help the EDDP learn what \textit{is not} a good arrangement of atoms. The energies for these structures are found with single-point calculations at the DFT level. The number of atoms in these initial structures is typically in the range of $2$-$20$. In the examples in Sec.~\ref{sec:case_studies}, the lowest number of atoms is $2$ for elemental lead, and the highest is $24$ for scandium hydride. 

It is possible, but not necessary, to add `marker' structures to the set. These will usually be structures the system is known to adopt, and manually including them helps ensure they are well-described. These structures are `shaken' (subjected to small random distortions). The DFT energies of the new set are calculated and added to the dataset. A first EDDP is trained on the random (and possible marker) structures and then refined in an iterative process. In each iteration the current EDDP is used to relax a set of newly generated random structures to their local minima. The structures corresponding to those minima are also shaken. The energies of these shaken structures and those at the minima are calculated using DFT and added to the dataset. A new EDDP is then trained for the next iteration. The training is terminated after a few, typically five, iterations. The default values for these parameters, along with all others important for the generation of EDDPs, are summarised in table \ref{tab:eddp_main_parameters}.

The cost function used to train the potential takes the form

\begin{equation}
    C=\frac{1}{S}\sum_s\left|E_s - \sum^{N_s}_iE(\mathbf{F}_{s,i}) \right|^p,
    \label{eq:costfunction}
\end{equation}
where the index \textit{s} runs over the total \textit{S} structures in the training dataset, $E(\mathbf{F}_{s,i})$ is the EDDP energy of atom \textit{i}, $E_s$ is the total DFT energy and $N_s$ is number of atoms in each structure. The exponent $p=1.25$, in a compromise between minimising the mean average and root mean square errors. Note that, in contrast to other schemes, forces are not included in the cost function. The low computational cost of energy calculations for small-cells allows for a denser sampling of the energy landscape and the quality of forces from the resulting potentials has proven to be sufficient for many applications (see sec. \ref{sec:case_studies} for examples of dynamical calculations). In particular, the shaken structures ensure sensible forces by providing information about the potential energy surface near the minima. 

On every iteration, the data is split into training, validation, and testing sets at a ratio of approximately 80:10:10. The cost function is minimised using the Levenberg-Marquardt optimisation algorithm~\cite{Levenberg-Marquardt-I, Levenberg-Marquardt-II} only on the \textit{training} set. The \textit{validation} set is used to implement early stopping~\cite{Early-stopping} in order to avoid over-fitting. Finally, the remaining error against DFT is calculated on the \textit{testing} set in order to assess the quality of the potential.

As discussed in Sec \ref{sec:main_features} and shown in figure \ref{fig:eddp}a, an ensemble of potentials is used in the EDDP scheme. The weighted average potential is generated \textit{via} non-negative least squares (NNLS)~\cite{NNLS} reusing and fitting to the validation dataset. This combines a small subset of all generated potentials with positive weights; most potentials will have weights of exactly 0, avoiding over-fitting problems arising from an unconstrained least-squares fit~\cite{Chris-EDDPs}. The NNLS combines only those potentials important and useful for describing the energy landscape well. This reduces the number of neural network predictions which must be carried out. Combining a number of potentials allows the size of the individual neural networks to be kept small. In our experience, the ensemble usually results in a smaller testing error than any individual potential.

The existence of many individual potentials allows for statistical analysis of the performance of the potential~\cite{Committee-NNs}. The (unweighted) standard deviation of the predicted energies across the subset of potentials selected by the NNLS is referred to as the `ensemble deviation'. Although not a direct measurement of the error with respect to DFT, it provides a useful indicator of how well-constrained the potential is in a given region of structure space. This can be used to bias the structures included in the dataset towards those which are not yet well-described, introducing an additional element of active learning. It can also be used to assess the transferability of potentials. It should be kept in mind that, in principle, it is possible for the deviation to be small even though the error compared to DFT is large\cite{Bias-variance-tradeoff}, however we are yet to observe such a case for any EDDP ensemble. Finally, we use the ensemble deviation to remove pathological structures from high-throughput searches. These pathological structures appear in regions where the potential is not well-constrained and are usually characterised by unreasonably small atomic separations. Such structures will have ensemble deviations of hundreds of eV per atom. This is orders of magnitude larger than well-described structures, whose deviations are of the order of a few meV per atom.

\begin{table}
\caption{\label{tab:eddp_main_parameters}Default values of the main parameters governing EDDP training.}
\begin{ruledtabular}
\begin{tabular}{ccc}
Parameter & Default \\
\hline
Radial cutoff, $r_{c}$ & $3.75$~\AA \\
Size of hidden layer & $5$ \\
Number of training cycles, n$_{\text{max}}$ & $5$ \\
Number of polynomials, $M$ & $5$ \\
Lowest exponent & $2$ \\
Highest exponent & $10$ \\
Number of initial random structures & $1000$ \\
Number of minima per cycle & $100$ \\
Number of shakes per minimum & $10$ \\
Shake amplitude & $0.02$~\AA \\
\end{tabular}
\end{ruledtabular}
\end{table}

\section{Implementation and Practical Use}
\label{sec:practical_use}

As originally implemented, the EDDP package is a set of \texttt{bash} scripts and \texttt{Fortran} codes. These are available under the GPL2 license~\cite{EDDP-website}, along with AIRSS~\cite{AIRSS-website}, which the EDDP package uses to generate the training set. The \textsc{CASTEP} DFT package~\cite{CASTEP}, with which the EDDP framework has been integrated, is available at no cost under academic license~\cite{getting-castep}.

\subsection{Generating EDDPs}
\label{sec:generating_eddps}

The \texttt{chain} script is distributed with the EDDP package and steps through the iterative training scheme discussed in Sec.~\ref{sec:how_to_train}. It is executed on a head node and launches jobs on a set of specified compute nodes it must be able to reach \textit{via} \texttt{ssh}. This is a convenient setup for small- to mid-size personal clusters not inhibited by wallclock time restrictions.

\begin{figure}[!bt]
\includegraphics[width=0.95\linewidth]{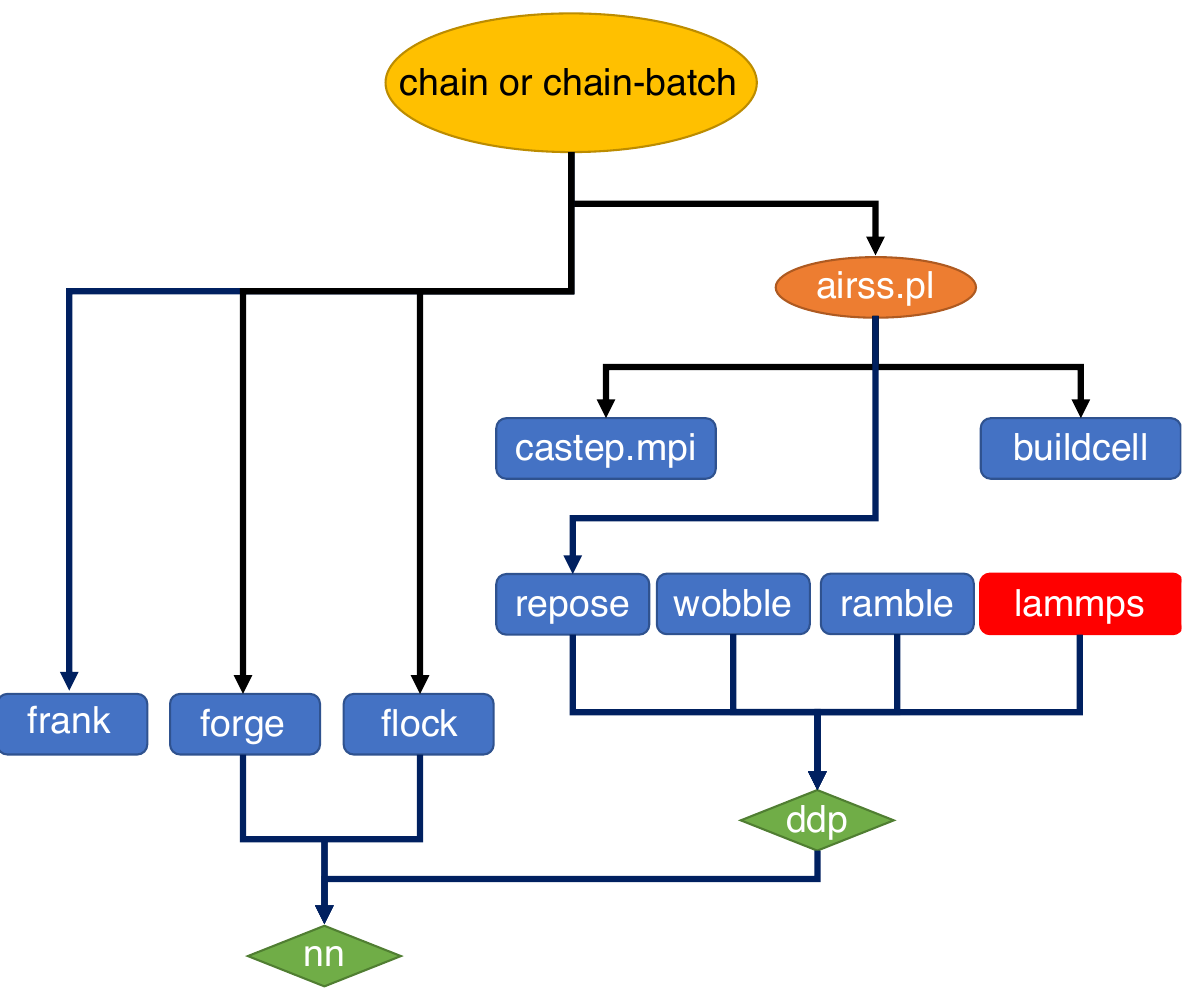}
\caption{\label{fig:eddp_codes_flowchart} The relationship between the different codes comprising the EDDP framework. Downward arrows indicate a call. Yellow (orange) ellipses represent \texttt{bash} (\texttt{perl}) scripts, blue (red) rectangles represent compiled \texttt{Fortran} (\textsc{C++}) programs, and green rhombi are \texttt{Fortran} modules. For details on the different codes, see the main text.}
\end{figure}

Figure \ref{fig:eddp_codes_flowchart} shows the relationship between \texttt{chain} and the other codes in the EDDP framework. It calls AIRSS, which generates random structures using \texttt{buildcell}, and then calculates their energies using DFT. These calculations are small and run independently, resulting in efficient parallelism and scalability. \texttt{frank}, \texttt{forge}, and \texttt{flock} are responsible for feature vector generation, neural network fit, and NNLS combination of the final potential respectively. \texttt{flock} and \texttt{forge} interface with the neural network via the \texttt{nn} module. \texttt{repose} is the supported geometry optimisation code; it is run through the AIRSS script to generate relaxed structures for the training dataset. All the codes described above are developed in-house and do not use any pre-existing neural-network libraries for \texttt{Fortran}. They currently make use of OpenMP parallelisation and Basic Linear Algebra Subprograms (BLAS) libraries, but detailed profiling and optimisation is yet to be implemented.

From a functionality perspective, the EDDP package provides a self-contained pipeline for data-set generation, potential training, geometry optimisation and dynamical simulations, via the \texttt{buildcell}, \texttt{repose}, \texttt{ramble} and \texttt{wobble} codes. The most unique feature of the EDDP package compared to other \texttt{Fortran} neural network potential implementations is the in-built integration with AIRSS for generation of the data-set of small random structures.

A Julia implementation (developed by one of us) is also available (\texttt{EDDP.jl})~\cite{eddp-jl-github}. This implementation has comparable performance to the \texttt{Fortran} version and allows easy integration with a range of optimisation algorithms, machine learning frameworks, and neural network implementations. It also interfaces with \texttt{python}, allowing packages such as the Atomic Simulation Environment (ASE)~\cite{ase-paper} and \texttt{phonopy}~\cite{Phonopy} to use the EDDPs.

High-performance computing (HPC) clusters with queueing systems are supported using the \texttt{ddp-batch} package, available separately on GitHub~\cite{ddp-scripts-github}. \texttt{chain-batch} is based on \texttt{chain}, and takes the same command line arguments, but instead of launching the various steps in the training process directly using \texttt{ssh}, it submits them as job scripts to the HPC queueing system. Users are able to specify the scheduler options separately for each of these jobs in a \texttt{.schedopt} file, allowing for the available compute resources and architectures to be employed optimally. \texttt{chain-batch} monitors the status of the HPC jobs and only progresses once the required number of calculations has been completed. If the job is interrupted by wallclock time limitations before then, it is automatically resubmitted. \texttt{chain-batch} is  transferable to different HPC systems; so far, it has successfully been used on the SGE-based Thomas cluster hosted at University College London, as well as the Slurm-based Cambridge Service for Data-Driven Discovery (CSD3) and the UK National Supercomputing Service ARCHER2.

\subsection{Running Calculations with EDDPs}
\label{sec:eddp_calculations}

\begin{table}
\caption{\label{tab:eddp_codes}Summary of simulation codes in the EDDP package.}
\begin{ruledtabular}
\begin{tabular}{cc}
Programme & Task \\\hline
\texttt{repose} & Local geometry optimisation\\
\texttt{wobble} & Finite-difference phonon calculations \\
\texttt{ramble} \& LAMMPS & Molecular dynamics \\
\end{tabular}
\end{ruledtabular}
\end{table}

Table~\ref{tab:eddp_codes} summarises the different codes which use the EDDPs to carry out different types of atomistic calculations. The EDDPs' original purpose, structure prediction, is enabled by \texttt{repose}, which performs local structural optimisation of atomic configurations and is interfaced with the AIRSS package. In addition, the EDDP suite of codes supports phonon calculations based on the finite differences method~\cite{Tomeu-finite-differences} through \texttt{wobble}, which calculates the vibrational energy, as well as phonon dispersions and densities of states. MD simulations are performed using \texttt{ramble}, as well as through a LAMMPS~\cite{Lammps} interface. These codes interact with the neural network via the \texttt{ddp} module, see figure \ref{fig:eddp_codes_flowchart}.

\begin{table}
\caption{\label{tab:benchmarks}Single-core performance of potentials used in this work.}
\begin{ruledtabular}
\begin{tabular}{cc}
System & CPU Time [$\mu$s/atom/step]\\\hline
Carbon (Fig. \ref{fig:carbon_pes}(b) in section \ref{sec:smoothness_transferability_carbon}) & 475\\
Lead (section \ref{sec:elements_lead}) & 324 \\
ScH$_{12}$ (section \ref{sec:binaries_ScH}) & 23781 \\
lcs-Zn(CN)$_{2}$ (section \ref{sec:ternaries_zinc_cyanide}) & 250 \\
\end{tabular}
\end{ruledtabular}
\end{table}

\texttt{repose}, \texttt{wobble}, and \texttt{ramble} are \texttt{Fortran} codes, supplied by the EDDP package. They are not fully-featured packages, but rather designed to be workhorses which carry out basic tasks and can easily be integrated into more complex workflows. The EDDP-enabled version of LAMMPS permits a wider range of dynamical simulations to be conducted. The EDDPs are implemented through a new `pair-style' and currently permit OpenMP parallelisation over a single node. Melting point calculations for lead (see Sec. \ref{sec:lead_phase_diagram}) have been computed separately and validated between LAMMPS and \texttt{ramble}.

Both \texttt{wobble} and \texttt{ramble} can automatically construct nearly cubic supercells with a specified number of atoms given any primitive, or base, unit cell. While methods exist to generate such supercells (see e.g. Erhart \textit{et al.}\cite{Erhart2015}), the approach outlined below is more robust to varying orientations of the basis vectors, and therefore more suitable to high-throughput calculations.

A structure with a primitive unit cell basis, $\mathbf{S_p}$, can be transformed to a supercell basis, $\mathbf{S_s}$, by matrix multiplication,
\begin{equation}
    \mathbf{S_s} = \mathbf{P} \mathbf{S_p},
\end{equation}
where $\mathbf{P}$ is an integer matrix. The aim is to choose $\mathbf{P}$ which generates the `most cubic' $\mathbf{S_s}$ for a given $\mathbf{S_p}$. In our approach, we define a `cost function', $\Delta$, which is agnostic to the unit cell orientation and contains the lattice parameters, $x_i$, angles, $\alpha_i$, and the target number of unit cells, $N_\textnormal{target}$,
\begin{equation}
	\Delta=\frac{\sqrt{\sum_i \left( x_i - \sum_j \frac{x_j}{3} \right)^2 }}{\sum_i \frac{x_i}{3}}\\
	- \sqrt{\sum_i \cos^2\alpha_i} + \frac{|N_\textnormal{target} - \det{(\mathbf{P})}|}{N_\textnormal{target}}.
\end{equation}
This cost function is minimised stochastically by applying random changes of $+1$, $0$, or $-1$ to the elements of $\mathbf{P}$. If this change lowers the cost function, then the new $\mathbf{P}$ is accepted. This procedure is continued until a stable $\mathbf{P}$ is found which will contain close to the target number of atoms and have a close-to-cubic lattice shape.

The method described here is distinct from the method of Erhart \textit{et al.}\cite{Erhart2015}, which instead attempts to minimise the off-diagonal terms of $\mathbf{S_s}$ and so is sensitive to the specific orientation of $\mathbf{S_p}$. Our method is well suited to high-throughput calculations which may encounter non-standard unit cell orientations.

\begin{table*}
\caption{\label{tab:eddp_knobs}Some of the most important parameters for obtaining high-quality EDDPs, along with commonly suitable values and strategies for setting them appropriately for particular cases.}
\begin{ruledtabular}
\begin{tabular}{cccc}
Parameter & Common values & How to determine it & Comments\\\hline
Radial cutoff $r_{c}$ & $\sim2\times$bond length & Simple hyperparameter search & It is essential to use a sufficiently large $r_{c}$. \\
&&& For $r_{c} > 5$ \AA, it may be beneficial \\
&&&to increase the number of polynomials slightly. \\
\\

Size of input dataset & 1000 random \& 5500 (shaken) & MAE$_\text{testing}$ >> MAE$_\text{training}$ & For systems with multiple elements, \\
& local minima structures & indicates that more data is necessary & a larger dataset will usually be required. \\
& (for single-element systems) &  & \\
\\
Number of polynomials $M$ & $5$-$7$ & Simple hyperparameter search & \\
\\
Range of minimum & Should encapsulate the & Known experimental structures or & It is important that this is in the right\\
atomic separations \& & range exhibited by the & simple AIRSS-\textsc{CASTEP} search &  range; the exact details are not crucial.\\
volumes per atom & local minima structures & & \\
\end{tabular}
\end{ruledtabular}
\end{table*}

\subsection{Benchmarks}
\label{sec:benchmarks}

Table \ref{tab:benchmarks} summarises the computational performance of the different potentials used in this work. The timing tests have been carried out on a single core of an Apple M1 Pro chip. The systems for which the timings were measured were chosen to reflect the practical applications in section \ref{sec:case_studies}. All benchmarks are 100-step MD simulations. The Pb benchmark is a solid-liquid coexistence simulation at $0$ GPa; the ScH$_{12}$ benchmark was carried out at 300 GPa; and the lcs-Zn(CN)$_{2}$ benchmark is a MOF at 0 GPa. The different densities explain the large spread of timings. The open structure of the Zn(CN)$_{2}$ MOF system gives it the lowest number density of all the systems studied. It therefore yields the fastest benchmark, in spite of the large feature vectors needed to describe the three elements. Conversely, ScH$_{12}$ at $300$ GPa is significantly denser than all other systems, leading to a larger computational overhead for calculation of the feature vectors.

For carbon, a cubic cell with 800 atoms at a density of 2 g/cm$^{3}$ was constructed in order to enable comparison with the potentials benchmarked by Qamar \textit{et al.}~\cite{Ralf-Drautz-carbon-ace}. We note that the timings presented in table \ref{tab:benchmarks} represent a snapshot of the current performance of the EDDP code suite. Further optimisations, particularly of the feature calculation and GPU porting, are in progress. Nevertheless, the benchmarks demonstrate that the EDDPs are capable of achieving costs competitive with ACE, and an order of magnitude lower than the established methods tested by Qamar \textit{et al.}~\cite{Ralf-Drautz-carbon-ace}. The benchmark suite is available~\cite{EDDP-benchmark}.

\subsection{Best Practices}
\label{sec:best_practices}

The `recipe' for generating an EDDP is straightforward, and the key input parameters have reasonable defaults. For some systems or applications, other values may give better results. In particular, we note that the default radial cutoff, $3.75$ \AA, is chosen to obtain fast potentials most suitable for structure search. If a more accurate and transferable potential is required, a higher cutoff must typically be chosen. This is the case for all examples presented in section~\ref{sec:case_studies}.

Table~\ref{tab:eddp_knobs} summarises (in roughly descending order of impact) some common strategies for obtaining well-behaved and transferable EDDPs. This is not intended to be an exhaustive list; the case studies described in section~\ref{sec:case_studies} present a wide range of different strategies for generating high-quality EDDPs. However, the parameters summarised in table~\ref{tab:eddp_knobs} will provide suitable starting points, especially for those still learning to use EDDPs. All EDDP training parameters can be optimised using hyperparameter searches. This does not require regeneration of the dataset and is hence computationally cheap, especially considering the lightweight nature of the neural network architecture.

\section{Case Studies}
\label{sec:case_studies}

\begin{table*}
\caption{\label{tab:training_parameters}The input parameters chosen for the EDDP training as well as the underlying DFT calculations in the different case studies.}
\begin{ruledtabular}
\begin{tabular}{ccccc}
& \multicolumn{4}{c}{System}\\
Parameter & Carbon & Lead & Scandium hydride & Zinc cyanide\\\hline
\multicolumn{5}{c}{DFT Parameters} \\
Energy cutoff [eV] & 600 & 600 & 600 & 600 \\
$\mathbf{k}$-point spacing [$\times 2\pi$ \AA$^{-1}$] & 0.05 & 0.017 & 0.03 & 0.04\\
XC functional & PBE~\cite{PBE} + TS dispersion correction~\cite{Tkatchenko2009} & PBEsol~\cite{PBEsol} & PBE & PBE + TS dispersion correction\\
\hline
\multicolumn{5}{c}{Structure Building Parameters}\\
Minimum interatomic distance [\AA] & $1$-$2$ & $2$-$4$ & Sc-Sc: $1.79$-$2.79$ & $0.4$-$4$ \\
& && H-H: $0.60$-$1.46$ & \\
&&& H-Sc: $1.20$-$2.32$ & \\
Volume per atom [\AA$^{3}$] & $4$-$11$ & $25$-$35$ & Sc: $6.9$-$16.1$ & $25$-$100$\\
&&& H: $1.2$-$4.3$ & \\
\hline
\multicolumn{5}{c}{EDDP Training Parameters}\\
$r_{c}$ [\AA] & $5.5$ & $7$ & $6$ & $6$\\
Number of exponents & $5$ & $6$ & $5$ & $5$\\
Highest body order & $3$ & $3$ & $3$ & $3$ \\
Number of nodes in hidden layer & $5$ & $5$ & $5$ & $5$\\
Number of random structures & $1000$ & $1000$ & $10 000$ & 20000\\
Number of cycles & 5 & 5 & 5 & 5\\
Number of local minima per cycle & $100$ & $100$ & $100$ & 110 \\
Number of shakes per local minimum & $10$ & $10$ & $10$ & 10 \\
Total number of structures & $6500$ & $6500$ & $15500$ & 32100 \\
Pressure range [GPa] & $0$ & $0$-$50$ & $50$-$400$ & $0$-$10$ \\
Number of EDDPs generated & $250$ & $280$ & $256$ & $405$\\
\hline
Number of EDDPs selected by NNLS & 19 & 14 & 19 & 45\\
MAE [meV/atom] & $64.88$ & $3.14$ & $25.33$ & $44.34$\\
\end{tabular}
\end{ruledtabular}
\end{table*}

In this section, we present a series of case studies highlighting different features and capabilities of the EDDPs. Carbon, with its diverse range of allotropes and chemical bonds, is an interesting and difficult test case for machine-learned potentials. Our EDDPs generate smooth potential energy surfaces in good agreement with DFT for a wide variety of different structures. Lead, as a heavy metal, is expensive to describe in DFT, due to the need for both a dense $\mathbf{k}$-point grid and the inclusion of spin-orbit coupling (SOC). Experimentally lead is well-understood, making it a suitable test case. Training an EDDP on high-quality first-principles data enables structure searching and the successful reproduction of its known phase diagram up to $20$ GPa.

Two further case studies, the scandium hydride system under pressure and a MOF, Zn(CN)$_2$, demonstrate that EDDPs are able to describe systems with more than a single element. For scandium hydride, a single potential can be used to robustly search a range of stoichiometries, across a pressure range of $50-400$ GPa. MD simulations reveal a prediction of superionicity at 350 GPa and 600 K in this system. In Zn(CN)$_2$, we compute a negative thermal expansion in good agreement with the experimentally observed value, without prior knowledge of the stable structures. These two case studies also reveal good practices for training an EDDP for variable stoichiometry.

All first-principles calculations presented in this work were carried out using the \textsc{CASTEP} plane-wave DFT package~\cite{CASTEP}. The EDDP package interfaces with the plane-wave DFT calculations via the AIRSS package, and only requires single-point energies. It can be straightforwardly adapted to different DFT codes, or different total energy methods altogether. The data that supports these case studies is available on MaterialsCloud~\cite{Materials-cloud-data}.

\subsection{\label{sec:smoothness_transferability_carbon}Smoothness and Transferability: Carbon}

\begin{figure}[ht]
\includegraphics[width=1.0\linewidth]{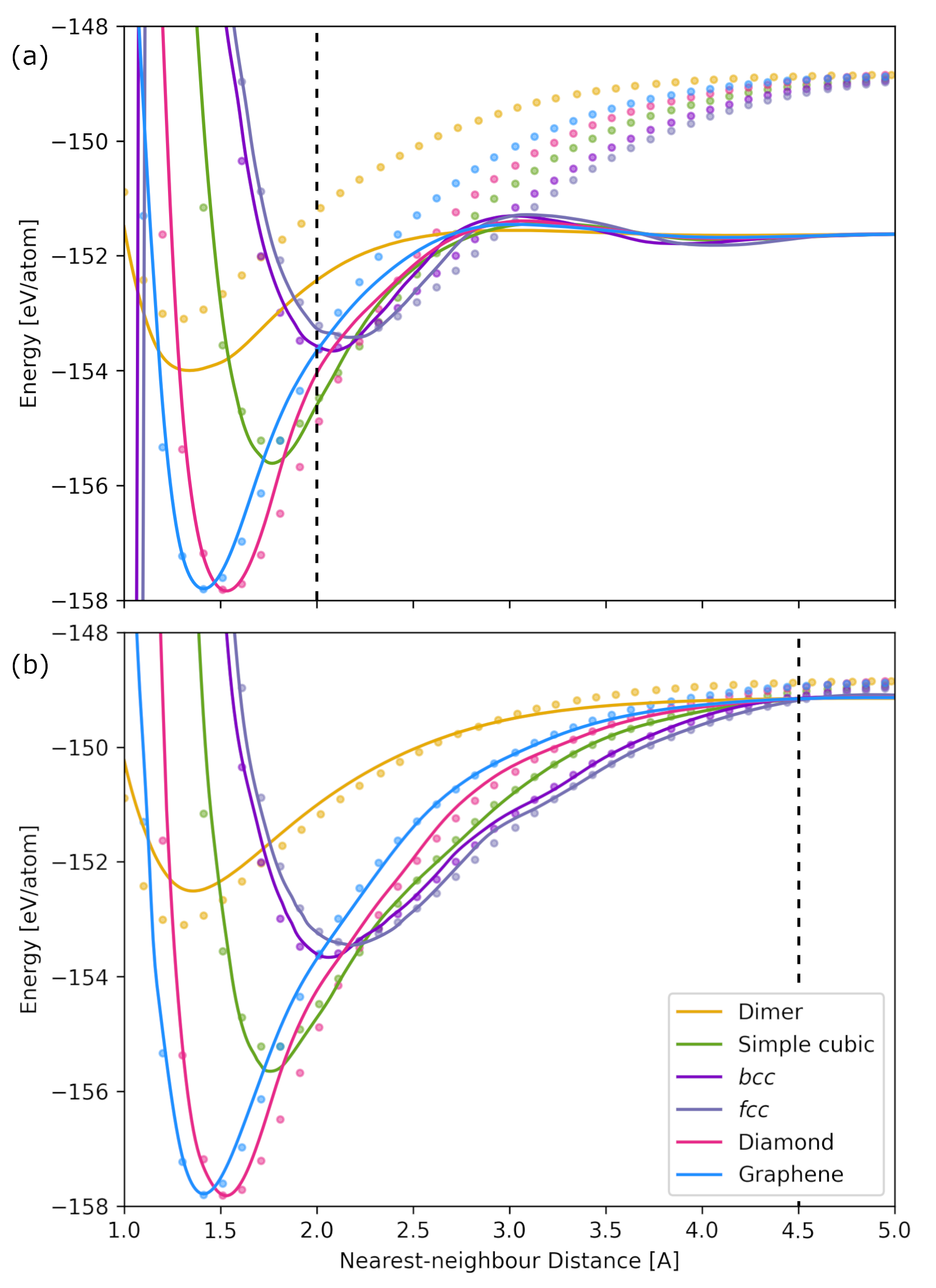}
\caption{\label{fig:carbon_pes} Potential energy curves of a number of carbon structures as a function of distance for (a) an EDDP trained on data with only short minimum atomic separations and (b) an EDDP with longer-range data added. Lines represent EDDP curves and dots represent DFT calculations. Dashed vertical lines indicate the largest nearest-neighbour distance in the respective training sets.}
\end{figure}

In the context of structure prediction, a machine-learned potential should be smooth, transferable, and robust. A smooth potential is well suited to global optimisation, allowing physically sensible energy minima to be accessed from less physical initial conditions. The potential must also be transferable to describe these diverse initial conditions with sufficient accuracy. 

Carbon is the essential element of organic life~\cite{Nature-building-blocks} and a building block in many diverse and complex molecules. The allotropes of pure carbon are also of continued scientific interest~\cite{SACADA-Carbon-database, Chris-Carbon-group-graph-theory} and machine-learned potentials have been applied to accelerate their prediction and characterisation~\cite{Carbon-GAP-20}. A difficult and illuminating test is to generate a potential for carbon which is both smooth and transferable to systems far away from idealised diamond- or graphite-like structures. While several transferable Carbon potentials have been developed, they vary significantly in smoothness, as discussed by Qamar \textit{et al.} in the comparison to their smooth ACE potential~\cite{Ralf-Drautz-carbon-ace}, and therefore suitability for structure search.

To highlight the comparative smoothness and transferability of a well-trained EDDP,  we train and compare two separate potentials for carbon. The training set for the first potential contains only structures with minimum nearest-neighbour atomic separations between $1$ and $2$ \AA. The second potential is trained on a set which contains a larger range of nearest-neighbour distances.

Table \ref{tab:training_parameters} summarises the parameters governing the generation of the first potential. $1000$ random structures were used in the training set with minimum interatomic distances of $1$-$2$ \AA. $5500$ local minima and shakes were added to the set in $5$ cycles of iterative training. 250 EDDPs were generated from this training data. The NNLS reduced this to 19 for the final potential, which had a testing mean average error (MAE) of $64.88$ meV/atom. This comparatively large error (the largest of all systems in this section) results from the diversity of carbon systems. In many EDDP training sets, a large segment of the error arises from badly described high-energy structures. In this case, however, the test set error when considering only structures within $1$ eV/atom of the ground state is still $63.55$ meV/atom.

The second potential was trained using the same parameters, but $500$ randomly generated structures with minimum C-C separations between $1.75$ and $4.5$ \AA$ $ and volumes per atom from $40$-$70$ \AA$ $ were added to the dataset generated above. $100$ identical copies of the isolated carbon atom were also included. This structure is added multiple times to ensure a suitable weighting in the fitting process. We emphasise that other than this, the dataset remained completely random. For instance, the isolated carbon dimer is not included in the training set. 250 EDDPs were again generated from this training data, of which the NNLS selected 14 for the final potential. The testing mean average error was slightly higher at $73.25$ meV/atom in this case. The testing error over structures no more than $1$ eV/atom from the ground state is reduced slightly, to $71.35$ meV/atom.

The potential energy curve for the EDDP trained on smaller interatomic distances is shown in Fig.~\ref{fig:carbon_pes}(a). With exception of the dimer, the potential accurately describes all structures around their respective minima, including the simple cubic crystal -- an uncommon structure for carbon at low pressure. Any unphysical oscillations in the curves are minor, and their smoothness for atypical structures is enhanced dramatically compared to the potentials tested by Qamar \textit{et al.}\cite{Ralf-Drautz-carbon-ace} - with the exception of ACE. This is true even in regions with nearest-neighbour distances larger than $2$ \AA$ $, yielding a `sensible' if not quantitatively correct extrapolation into this region. A shallow false minimum exists between $3.5$ and $4$ \AA$ $, but this can be avoided in structure searches applying chemically sensible constraints.

Fig.~\ref{fig:carbon_pes}(b) shows that by augmenting the dataset, the correct description of longer-range interactions can be achieved. The curve now accurately predicts energies at long range, making the potential more suitable for applications such as surface or defect structure searches. No false minima exist in this EDDP.

From this example, we see that that EDDPs are, by construction, smooth. We also show that it is trivial to produce an EDDP which is suitable for the most important regions of structure space - around the minima of the potential energy curves - and straightforward to improve the description of longer-range interactions.

In order to enable comparisons with other carbon potentials in the literature\cite{Ralf-Drautz-carbon-ace, Carbon-PANNA}, we calculated the lattice constant and bulk modulus of diamond, using both the EDDP used to generate Fig.~\ref{fig:carbon_pes}(b), and a new EDDP trained (with the same hyperparameters) on the dataset curated by Qamar \textit{et al.}\cite{Ralf-Drautz-carbon-ace}. For the sake of brevity, these two potentials are referred to as EDDPs I and II respectively. 

\begin{table}
\caption{\label{tab:carbon_properties}The properties of Carbon calculated by a variety of different machine-learned potentials. Values in brackets are the corresponding DFT values.}
\begin{ruledtabular}
\begin{tabular}{ccc}\\
Potential & Lattice Parameter [\AA] & Bulk modulus [GPa]\\\hline
EDDP I & 3.502 (3.572) & 580 (431) \\
EDDP II & 3.570 (3.572) & 433 (431) \\
ACE\cite{Ralf-Drautz-carbon-ace} & N.A. & 429 (435) \\
PANNA\cite{Carbon-PANNA} & 3.576 (3.584) & 431 (425) \\
\end{tabular}
\end{ruledtabular}
\end{table}

Table \ref{tab:carbon_properties} summarises the properties predicted by these EDDPs, as well as those predicted by ACE\cite{Ralf-Drautz-carbon-ace} and PANNA\cite{PANNA-potential, Carbon-PANNA}, and compares them to the corresponding DFT values. EDDP I disagrees most strongly with DFT. The dataset used to generate EDDP I is much smaller than those used for the ACE and the PANNA potentials, and the EDDP is trained only on the energies. It is designed for transferability and cost-effectiveness to accelerate searches and is not in any way refined to predict properties for diamond. EDDP II, however, predicts properties in excellent agreement with DFT. This demonstrates that the EDDP formalism can produce significantly more accurate potentials when necessary.

Overall, this brief comparison highlights the flexibility of the EDDPs, which can produce both cheap and transferable potentials for searching, as well as accurate potentials competitive with state-of-the-art methods. The balance between the two extremes can be tuned to fulfil the specific task at hand by adapting the training set and hyperparameters.

\subsection{\label{sec:elements_lead}(Heavy) Elements: Lead}

Lead (Pb), with atomic number 82, is relatively common in nature. It is the the heaviest element with stable isotopes and the end-product of the three most common radioactive decay chains~\cite{Radioactive-decay}. Its ground state structures have been studied extensively up to hundreds of GPa. At ambient temperature, the sequence is \textit{fcc} $\rightarrow$ \textit{hcp} $\rightarrow$ \textit{bcc} with increasing pressure~\cite{McMahon-Nelmes-metal-structures}. The melt curve has also been investigated both experimentally~\cite{Lead-melting-DAC, Lead-melting-shockwave, Lead-melting-Bridgman} and with DFT~\cite{Lead-melting-DFT}.

Spin-orbit coupling, a relativistic effect, is an important contribution to the behaviour of heavy elements, as it is proportional to the nuclear charge~\cite{Relativistic-QM}. In Pb, it shifts the phase boundaries, renormalises the phonons, and increases the superconducting transition temperature~\cite{Pb-phases-SOC, Pb-SOC-phonons, Pb-SOC-superconductivity}. However, the additional computational expense associated with SOC~\cite{Giustino-DFT} prohibits high-throughput DFT searches and \textit{ab initio} MD. This difficulty is compounded by the need for dense $\mathbf{k}$-point grids to obtain converged DFT energies in metals.

Accelerating such calculations is therefore an important application of machine-learned potentials, enabling structure prediction for previously inaccessible regions of the periodic table. Here, we train an EDDP for Pb, with SOC included in the training dataset, and deploy it for structure searching, phonon calculations, and MD simulations. The resulting pressure-temperature phase diagram up to $20$ GPa and $2500$ K shows good agreement with available experimental data. Compared to DFT, the reductions in computational cost amount to 4-5 orders of magnitude for the structure searches and 7 orders of magnitude for the lattice dynamics.

\subsubsection{\label{sec:lead_training}Training}

Table \ref{tab:training_parameters} summarises the training parameters used to generate the Pb potential. A norm-conserving pseudopotential (NCP) with two projectors on both the $6$s and $5$d states, and one on the $6$p state, was constructed. The \textsc{CASTEP} pseudopotential string is \texttt{3|2.4|12|14|26|60NN:61N:52NN(qc=7)}. The potential's error over the whole testing set is $3.14$ meV/atom. When considering only the low-energy structures at most 1 eV/atom from the ground state, this is further reduced by more than half to $1.31$ meV/atom. A separate potential was trained on DFT energies not including SOC. In this case, the default C19 ultrasoft pseudopotential was used but all other parameters were left unchanged. All calculations presented below include SOC unless otherwise stated.

\subsubsection{\label{sec:lead_searching_enthalpy}Benchmark Search and Enthalpy Curves}

Table~\ref{tab:lead_searches} compares the number of $8$-atom Pb structures found with DFT and the EDDP per hour. The EDDP is very nearly $5$ orders of magnitude faster than DFT, and easily finds the favourable \textit{fcc} and \textit{hcp} structures. Even accounting for the computational expense of generating the DFT training data --- which uses only single-point energy calculations --- the total computational cost of a thorough structure search is significantly lower with the EDDP. This underlines starkly the challenge of carrying out \textit{ab initio} structure prediction for heavy metals at the level of accuracy required.

\begin{table}
\caption{\label{tab:lead_searches}The number of $8$-atom structures found per hour per core in the SOC energy landscape of lead using both DFT and the EDDP on $112$ CPU cores.}
\begin{ruledtabular}
\begin{tabular}{cc}
Method & Number of structures per hour per core \\
\hline
DFT & $0.0004$ \\
EDDP & $37.107$ \\
\end{tabular}
\end{ruledtabular}
\end{table}

\begin{figure}[htpb]
\includegraphics[width=1.0\linewidth]{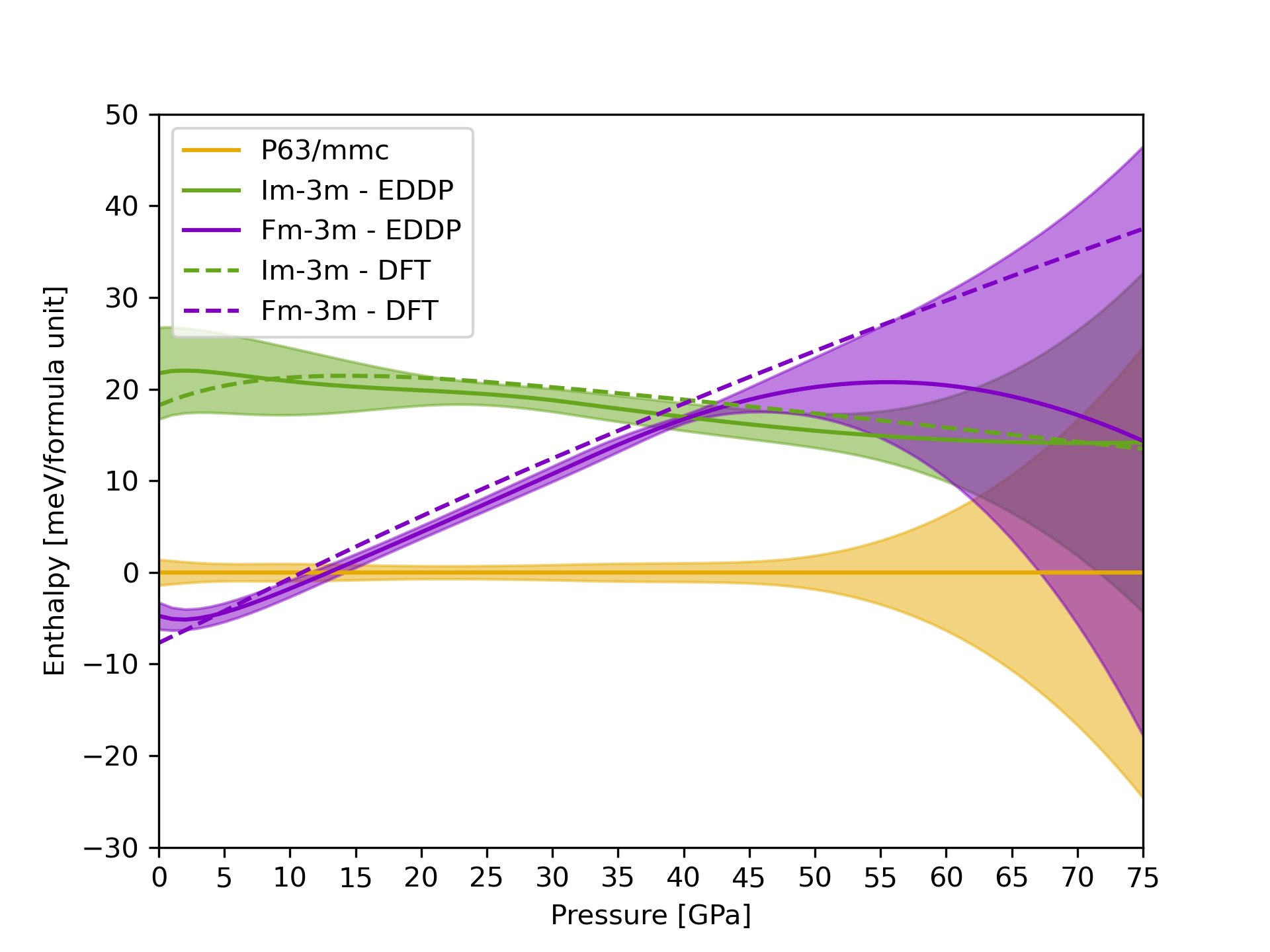}
\caption{\label{fig:pb_enthalpy_dev} Enthalpies and deviations up to $75$ GPa of the \textit{fcc} and \textit{bcc} structures of lead, relative to the \textit{hcp} enthalpy. The shaded regions indicate the EDDP ensemble deviations.}
\end{figure}

Next, we use DFT and the EDDPs to calculate the static-lattice enthalpies of the three lowest-pressure known ground states of lead, \textit{fcc}, \textit{hcp}, and \textit{bcc}, up to $75$ GPa. The results are shown in Fig.~\ref{fig:pb_enthalpy_dev}, relative to the \textit{hcp} enthalpy. The EDDP shows good agreement with DFT for all structures' enthalpies up to $\sim 50$ GPa and correctly predicts the \textit{fcc} $\rightarrow$ \textit{hcp} transition pressure. Fig.~\ref{fig:pb_enthalpy_dev} also shows the ensemble deviation of the energy estimations as a shaded region around each curve. Beyond $\sim50$ GPa (the upper limit of the training data), the DFT and EDDP energies begin to diverge, just as the deviations increase. This indicates that ensemble deviations are good predictors of the model's uncertainty.

The errors in the EDDP predictions are slightly higher at pressures near the edge of the training data ($1.5$-$2$ meV below $\sim5$ GPa and above $\sim45$ GPa, versus $1$ meV in the range of $5$-$45$ GPa). This results from the training set being better constrained in the intermediate region of the training pressure range and suggests that training should include a wider range of pressures than those to be studied. In addition, the \textit{bcc} structure of Pb is less well-described below $15$ GPa, since it is unstable in this regime. Hence it is not found in the training set at the corresponding volumes. The ensemble deviations faithfully reflect these features.

\subsubsection{\label{sec:lead_phonons}Lattice Dynamics}

In this section we compare phonon calculations of the ground state (\textit{fcc}) lead structure between DFT and the EDDP at $0$ GPa. The DFT calculations use the finite-difference \texttt{Caesar} code, which accelerates phonon calculations by constructing a number of small non-diagonal supercells~\cite{Tomeu-nondiagonal-supercells}. An $8\times 8 \times 8$ $\mathbf{q}$-point grid is found necessary to obtain a phonon dispersion in qualitative agreement with experiment.

Phonon calculations using DFT are very sensitive to small changes in the calculated electronic structure. To obtain well converged calculations, we required higher precision parameters. A plane-wave cutoff of $1000$ eV, $\mathbf{k}$-point spacing of $0.01\times 2\pi$ \AA $^{-1}$ and an self-consistent field tolerance of $10^{-8}$ eV/atom were used. Furthermore, we increased the `standard' and `fine' Fourier transform grids to include plane waves up to $2\times G_\text{max}$ and $2.5\times G_\text{max}$, where $G_\text{max}$ is the diameter of the reciprocal space cut off sphere.

\begin{figure}[htpb]
\includegraphics[width=0.95\linewidth]{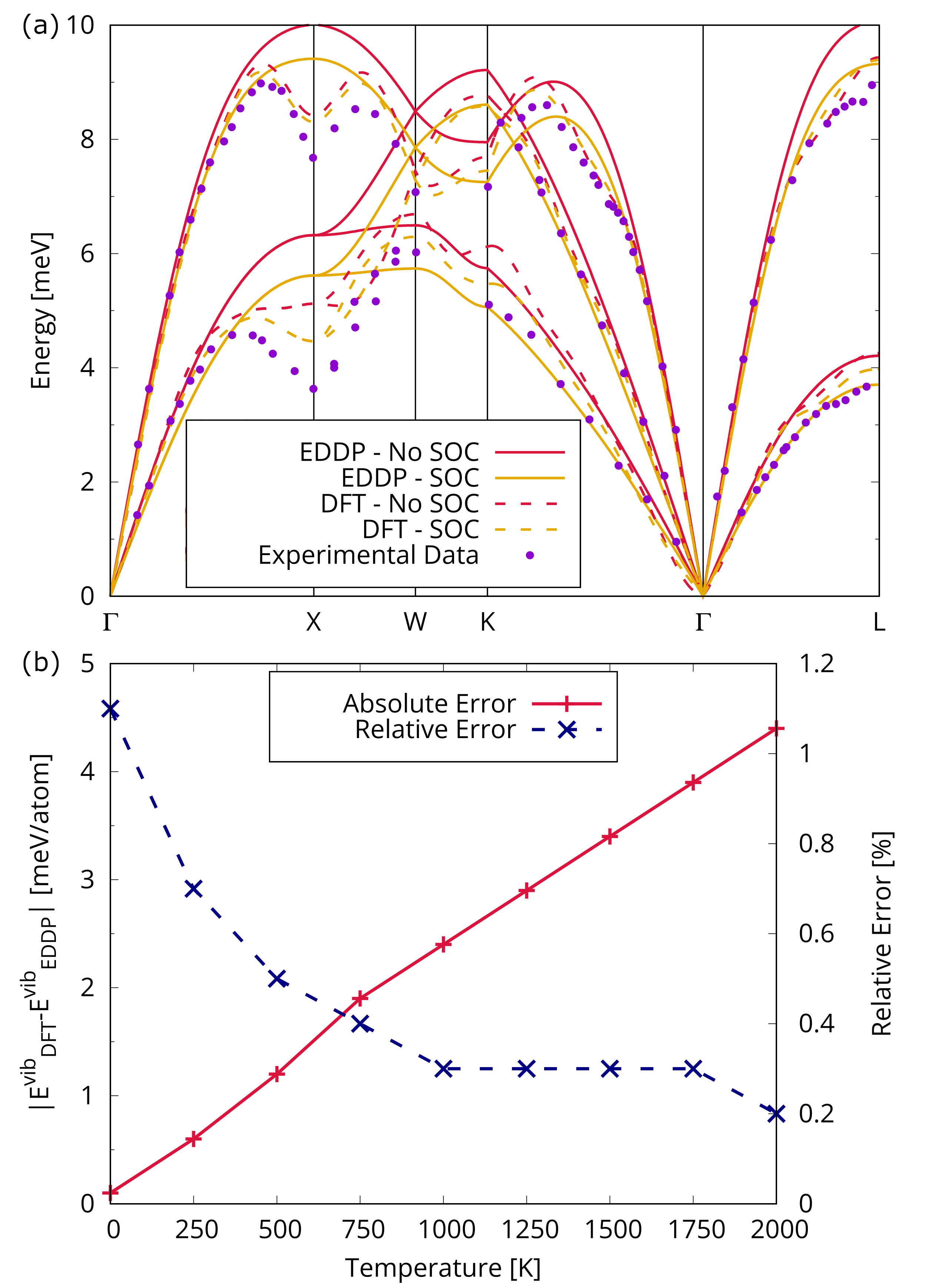}
\caption{\label{fig:pb_phonons} (a) The phonon dispersion of \textit{fcc} lead, calculated without and with SOC using DFT as well as the EDDP, and compared to experimental data from Brockhouse \textit{et al.}\cite{Lead-phonons-experiment}. (b) The error in the EDDP prediction of the vibrational energy compared to DFT as a function of temperature.}
\end{figure}

Fig.~\ref{fig:pb_phonons}(a) compares the phonon dispersions calculated using the EDDP with those from DFT and experiment. The theoretical dispersions are calculated both with and without SOC. In both cases the EDDPs reproduce the main features of the phonon dispersion, with particularly good agreement at low frequencies. We find poorer agreement between the EDDP and DFT results when SOC is not included in the calculations.

The EDDP fails to capture the Kohn anomaly at the X point, both with and without SOC. Kohn anomalies are uncommon, and result from a rapid change of the screening of certain lattice vibrations by the electrons~\cite{Kohn-anomaly}. Kohn anomalies occur along wavevectors which connect different momenta on the Fermi surface, the so-called nesting vectors. It is perhaps not surprising that a machine-learned potential, which integrates out the details of the electronic structure, struggles to capture such a subtle effect of the electron-phonon coupling. Kohn anomalies are strongly localised in reciprocal space; in real space they only occur for very specific extended configurations. It appears that the EDDP simply averages over this anomalous point in the electronic structure. This is not a fundamental limitation of machine-learned potentials. Recently, Wang \textit{et al.} presented an MTP which successfully reproduced the Kohn anomaly in $\alpha$-Uranium~\cite{Uranium-phonons-MTP}. This potential was trained on data collected from MD runs of a supercell of $\alpha$-Uranium at a range of pressures and temperatures.

While the EDDP does not capture all the details of the phonon dispersion, it produces good agreement for the total vibrational energy (evaluated by integrating over all phonon modes). This is essential for accurate thermodynamic calculations. Fig.~\ref{fig:pb_phonons}(b) summarises the error in the vibrational energies predicted by the EDDP compared to those from DFT, including SOC. The EDDP results are within $5$ meV/atom of the DFT prediction for all temperatures considered. The zero-point energies agree almost exactly, with an error of $0.1$ meV/atom. At $2000$ K, the EDDP phonon energy is $4.4$ meV/atom lower than that from DFT. The relative error decreases with temperature and is less than $1$\% from $250$ K. This is remarkable precision for a potential trained only on the energies, at a fraction of the computational cost of the DFT calculations. For these phonon calculations, $42882$ core hours were required for DFT compared to only $0.01$ with the EDDP, a speed-up of more than $6$ orders of magnitude.

\subsubsection{\label{sec:lead_phase_diagram}Phase Diagram up to 20 GPa and 2500 K}

\begin{figure*}[!bht]
\includegraphics[width=0.95\linewidth]{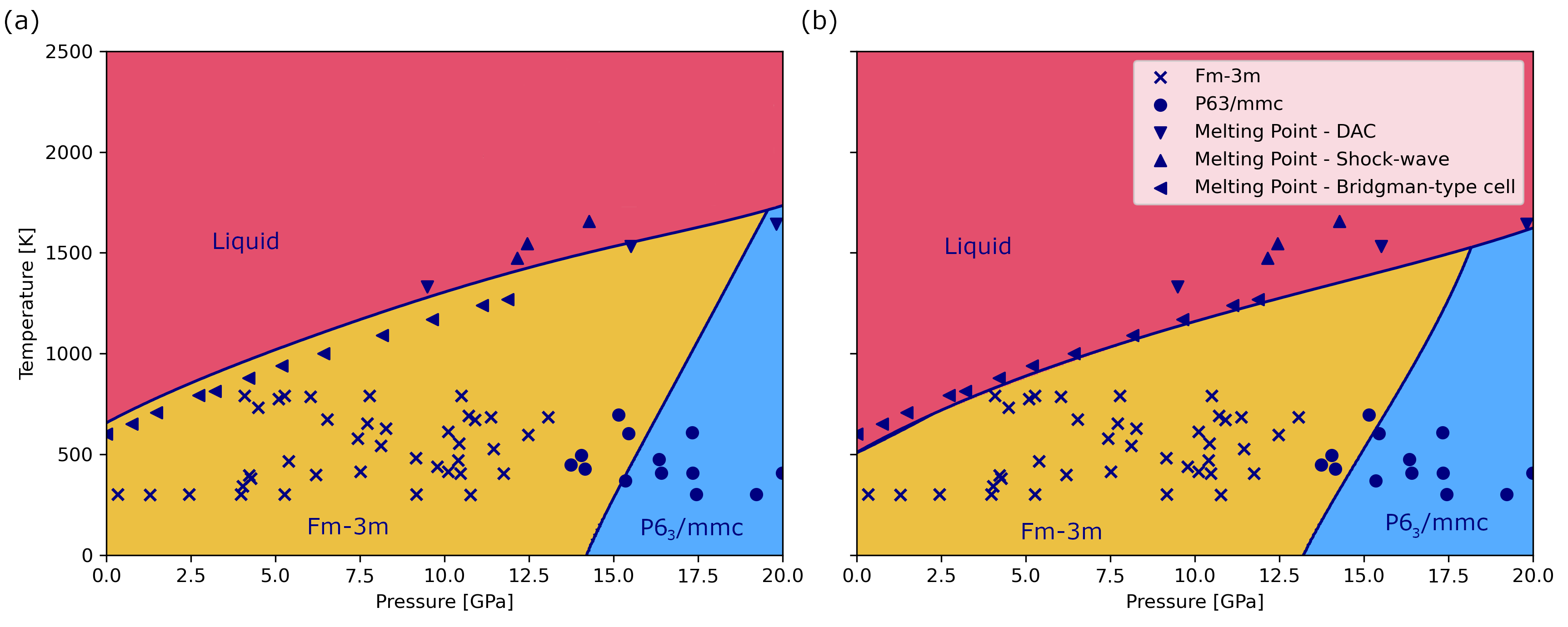}
\caption{\label{fig:pb_eddp_pd} Phase diagram of Pb up to $20$ GPa and $2500$ K, calculated using EDDPs trained on DFT data (a) without and (b) with SOC. The symbols indicate experimental data obtained from a variety of sources. The data for the \textit{fcc} and \textit{hcp} phases are taken from Kuznetsov \textit{et al}~\cite{Lead-FCC-HCP-experiment}. For the melting curve, the DAC data are from Godwal \textit{et al.}~\cite{Lead-melting-DAC}, the shock-wave data from Partouche-Sebban \textit{et al.}~\cite{Lead-melting-shockwave}, and the Bridgman cell data from Errandonea~\cite{Lead-melting-Bridgman}.}
\end{figure*}

The phase diagram was computed beginning with static-lattice enthalpies for the \textit{fcc} and \textit{hcp} phases, calculated in $1$ GPa steps from $0$ to $20$ GPa. For the relaxed structures at each of these pressures, a harmonic phonon calculation was then carried out. Thermal expansion was accounted for using the quasi-harmonic approximation (QHA)~\cite{Dove-Lattice-Dynamics, Quasi-harmonic-approximation}. This process requires at least several dozen phonon calculations. At the DFT level of theory, a single phonon calculation of the requisite quality would take hours of wallclock time on a CPU node. Using the EDDP, all calculations are completed in a matter of minutes.

The melting points (T$_{\text{m}}$) are calculated using the coexistence molecular dynamics method~\cite{Coexistence-melting-I, Coexistence-melting-II, Coexistence-melting-III, Coexistence-melting-IV}. Simulation cells containing $1000$ `solid' and $1000$ `molten' atoms are used. Their motion is simulated using NpH molecular dynamics at pressure steps of $2.5$ GPa, covering again the pressure range from $0$ to $20$ GPa. The runs are allowed to equilibrate, and the temperature is then averaged over at least $75$ ps. The melt curve is interpolated between these explicit calculations using a polynomial fit.

Fig.~\ref{fig:pb_eddp_pd} shows the pressure-temperature phase diagram of lead (a) without and (b) with SOC, including experimental data from a variety of sources~\cite{Lead-melting-DAC, Lead-melting-shockwave, Lead-melting-Bridgman, Lead-FCC-HCP-experiment}. In both cases the P-T trends of the phase boundaries are in good agreement with available data. Including SOC causes a shift in the transition temperatures and pressures: $T_m$ is decreased by $\sim 100$ K and the \textit{fcc} $\rightarrow$ \textit{hcp} transition pressure is decreased by $\sim 2$ GPa.

For $T_m$, this leads to better agreement with the Bridgman-type cell melting point measurements of Errandonea~\cite{Lead-melting-Bridgman} (the most modern experimental data available). The reduction of $T_{m}$ when SOC is taken into account arises from the softening of the phonon modes~\cite{Lindemann-criterion, Lindemann-Gruneisen-Laws}. This relationship is more pronounced with the EDDP than with DFT (Fig. \ref{fig:pb_phonons}), suggesting the SOC-induced shift in $T_m$ is likely overestimated with the EDDP. The EDDP overestimates the experimental \textit{fcc} $\rightarrow$ \textit{hcp} transition pressure in both cases, but the disparity is reduced to $\sim 1$ GPa with SOC included. The \textit{fcc}-\textit{hcp}-liquid triple point is located at around 18.1 GPa and 1528 K. This exceeds the estimate of Errandonea, who places the triple point at 15 GPa and 1480 K~\cite{Lead-melting-Bridgman}.

In summary, an EDDP has been generated that can successfully cover the pressure range from $0$ to $50$ GPa in lead. The potential gives phonon dispersions in qualitative agreement with experiment and DFT, but at specific $\mathbf{k}$-points fails to capture all of the electronic structure effects seen in DFT. Vibrational energies are nevertheless reproduced with errors of less than $5$ meV/atom for the \textit{fcc} crystal even at $2000$ K. Relative errors are around $1$\%. The reduction in computational cost offered by the EDDP compared to DFT allows for prediction of the phase diagram of lead in the $0$ to $20$ GPa pressure range. The coexistence MD required the EDDP, trained only on a variety of small crystal structures, to adequately describe $1000$-atom solid phases, the liquid, and the interaction between the two. The resulting melting curve is in good agreement with experiment. 

\subsection{Binaries: Scandium Hydride}
\label{sec:binaries_ScH}

Superhydrides have garnered substantial interest due to their intriguing characteristics, including high temperature superconductivity and hydrogen diffusivity~\cite{Chris-superconducting-hydride-review}. A prominent example is LaH${}_{10}$, which has been shown in experiment to exhibit a superconducting critical temperature above 260 K at 200 GPa~\cite{Eremets-LaH10, Hemley-LaH10-experiment, Chris-clathrate-hydrides} and is predicted to have a high hydrogen diffusion coefficient of 1.7 x 10$^{4}$ cm$^{2}$/s at 170 GPa and 1500 K\cite{LaH10-Russell,LaH10-Paul}. 

Structure prediction has played an important role in the superhydride story. In fact, silane was the first system studied with AIRSS~\cite{Chris-Silane}, followed shortly by aluminium hydride~\cite{Pickard2007-aluminium-hydride}. In recent years, almost the entire periodic table of binary hydrides have been the target of a structure search~\cite{Chris-superconducting-hydride-review}, saturating what can be found using \textit{ab-initio} searches. These studies are often limited partly by the relatively small system sizes accessible using DFT and partly by the limited number of structures calculated during the searches - Have we truly found the minimum energy structure? The lack of certainty with which we can answer this question is perhaps indicated by the discrepancy between the large number of predicted superconducting hydrides and the small number that have been successfully synthesised. To overcome this limitation, larger systems must be explored adequately. A recent structure search on silane with up to 16 f.u. identified a new low enthalpy Pa-3 silane structure, which is not accessible with small-system searches. This highlights the importance of exploring larger systems to identify the true minimum energy structure~\cite{Chris-EDDPs}. Here, we demonstrate how EDDPs can be used to accelerate these searches, include larger unit cells and sample more stoichiometries. 

Scandium hydride has been subject to \textit{ab initio} searches and several stable structures have been predicted, including ScH$_9$ which has a $T_c$ above 160 K at 300 GPa~\cite{ScH-Ashcroft}. We train a single EDDP to cover a range of pressures and stoichiometries. We then use this potential to search a wider range of stoichiometries and larger unit cells to rediscover these stable structures. We  demonstrate that a refined potential can be used for MD simulations of ScH${}_{12}$, which exhibits superionicity.

\begin{figure}[!htb]
\includegraphics[width=1.0\linewidth]{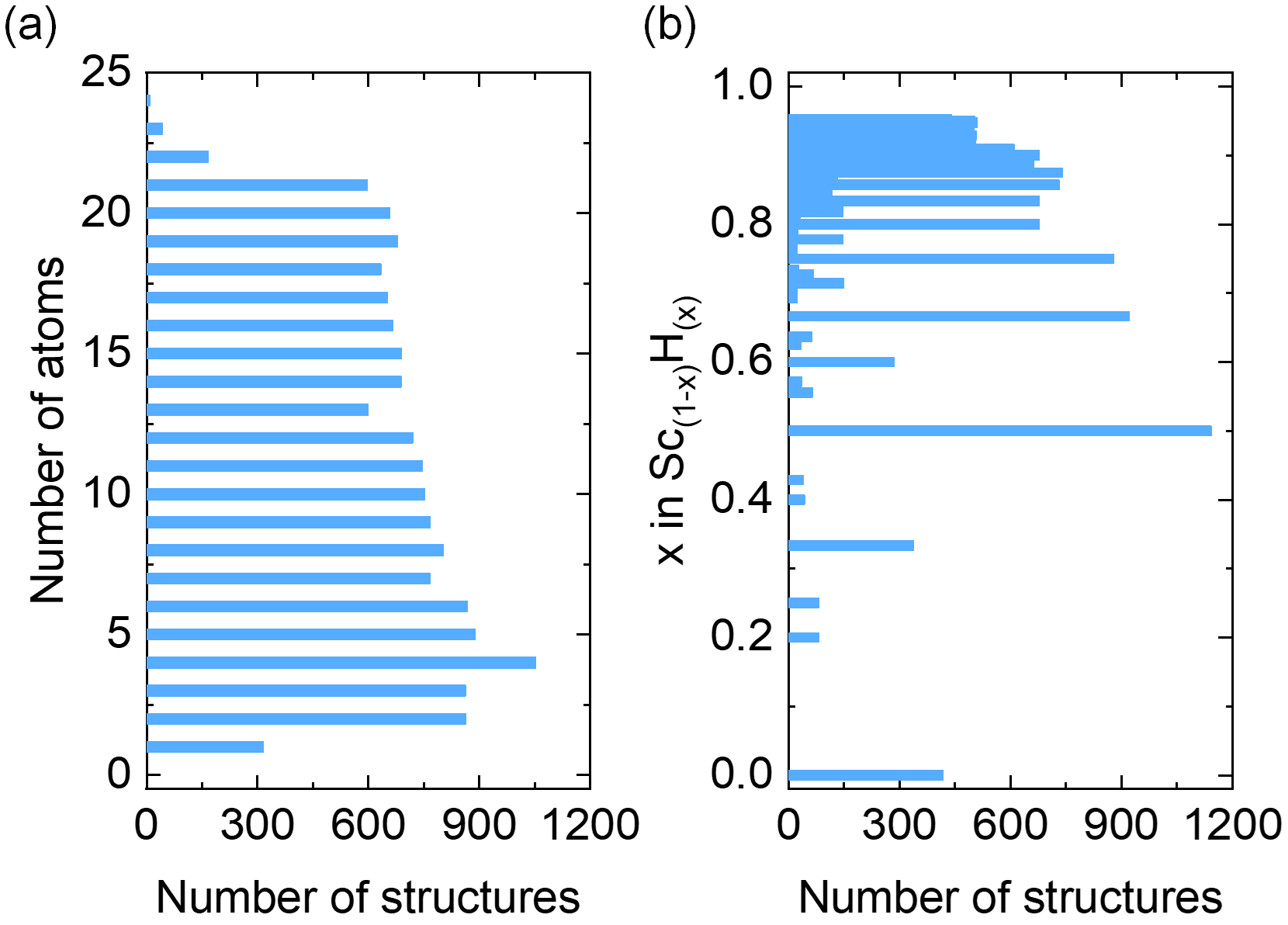}
\caption{\label{fig:sch_dataset} Distribution of structures in the scandium hydride training data; by (a) number of atoms in the unit cell and (b) Distribution of stoichiometries in the dataset.}
\end{figure}

Table \ref{tab:training_parameters} summarises the training parameters used to generate the Sc-H potential. A key step in generating EDDPs for binaries is to include a diverse range of stoichiometries in the training data. In this case stoichiometries of Sc${}_x$H${}_y$, where x = $1$-$4$ and y = $0$-$20$, were included in the dataset with the distribution shown in Fig. \ref{fig:sch_dataset}. 

During iterative training, local minima structures are found and included in the training data. However, the potential can often find structures multiple times --- particularly if the surrounding energy basin is large --- leading to over-fitting and poor transferability. To account for this, only local minima with an ensemble deviation greater than 0.02 eV/atom are added to the dataset. With this constraint, each iteration provides new environments to the training data, improving the potential. Fig.~\ref{fig:sch_dataset} shows the resulting distribution of structures in the training data in terms of both their number of atoms and their stoichiometries. The range of stoichiometries and system sizes generated during training is similar to the range used in \textit{ab initio} searches of binary systems. However, with our method, DFT geometry optimisations are not required. The structures are relaxed entirely using EDDPs. In this way, an AIRSS search is performed during training, with no DFT geometry optimisations.

\subsubsection{Structure Search}

Searching with EDDPs can be carried out using the following three-step procedure: searching with the potential to generate a large number of structures, screening the best structures, and finally performing full DFT geometry optimisations to arrive at DFT ground states. 

\begin{figure*}[bt]
\includegraphics[width=1.0\linewidth]{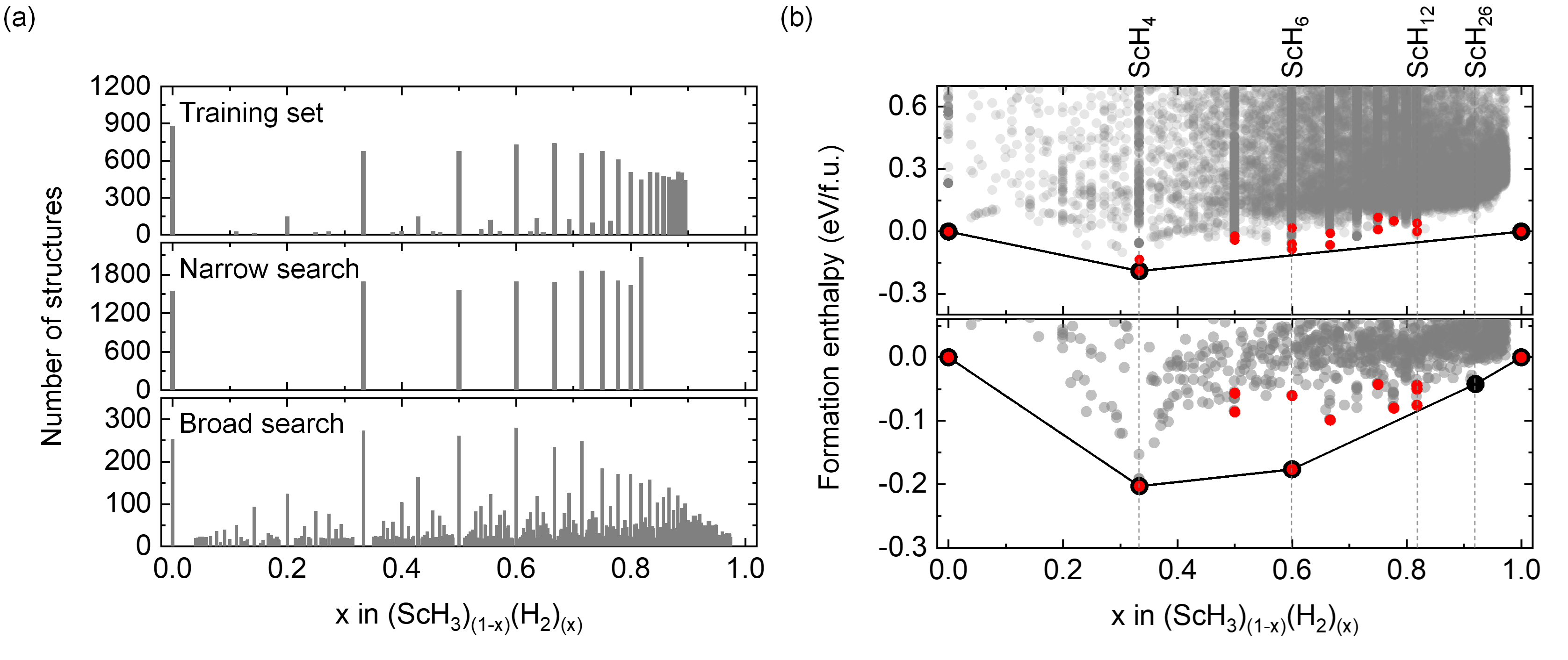}
\caption{\label{fig:sch_convexhull} (a) Stoichiometry distributions of the training dataset and local minimum structures obtained from narrow and broad structure searches. (b) Convex hull diagram of scandium hydride constructed based on (top) DFT single point energies for structures searched at 250 GPa and (bottom) DFT-relaxed structures. Formation enthalpies are calculated with respect to ScH$_3$ and solid H$_2$. Black dots represent stable phases on the convex hull, gray dots represent phases above the convex hull, and red dots represent previously reported structures or similar structures that have been rediscovered.}
\end{figure*}

To search with this potential, we first use AIRSS and \texttt{repose} to search for structures containing Sc${}_x$H${}_y$ where x = $1$ and y = $1$-$12$ with 1-4 f.u. for a deep, narrow search and x = $1$-$8$ and y = $1$-$80$ for a broad search at 250~GPa. We used the same constraints in volumes and atomic separation for searching as we did for training (listed in table~\ref{tab:training_parameters}) but with an additional constraint of between 1 and 48 randomly chosen symmetry operations. This resulted in 39645 local minimum structures. We emphasise here that many stoichiometries in the broad structure search were not present in the training data and had not previously been accessible using DFT, as shown in Fig.~\ref{fig:sch_convexhull}(a).

For screening, single-point DFT calculations are performed on structures. We then re-rank the structures based on DFT enthalpies at 250 GPa, which shows reasonable -- but not perfect -- agreement with the energy rankings from the EDDP. Consistent rankings are not important here, so long as all the relevant structures are low enough in energy to be carried forward to this stage. Finally, structures which remain within an enthalpy window of 0.01 eV/atom after the re-ranking are retained for a full geometry optimisation. This typically does not require many optimisation steps since the structure is already close to the energy minimum.

\begin{figure*}[tb]
\includegraphics[width=1.0\linewidth]{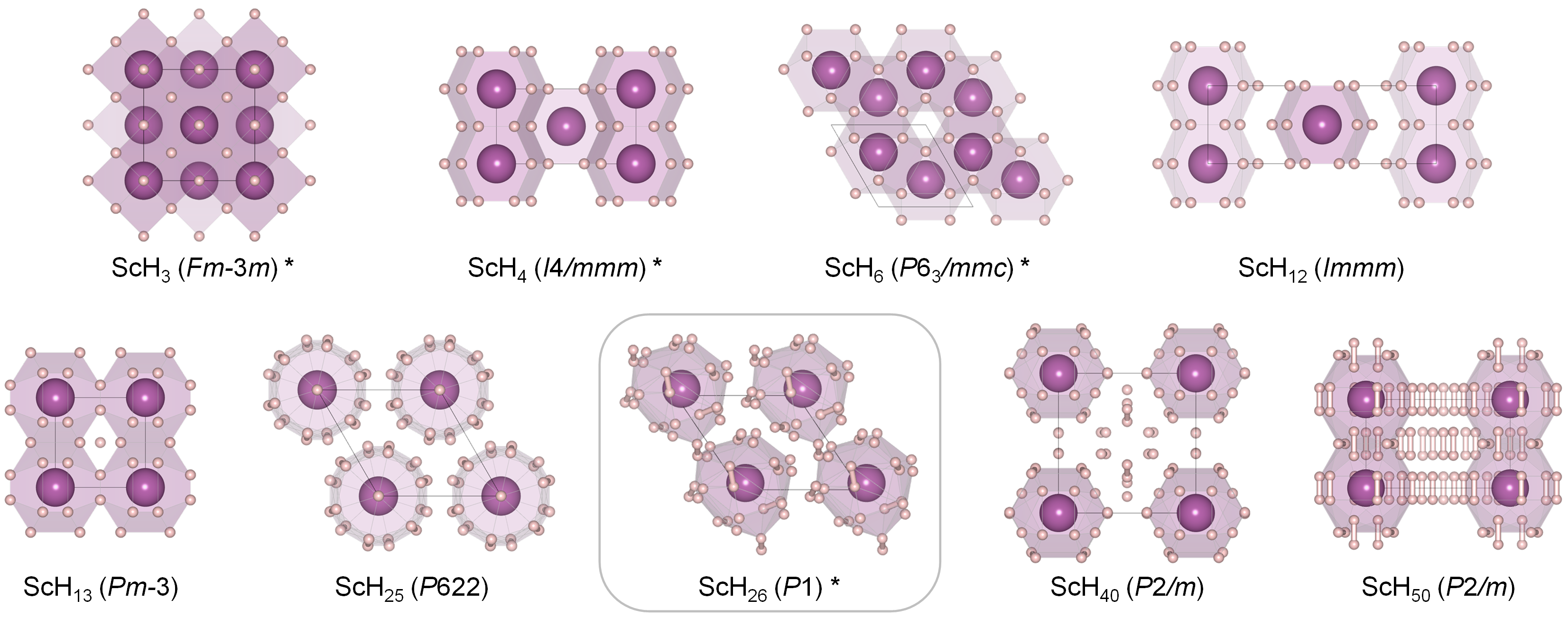}
\caption{\label{fig:sch_structures} Stable and metastable superhydride structures. The first row includes some rediscovered structures previously reported. The second row includes some previously unreported superhydride structures that have been identified in a broad structure search. An asterisk indicates the structure is on the convex hull. The gray box indicates the unreported superhydride structure on the convex hull.}
\end{figure*}

By this procedure, many previously reported scandium hydride structures were rapidly rediscovered. In Fig.~\ref{fig:sch_convexhull}(b) we plot the resulting convex hull where we see that \textit{Fm-3m} ScH$_{3}$, \textit{I4/mmm} ScH$_{4}$, and \textit{P6$_{3}$/mmc} ScH$_{6}$ are stable at 250 GPa. These results are in agreement with the findings of Ye \textit{et al.}\cite{ScH-Ashcroft}. Additionally, our extensive broad search has identified a previously unreported superhydride on the hull, ScH$_{26}$, highlighted in Fig.~\ref{fig:sch_structures}. Finite-difference phonon calculations at the DFT level confirmed the dynamical stability of this structure on a $3\times3\times3$ $\mathbf{q}$-point grid.

\subsubsection{Superionicity}

\begin{figure}[!tpb]
\includegraphics[width=1.0\linewidth]{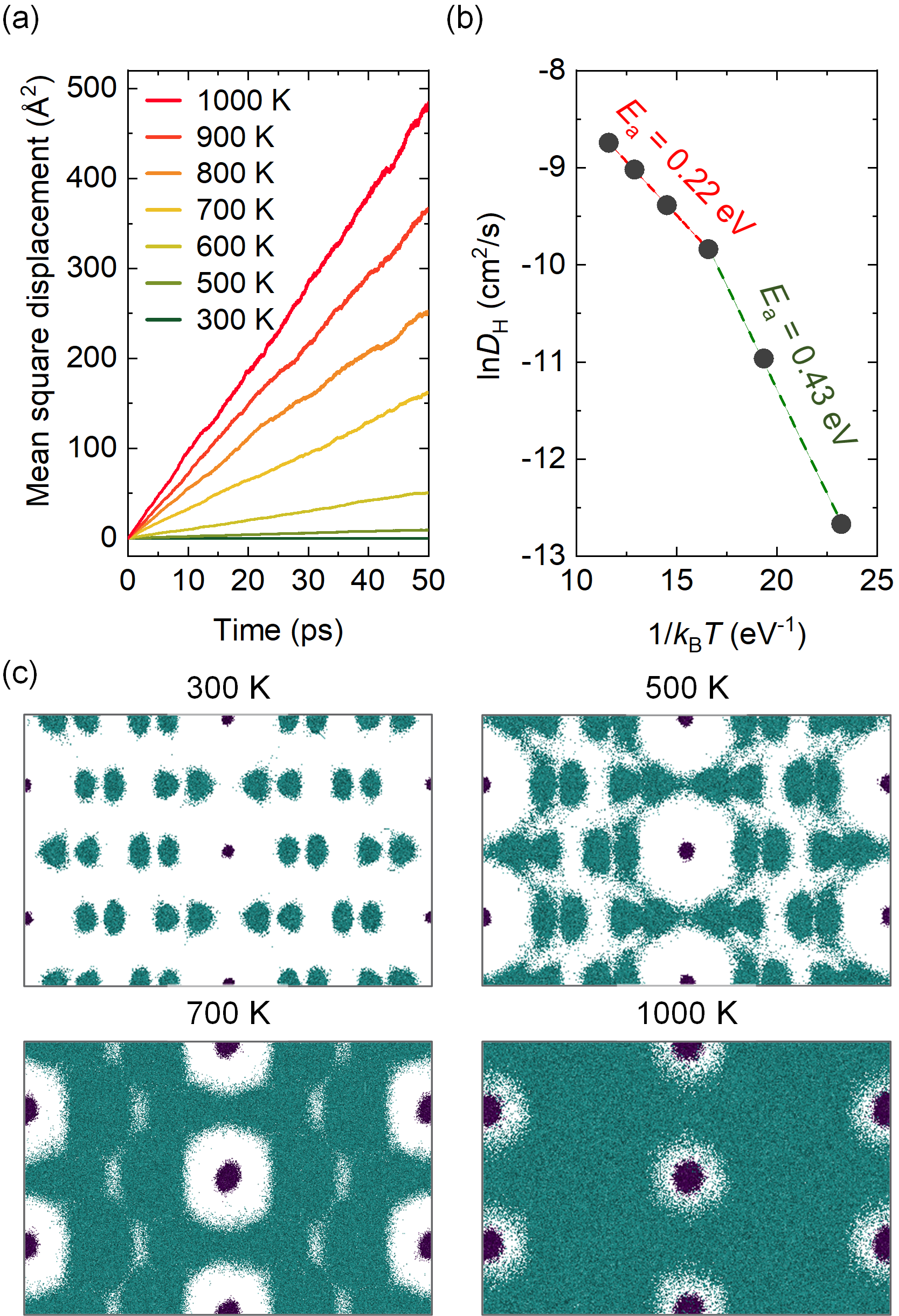}
\caption{\label{fig:sch_superionicity} Hydrogen diffusion in \textit{Immm} ScH$_{12}$ at 350 GPa. (a) Mean square displacement of hydrogen atoms at different temperatures. (b) Diffusion coefficient as a function of inverse temperature. (c) Diffusion path of hydrogen for 50 ps at 300, 500, 700, and 1000 K. Green and purple lines correspond to the trajectories of hydrogen and scandium, respectively.}
\end{figure}

\textit{Immm} ScH${}_{12}$ is one of the stable superhydrides above 325 GPa\cite{ScH-Ashcroft}. To further explore the hydrogen diffusion in the Sc-H system using MD, we generated a refined EDDP, targeting a single stoichiometry with \textit{Immm} ScH${}_{12}$ and \textit{C2/c} H${}_{2}$ as marker structures. The refinement process involved adding 2000 Immm ScH${}_{12}$ structures and 2000 \textit{C2/c} H${}_{2}$ structures to the training dataset with the atoms randomly perturbed by an amplitude of 0.2 \AA. Additionally, the unit cell vectors are scaled by randomly chosen factors between -0.02 and +0.02.

We used \texttt{ramble} to carry out MD simulations on nearly-cubic supercells containing around 600 atoms using the method described in section~\ref{sec:eddp_calculations}. The simulations were performed for 60 ps at temperatures of 300, 500, 600, 700, 800, 900, and 1000 K, and a pressure of 350 GPa. The diffusion of hydrogen was analyzed for the last 50 ps of each trajectory.

At 300 K, both Sc and H atoms vibrate around their mean positions without any diffusion occurring during the 60 ps simulation timescale. The H atoms begin diffusing at $500$ K. We calculated the mean square displacement (MSD) averaged over all H atoms <$\Delta r^2$>. As shown in Fig.~\ref{fig:sch_superionicity}(a), hydrogen atoms become increasingly diffusive with temperature. The diffusion coefficient of hydrogen $D_{\textnormal{H}}$ was then calculated based on the Einstein relation <$\Delta r^2$> = 6$D_{\textnormal{H}}t$, assuming a three-dimensional random walk. We obtain a value of $D=$ $3.2\times 10^{-6} \ \text{cm}^{2}/\text{s}$ at 500 K, which increases to $1.6\times 10^{-4} \ \text{cm}^{2}/\text{s}$ at 1000 K.

Based on the temperature dependent diffusion coefficients shown in Fig.~\ref{fig:sch_superionicity}(b), the activation energy $E_{a}$ of hydrogen diffusion was estimated by fitting the data to the Arrhenius equation
\begin{equation}
    \label{eq:arrhenius_equation}
    D_{\textnormal{H}} = D_{0}e^{-E_{a}/k_{B}T}
\end{equation}
The analysis of temperature-dependent diffusion coefficients reveals the manifestation of Non-Arrhenius behavior in ScH$_{12}$, akin to observations in other superionic conductors\cite{Arya2021-Non-Arrhenius-AgCrSe2, Martin1996-Non-Arrhenius, Ong2021-Non-Arrhenius}. Notably, two distinct regimes with an approximate transition temperature of 700 K are observed. Below 700 K, hydrogen atoms begin to move between lattice sites and $E_a = 0.43$ eV. Above 700 K, $E_a = 0.22$ eV as the hydrogen sub-lattice melts. These regimes can be differentiated in the trajectories in Fig.~\ref{fig:sch_superionicity}(c). In the sub-lattice melting regime, the activation energy is significantly lower, particularly when compared to LaH$_{10}$, which after sub-lattice melting has an activation energy of 0.44 eV above 1000 K (albeit at a lower pressure of 163~GPa)~\cite{LaH10-Paul}.

We have demonstrated that a single potential can predict stable structures across a wider range of stoichiometries than typically searched with DFT and rediscover the known phases. Refinement of this potential allows for molecular dynamics at larger length- and time-scales than previous work, leading to a prediction of a two-stage superionic transition; from site hopping to sub-lattice melting.

\subsection{Ternaries: Zinc Cyanide}
\label{sec:ternaries_zinc_cyanide}

\begin{figure*}[tpb]
	\centering
	\includegraphics[width=0.85\linewidth]{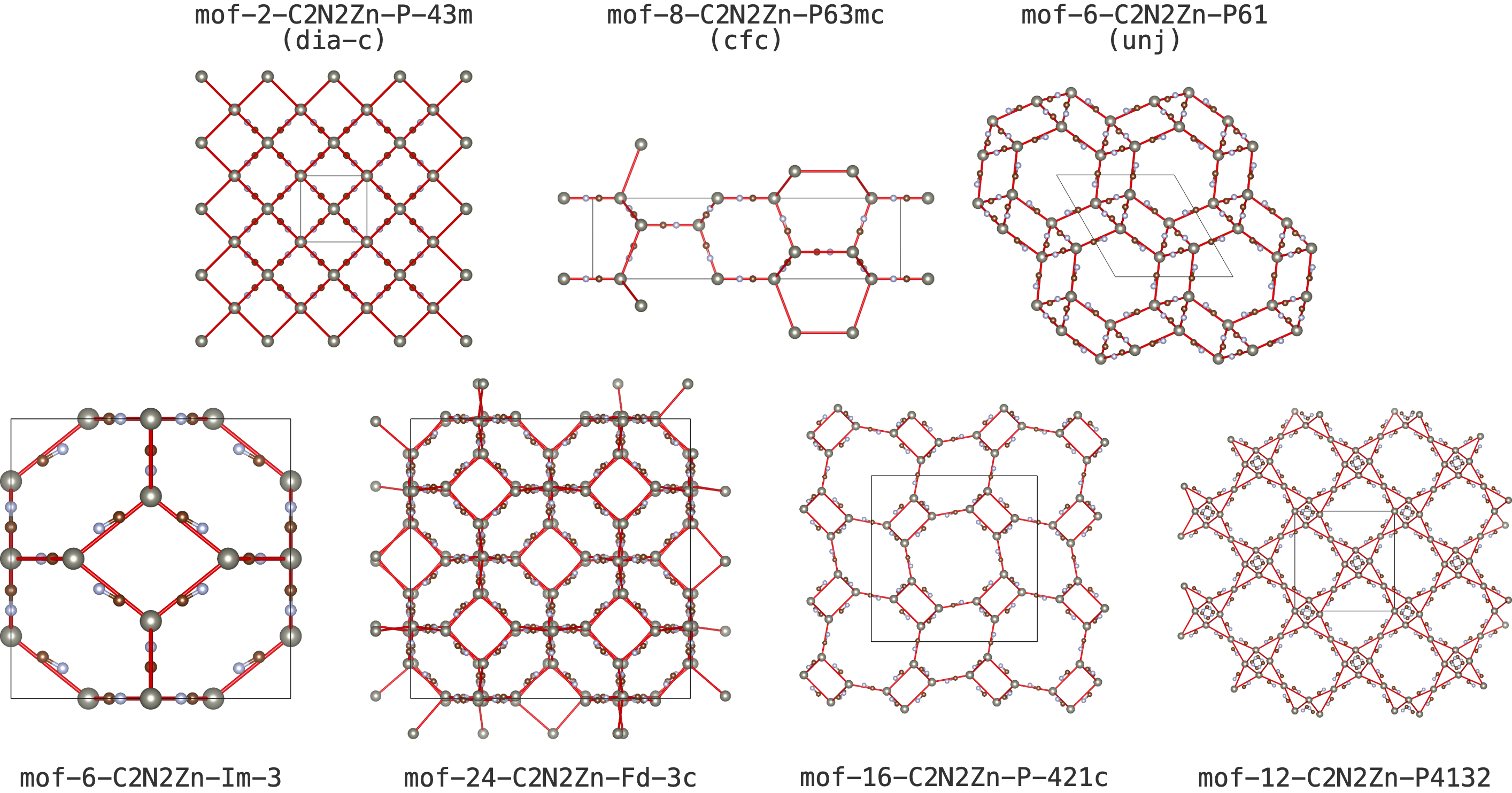}
	\caption{Crystal structures of Zn(CN)$_2$ polymorphs. The top row includes some previously calculated polymorphs and the second row includes low energy structures found by structure search using EDDPs.}
	\label{fig:mof-structures}
\end{figure*}

\begin{figure}[!htpb]
	\centering
	\includegraphics[width=0.85\linewidth]{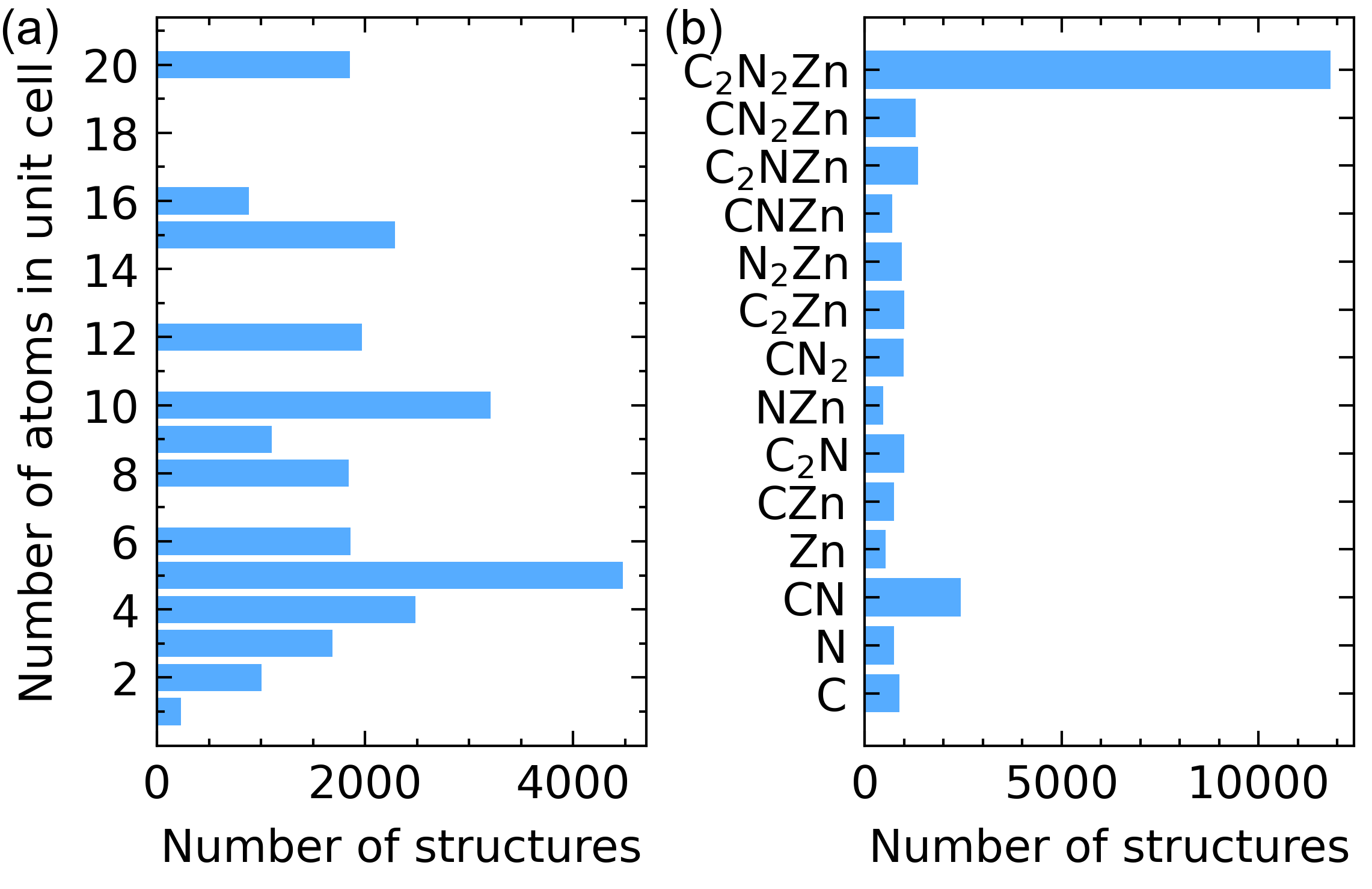}
	\caption{Distribution of structures in the Zn(CN)$_2$ training data; by (a) number of atoms in the unit cell and (b) stoichiometry. }
	\label{fig:mof-training}
\end{figure}

MOFs are a class of materials with a wide range of potential applications from hydrogen storage to drug delivery systems. They consist of metal ions connected to organic molecular `linkers' in often very complex, low-density, structures with large unit cells, making them difficult to study using plane-wave DFT. MOFs, therefore, are a difficult but desirable system to model using machine-learned potentials. Here, we choose a chemically simple MOF, zinc cyanide, which contains three elements, to demonstrate the use of EDDPs in such systems. 

Zinc cyanide, Zn(CN)${}_2$, has a disordered cubic crystal structure, with Zn atoms connected tetrahedrally by CN `linkers'~\cite{Williams1997}. The ordered approximation of the structure is analogous to two interpenetrated diamond-like sublattices, hereon labelled `dia-c' (see Fig.~\ref{fig:mof-structures}). Notably, this structure of Zn(CN)${}_2$ possesses a large negative thermal expansion with a coefficient\cite{Chapman2007} $\alpha=\frac{1}{V}\frac{dV}{dT} = -17 \times 10 ^{-6}$~K$^{-1}$, which has been attributed to the population of low-energy phonons~\cite{Goodwin2005}.

Other polymorphs of Zn(CN)${}_2$ have been synthesised in the presence of methanol-ethanol-water mixtures under pressure~\cite{Lapidus2013}. The small solvent molecules are are able to fill the vacancies formed by lower density Zn(CN)$_2$ structures like diamond and Lonsdaleite (dia, lon) which form at higher pressure. These structures, along with several `hand-crafted' zeolitic structures, have been modeled using an empirical force-field potential~\cite{Fang2013} to study thermal expansion~\cite{Trousselet2015}. 

In this example, we will reproduce these results without any assumed prior knowledge of the structure of Zn(CN)${}_2$ or any polymorphs, by first searching for low-energy, low density Zn(CN)${}_2$ structures, which could in principle be synthesised with other solvents. We then calculate the thermal expansion of these empty frameworks using molecular dynamics. We show that a general `all purpose' Zn + 2C + 2N potential can be used for structure searching and for molecular dynamics with reasonable accuracy. 

\subsubsection{Training}
\begin{figure}[tpb]
	\centering
	\includegraphics[width=0.95\linewidth]{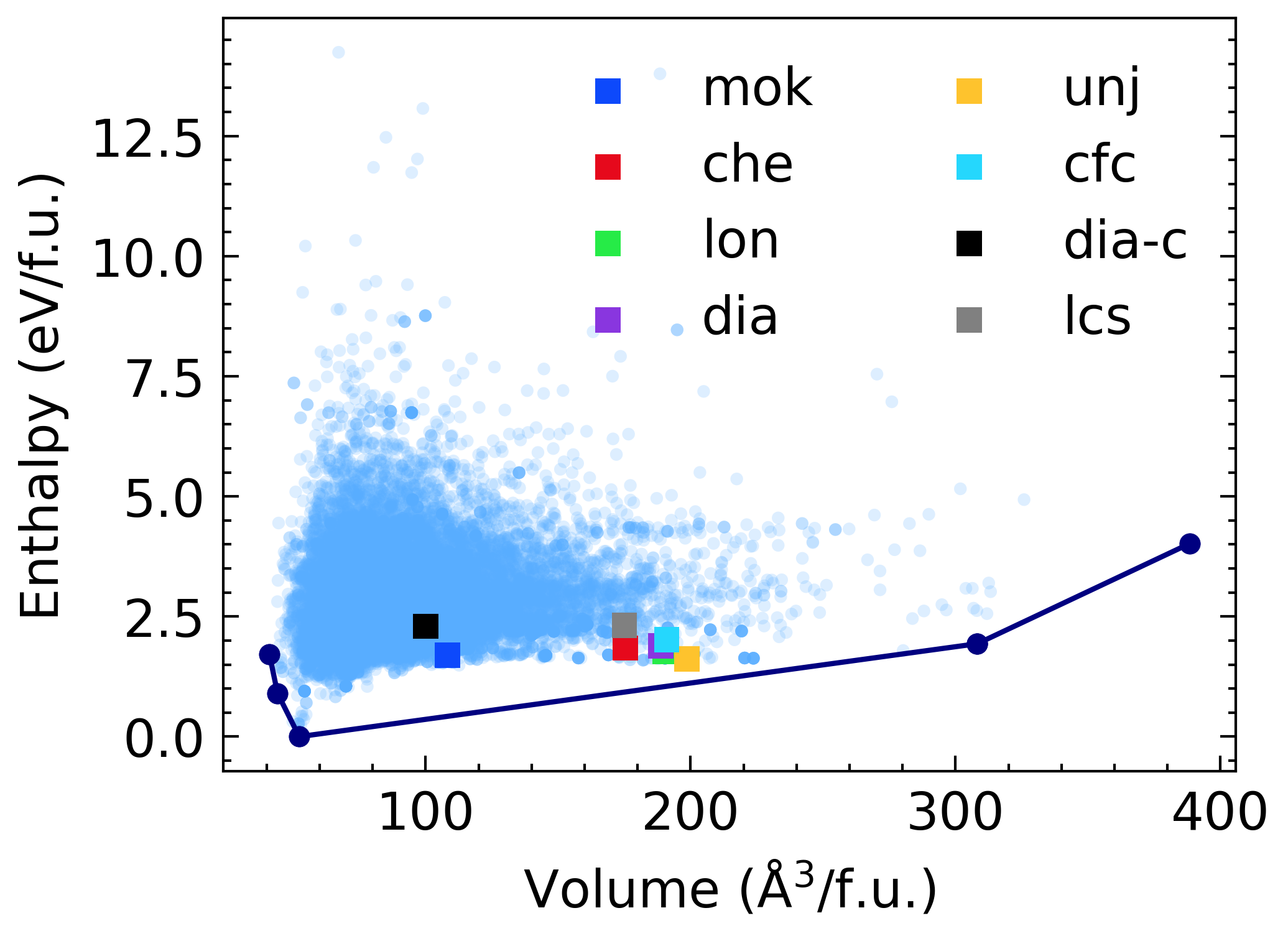}
	\caption{Range of volume - enthalpy values for structures found by random structure search with the Zn(CN)$_2$ EDDP. Square symbols indicate the named structure types.}
	\label{fig:mof-searching}
\end{figure}

Training potentials for ternary compositions can be more difficult than for a single species or binaries; the combinatoric variation of the local environment requires exponentially longer feature vectors --- they contain 498 components in this case, in contrast with 43 in lead and 172 in scandium hydride ---  and significantly more training data. We retain the principle of training potentials on small unit cells. We further emphasise a distinction between the local stoichiometry --- the atoms contained within a cut-off radius --- and the global stoichiometry of the crystal. Hence, for small-cell training, the variety of the training data can be enhanced by sampling structures with relatively few atoms, but with a range of global stoichiometries.

In this case, the training data consists of Zn$_l$C$_m$N$_n$ where $l=0,1$, $m=0,1,2$, $n=0,1,2$, with a weighting towards $l=1$, $m=2$, $n=2$, and up to 4 f.u. per cell. The resulting distribution of data is shown in Fig.~\ref{fig:mof-training}. Table \ref{tab:training_parameters} summarises the training parameters used to generate the Zn(CN)${}_2$ potential. Similarly to the carbon example, longer-range dispersion effects are implicitly incorporated in the generation of EDDPs by including them in the DFT dataset.

\subsubsection{Results}

\begin{figure*}[tpb]
	\centering
	\includegraphics[width=0.95\linewidth]{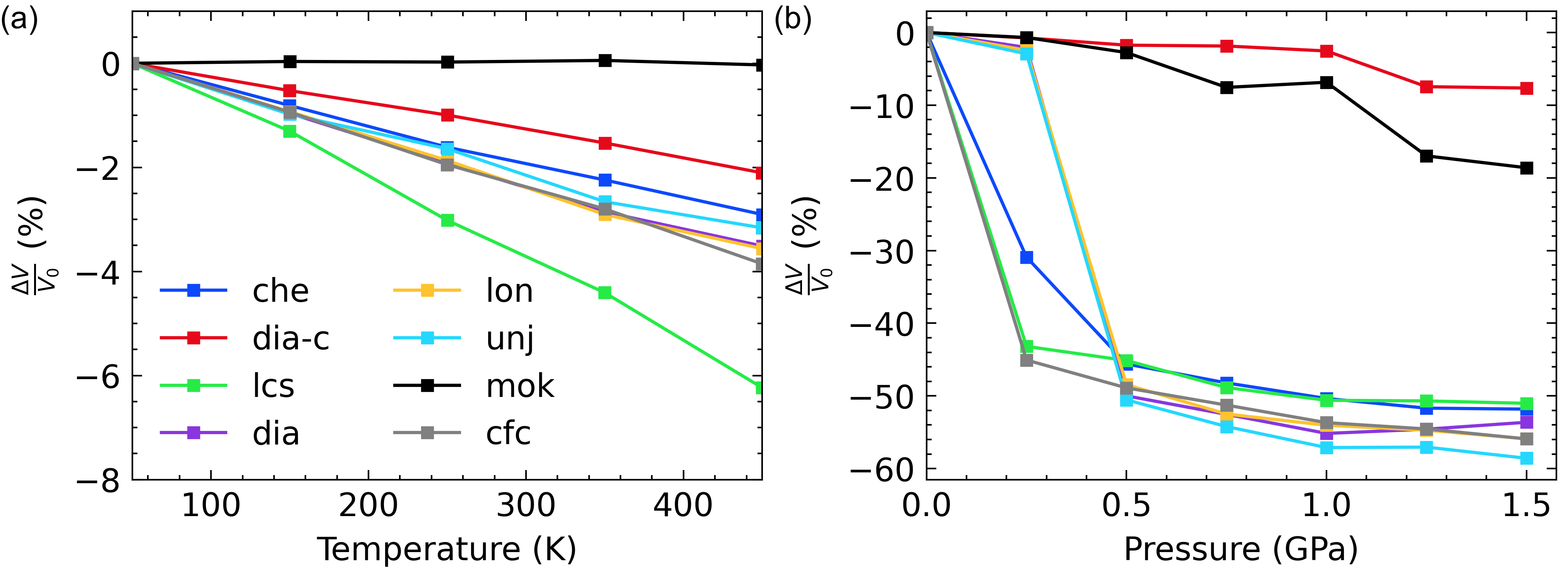}
	\caption{Relative volume change with temperature at 0\,GPa (a) and with pressure at 300\,K (b) for a range of metastable and hypothetical Zn(CN)${}_2$ structures.}
	\label{fig:mof-md}
\end{figure*}

\begin{table}[bp]
    \centering
    \caption{Thermal expansion coefficients for the structures shown in Fig.~\ref{fig:mof-md}, calculated with TS and MBD dispersion corrections. Experimental values\cite{Trousselet2015}, where available, are in parenthesis.}\label{tab:mof-NTE}
\begin{ruledtabular}
    \begin{tabular}{cccc}
    ID & label & $\alpha^\text{MBD}$ ($10^{-6}$ K$^{-1}$) & $\alpha^\text{TS}$ ($10^{-6}$ K$^{-1}$) \\ \hline
 \texttt{mof-152-C2N2Zn-I213} &       che &     -20.66 &     -24.88 \\ 
  \texttt{mof-2-C2N2Zn-P-43m} &     dia-c &     -14.21 (-17.46)&     -17.26 (-17.46)\\ 
 \texttt{mof-24-C2N2Zn-I-43d} &       lcs &     -30.68 &     -50.57 \\ 
   \texttt{mof-24-C2N2Zn-R3m} &       dia &     -25.33 &     -30.46 \\ 
  \texttt{mof-4-C2N2Zn-P63mc} &       lon &     -25.58 &     -30.70 \\ 
    \texttt{mof-6-C2N2Zn-P61} &       unj &     -21.90 &     -27.66 \\ 
     \texttt{mof-8-C2N2Zn-Cc} &       mok &       2.04 &       0.13 \\ 
  \texttt{mof-8-C2N2Zn-P63mc} &       cfc &     -25.72 &     -31.84 \\ 
    \end{tabular}
    \end{ruledtabular}
\end{table}

We begin by performing a random structure search with the EDDP, using AIRSS and \texttt{repose}, generating around 40,000 Zn(CN)$_2$ structures with a target volume between 100 and 300\,\AA$^{3}$/f.u., with each containing between 1 and 20 f.u. and 1-48 symmetry operations. Fig.~\ref{fig:mof-searching} shows these structures' enthalpy-volume distribution. We find a high concentration of points below 100 \AA$^{3}$/f.u., which have a higher density than is typical for a Zn(CN)$_2$ framework. These are structures which, despite being initialised with large unit cells, condensed into much denser phases during geometry optimisation. This demonstrates a difficulty with low-density structure prediction, as there are several accessible routes to condensed states, but much fewer routes to low-density phases. With EDDP structure searching, we can overcome this issue simply by performing very extensive searches. This is also a demonstration of the transferability of the EDDP, which --- in contrast to an empirically derived potential ---  is able to describe the layered hexagonal and tetrahedral carbon nitride-like environments typical of the condensed phases.

Several low-density zeolite-like structures were predicted in the search. We were able to rediscover the polymorphs proposed by Trousselet \textit{et al.} (which are indicated by squares in Fig.~\ref{fig:mof-searching}) as low-energy states close to the enthalpy-volume convex hull. We also find several other novel stable polymorphs with similar volumes and energies, the structures of which are shown in Fig.~\ref{fig:mof-structures}. These new structures include some which are similar to known clathrate types (\texttt{mof-6-C2N2Zn-Im-3} and \texttt{mof-24-C2N2Zn-Fd-3c}) and other novel complex networks (such as \texttt{mof-16-C2N2Zn-P-421c} and \texttt{mof-12-C2N2Zn-P4132}).

We then explored the the behaviour of some of the low-density polymorphs at finite temperature. Specifically, we used the structures calculated by Trousselet \textit{et al.} with the labels che, dia-c, lcs, dia, lon, unj, mok and cfc. These are tabulated, along with their structure ID, in table~\ref{tab:mof-NTE}. 

We used \texttt{ramble} to run MD at 50, 150, 250, 350, 450 K at ambient pressure with supercells containing around 500 atoms. These supercells were made to be `nearly-cubic' using the method described in section~\ref{sec:eddp_calculations}. Fig.~\ref{fig:mof-md}(a) shows volume change with temperature for these structures. Linear fits to the data show an expansion coefficient, $\alpha=-17.26\times10 ^{-6}$~K$^{-1}$ for the `dia-c', which is remarkably close to the experimental and previously calculated values of -17.4  and -16.9 $\times 10 ^{-6}$~K$^{-1}$ respectively. A second potential trained using the MBD dispersion correction scheme\cite{Tkatchenko2012} gave qualitatively similar results, as shown in table~\ref{tab:mof-NTE}, but a value of $\alpha=-14.21\times10 ^{-6}$~K$^{-1}$. The behaviour of the metastable hypothetical structures is in good agreement with those calculated using a parameterised force-field\cite{Trousselet2015}. 

We also ran MD simulations at $300$ K for a range of pressures. The volume changes are shown in Fig.~\ref{fig:mof-md}(b). We see that the denser structures, dia-c and mok, remain stable, whereas the more open structures undergo a structural transition above 0.25\,GPa. This is also in line with results obtained by Trousselet \textit{et al.}. 

With this example, we highlight the feasibility of generating a good potential for a ternary system and extracting physically sensible results. We should note that in calculating the thermal expansion, a well-parameterised empirical potential is clearly sufficient to model the dynamics. However, as demonstrated by our structure search, a machine-learned potential is transferable to more unusual systems, for example with broken cyanide bonds or layered C-N networks. This means EDDPs can be exploited for exploration-based studies --- where calculations can be performed on systems without any prior assumptions --- to make new discoveries. 

While this ternary system is feasible, typical MOFs contain more than 3 or 4 elements. Chemical complexity becomes a limiting factor for machine-learned potentials and EDDPs are no exception. The EDDP feature vector scales such that a naive extension to systems with more elements will lead to overly sparse feature vectors and a consequential bottleneck in calculation speed. Recent work by Ceriotti and co-workers has attempted to address this issue by considering `alchemical' correlations~\cite{Willatt2018}. To continue increasing the chemical complexity, it may be necessary to adopt a similar approach.

\section{Conclusion}
\label{sec:conclusion}

In this work, we have demonstrated that the EDDPs can significantly accelerate structure prediction, but also have applications beyond this initial intent. This is achieved without requiring a sophisticated (and costly) neural network architecture and relies only on single-point DFT energy calculations of small cells, making them convenient and cheap to train. The EDDPs' smoothness means they are well-behaved over wide regions of structure space. They possess a good degree of size transferability, allowing potentials trained on cells containing $24$ atoms or less to model much larger systems successfully. For pure elements, only a few thousand DFT calculations on small cells are required to achieve meV-level accuracy, excepting carbon, a particularly difficult system.

In the cases of both scandium hydride and zinc cyanide, a single EDDP each is capable of successfully searching a wide range of stoichiometries. This is evidenced by the reproduction of known results in these two test cases. An EDDP-enabled MD simulation of Immm-ScH${}_{12}$ predicts hydrogen sublattice melting and superionicity above $700$ K at $350$ GPa. This phenomenon has been demonstrated using AIMD for other superhydrides under high pressure. By combining high-throughput phonon calculations and coexistence MD, an EDDP was used to generate a phase diagram for lead. This was done taking SOC into account, which brings the melt curve into better agreement with experiment.

The EDDPs were originally termed `ephemeral' to indicate the intention to design a custom potential suitable for a given application and then discard it. Instead of expending considerable effort and computational power training a `perfect' potential, it can be preferable to train a sufficient potential and spend the available resources on using it.

This philosophy has not entirely been superseded - our intent is not to present a new set of benchmark potentials for the systems discussed. Nevertheless, the ease with which EDDPs can be trained means that for a given application, often several dozen potentials can be generated in order to assess the best combination of parameters. Furthermore, once a satisfactory potential has been generated, the robust nature and smoothness of the EDDPs means they are applicable more widely than originally anticipated. The `ephemeral' data-derived potentials have proven more long-lived than expected.

\begin{acknowledgements}
We thank Michele Simoncelli for his careful reading of and helpful feedback on an early version of this manuscript. We also express our appreciation to Siyu Chen and Bartomeu Monserrat for helpful discussions. P.T.S. gratefully acknowledges funding from the Department of Materials Science and Metallurgy at the University of Cambridge, and from a Trinity Hall research studentship. W.C.W. was supported by the Schmidt Science Fellows in partnership with the Rhodes Trust. W.C.W. and C.J.P. acknowledge support from the EPSRC (Grant EP/V062654/1), P.I.C.C. and C.J.P. acknowledge support from the EPSRC (Grant EP/S021981/1) and C.J.P. further acknowledges EPSRC support for the UKCP consortium (Grant EP/P022596/1). This work was supported by the Faraday Institution (Grant No. FIRG017) and used the MICHAEL computing facilities. We are grateful for computational support from the UK Materials and Molecular Modelling Hub, which is partially funded by EPSRC (EP/T022213/1, EP/W032260/1 and EP/P020194/1), for which access was obtained via the UKCP consortium. This work was performed using resources provided by the Cambridge Service for Data Driven Discovery (CSD3) operated by the University of Cambridge Research Computing Service (www.csd3.cam.ac.uk), provided by Dell EMC and Intel using Tier-2 funding from the EPSRC (capital grant EP/T022159/1), and DiRAC funding from the STFC (www.dirac.ac.uk). This work used the ARCHER2 UK National Supercomputing Service (https://www.archer2.ac.uk).

\textit{For the purpose of open access, the author has applied a Creative Commons Attribution (CC BY) licence to any Author Accepted Manuscript version arising from this submission.}
\end{acknowledgements}

\section*{Author Declarations}

\subsection*{Conflict of Interest}

The authors have no conflicts to disclose.

\section*{Data Availability}

The data that support the findings of this study are openly available on MaterialsCloud at \url{https://doi.org/10.24435/materialscloud:44-c5}.

\section*{References}

\bibliography{bibliography,EDDP-ljc}

\providecommand{\noopsort}[1]{}\providecommand{\singleletter}[1]{#1}%
\begin{thebibliography}{155}%
\makeatletter
\providecommand \@ifxundefined [1]{%
 \@ifx{#1\undefined}
}%
\providecommand \@ifnum [1]{%
 \ifnum #1\expandafter \@firstoftwo
 \else \expandafter \@secondoftwo
 \fi
}%
\providecommand \@ifx [1]{%
 \ifx #1\expandafter \@firstoftwo
 \else \expandafter \@secondoftwo
 \fi
}%
\providecommand \natexlab [1]{#1}%
\providecommand \enquote  [1]{``#1''}%
\providecommand \bibnamefont  [1]{#1}%
\providecommand \bibfnamefont [1]{#1}%
\providecommand \citenamefont [1]{#1}%
\providecommand \href@noop [0]{\@secondoftwo}%
\providecommand \href [0]{\begingroup \@sanitize@url \@href}%
\providecommand \@href[1]{\@@startlink{#1}\@@href}%
\providecommand \@@href[1]{\endgroup#1\@@endlink}%
\providecommand \@sanitize@url [0]{\catcode `\\12\catcode `\$12\catcode
  `\&12\catcode `\#12\catcode `\^12\catcode `\_12\catcode `\%12\relax}%
\providecommand \@@startlink[1]{}%
\providecommand \@@endlink[0]{}%
\providecommand \url  [0]{\begingroup\@sanitize@url \@url }%
\providecommand \@url [1]{\endgroup\@href {#1}{\urlprefix }}%
\providecommand \urlprefix  [0]{URL }%
\providecommand \Eprint [0]{\href }%
\providecommand \doibase [0]{http://dx.doi.org/}%
\providecommand \selectlanguage [0]{\@gobble}%
\providecommand \bibinfo  [0]{\@secondoftwo}%
\providecommand \bibfield  [0]{\@secondoftwo}%
\providecommand \translation [1]{[#1]}%
\providecommand \BibitemOpen [0]{}%
\providecommand \bibitemStop [0]{}%
\providecommand \bibitemNoStop [0]{.\EOS\space}%
\providecommand \EOS [0]{\spacefactor3000\relax}%
\providecommand \BibitemShut  [1]{\csname bibitem#1\endcsname}%
\let\auto@bib@innerbib\@empty
\bibitem [{\citenamefont {Schmidt}\ \emph {et~al.}(2019)\citenamefont
  {Schmidt}, \citenamefont {Marques}, \citenamefont {Botti},\ and\
  \citenamefont {Marques}}]{Machine-learning-materials-general-review}%
  \BibitemOpen
  \bibfield  {author} {\bibinfo {author} {\bibfnamefont {J.}~\bibnamefont
  {Schmidt}}, \bibinfo {author} {\bibfnamefont {M.~R.~G.}\ \bibnamefont
  {Marques}}, \bibinfo {author} {\bibfnamefont {S.}~\bibnamefont {Botti}}, \
  and\ \bibinfo {author} {\bibfnamefont {M.~A.~L.}\ \bibnamefont {Marques}},\
  }\bibfield  {title} {\enquote {\bibinfo {title} {Recent advances and
  applications of machine learning in solid-state materials science},}\ }\href
  {\doibase 10.1038/s41524-019-0221-0} {\bibfield  {journal} {\bibinfo
  {journal} {npj Computational Materials}\ }\textbf {\bibinfo {volume} {5}},\
  \bibinfo {pages} {83} (\bibinfo {year} {2019})}\BibitemShut {NoStop}%
\bibitem [{\citenamefont {Kulik}\ \emph {et~al.}(2022)\citenamefont {Kulik},
  \citenamefont {Hammerschmidt}, \citenamefont {Schmidt}, \citenamefont
  {Botti}, \citenamefont {Marques}, \citenamefont {Boley}, \citenamefont
  {Scheffler}, \citenamefont {Todorovi{\'c}}, \citenamefont {Rinke},
  \citenamefont {Oses}, \citenamefont {Smolyanyuk}, \citenamefont {Curtarolo},
  \citenamefont {Tkatchenko}, \citenamefont {Bart{\'o}k}, \citenamefont
  {Manzhos}, \citenamefont {Ihara}, \citenamefont {Carrington}, \citenamefont
  {Behler}, \citenamefont {Isayev}, \citenamefont {Veit}, \citenamefont
  {Grisafi}, \citenamefont {Nigam}, \citenamefont {Ceriotti}, \citenamefont
  {Sch{\"u}tt}, \citenamefont {Westermayr}, \citenamefont {Gastegger},
  \citenamefont {Maurer}, \citenamefont {Kalita}, \citenamefont {Burke},
  \citenamefont {Nagai}, \citenamefont {Akashi}, \citenamefont {Sugino},
  \citenamefont {Hermann}, \citenamefont {No{\'e}}, \citenamefont {Pilati},
  \citenamefont {Draxl}, \citenamefont {Kuban}, \citenamefont {Rigamonti},
  \citenamefont {Scheidgen}, \citenamefont {Esters}, \citenamefont {Hicks},
  \citenamefont {Toher}, \citenamefont {Balachandran}, \citenamefont {Tamblyn},
  \citenamefont {Whitelam}, \citenamefont {Bellinger},\ and\ \citenamefont
  {Ghiringhelli}}]{Machine-learning-roadmap}%
  \BibitemOpen
  \bibfield  {author} {\bibinfo {author} {\bibfnamefont {H.~J.}\ \bibnamefont
  {Kulik}}, \bibinfo {author} {\bibfnamefont {T.}~\bibnamefont
  {Hammerschmidt}}, \bibinfo {author} {\bibfnamefont {J.}~\bibnamefont
  {Schmidt}}, \bibinfo {author} {\bibfnamefont {S.}~\bibnamefont {Botti}},
  \bibinfo {author} {\bibfnamefont {M.~A.~L.}\ \bibnamefont {Marques}},
  \bibinfo {author} {\bibfnamefont {M.}~\bibnamefont {Boley}}, \bibinfo
  {author} {\bibfnamefont {M.}~\bibnamefont {Scheffler}}, \bibinfo {author}
  {\bibfnamefont {M.}~\bibnamefont {Todorovi{\'c}}}, \bibinfo {author}
  {\bibfnamefont {P.}~\bibnamefont {Rinke}}, \bibinfo {author} {\bibfnamefont
  {C.}~\bibnamefont {Oses}}, \bibinfo {author} {\bibfnamefont {A.}~\bibnamefont
  {Smolyanyuk}}, \bibinfo {author} {\bibfnamefont {S.}~\bibnamefont
  {Curtarolo}}, \bibinfo {author} {\bibfnamefont {A.}~\bibnamefont
  {Tkatchenko}}, \bibinfo {author} {\bibfnamefont {A.~P.}\ \bibnamefont
  {Bart{\'o}k}}, \bibinfo {author} {\bibfnamefont {S.}~\bibnamefont {Manzhos}},
  \bibinfo {author} {\bibfnamefont {M.}~\bibnamefont {Ihara}}, \bibinfo
  {author} {\bibfnamefont {T.}~\bibnamefont {Carrington}}, \bibinfo {author}
  {\bibfnamefont {J.}~\bibnamefont {Behler}}, \bibinfo {author} {\bibfnamefont
  {O.}~\bibnamefont {Isayev}}, \bibinfo {author} {\bibfnamefont
  {M.}~\bibnamefont {Veit}}, \bibinfo {author} {\bibfnamefont {A.}~\bibnamefont
  {Grisafi}}, \bibinfo {author} {\bibfnamefont {J.}~\bibnamefont {Nigam}},
  \bibinfo {author} {\bibfnamefont {M.}~\bibnamefont {Ceriotti}}, \bibinfo
  {author} {\bibfnamefont {K.~T.}\ \bibnamefont {Sch{\"u}tt}}, \bibinfo
  {author} {\bibfnamefont {J.}~\bibnamefont {Westermayr}}, \bibinfo {author}
  {\bibfnamefont {M.}~\bibnamefont {Gastegger}}, \bibinfo {author}
  {\bibfnamefont {R.~J.}\ \bibnamefont {Maurer}}, \bibinfo {author}
  {\bibfnamefont {B.}~\bibnamefont {Kalita}}, \bibinfo {author} {\bibfnamefont
  {K.}~\bibnamefont {Burke}}, \bibinfo {author} {\bibfnamefont
  {R.}~\bibnamefont {Nagai}}, \bibinfo {author} {\bibfnamefont
  {R.}~\bibnamefont {Akashi}}, \bibinfo {author} {\bibfnamefont
  {O.}~\bibnamefont {Sugino}}, \bibinfo {author} {\bibfnamefont
  {J.}~\bibnamefont {Hermann}}, \bibinfo {author} {\bibfnamefont
  {F.}~\bibnamefont {No{\'e}}}, \bibinfo {author} {\bibfnamefont
  {S.}~\bibnamefont {Pilati}}, \bibinfo {author} {\bibfnamefont
  {C.}~\bibnamefont {Draxl}}, \bibinfo {author} {\bibfnamefont
  {M.}~\bibnamefont {Kuban}}, \bibinfo {author} {\bibfnamefont
  {S.}~\bibnamefont {Rigamonti}}, \bibinfo {author} {\bibfnamefont
  {M.}~\bibnamefont {Scheidgen}}, \bibinfo {author} {\bibfnamefont
  {M.}~\bibnamefont {Esters}}, \bibinfo {author} {\bibfnamefont
  {D.}~\bibnamefont {Hicks}}, \bibinfo {author} {\bibfnamefont
  {C.}~\bibnamefont {Toher}}, \bibinfo {author} {\bibfnamefont {P.~V.}\
  \bibnamefont {Balachandran}}, \bibinfo {author} {\bibfnamefont
  {I.}~\bibnamefont {Tamblyn}}, \bibinfo {author} {\bibfnamefont
  {S.}~\bibnamefont {Whitelam}}, \bibinfo {author} {\bibfnamefont
  {C.}~\bibnamefont {Bellinger}}, \ and\ \bibinfo {author} {\bibfnamefont
  {L.~M.}\ \bibnamefont {Ghiringhelli}},\ }\bibfield  {title} {\enquote
  {\bibinfo {title} {Roadmap on machine learning in electronic structure},}\
  }\href {\doibase 10.1088/2516-1075/ac572f} {\bibfield  {journal} {\bibinfo
  {journal} {Electronic Structure}\ }\textbf {\bibinfo {volume} {4}},\ \bibinfo
  {pages} {023004} (\bibinfo {year} {2022})}\BibitemShut {NoStop}%
\bibitem [{\citenamefont {Bhadeshia}, \citenamefont {MacKay},\ and\
  \citenamefont {Svensson}(1995)}]{Bhadeshia-NN-alloy-properties}%
  \BibitemOpen
  \bibfield  {author} {\bibinfo {author} {\bibfnamefont {H.~K. D.~H.}\
  \bibnamefont {Bhadeshia}}, \bibinfo {author} {\bibfnamefont {D.~J.~C.}\
  \bibnamefont {MacKay}}, \ and\ \bibinfo {author} {\bibfnamefont {L.-E.}\
  \bibnamefont {Svensson}},\ }\bibfield  {title} {\enquote {\bibinfo {title}
  {Impact toughness of {C}--{Mn} steel arc welds -- {Bayesian} neural network
  analysis},}\ }\href {\doibase 10.1179/mst.1995.11.10.1046} {\bibfield
  {journal} {\bibinfo  {journal} {Materials Science and Technology}\ }\textbf
  {\bibinfo {volume} {11}},\ \bibinfo {pages} {1046--1051} (\bibinfo {year}
  {1995})}\BibitemShut {NoStop}%
\bibitem [{\citenamefont {Sutton}\ \emph {et~al.}(2019)\citenamefont {Sutton},
  \citenamefont {Ghiringhelli}, \citenamefont {Yamamoto}, \citenamefont
  {Lysogorskiy}, \citenamefont {Blumenthal}, \citenamefont {Hammerschmidt},
  \citenamefont {Golebiowski}, \citenamefont {Liu}, \citenamefont {Ziletti},\
  and\ \citenamefont
  {Scheffler}}]{Machine-learning-property-prediction-competition}%
  \BibitemOpen
  \bibfield  {author} {\bibinfo {author} {\bibfnamefont {C.}~\bibnamefont
  {Sutton}}, \bibinfo {author} {\bibfnamefont {L.~M.}\ \bibnamefont
  {Ghiringhelli}}, \bibinfo {author} {\bibfnamefont {T.}~\bibnamefont
  {Yamamoto}}, \bibinfo {author} {\bibfnamefont {Y.}~\bibnamefont
  {Lysogorskiy}}, \bibinfo {author} {\bibfnamefont {L.}~\bibnamefont
  {Blumenthal}}, \bibinfo {author} {\bibfnamefont {T.}~\bibnamefont
  {Hammerschmidt}}, \bibinfo {author} {\bibfnamefont {J.~R.}\ \bibnamefont
  {Golebiowski}}, \bibinfo {author} {\bibfnamefont {X.}~\bibnamefont {Liu}},
  \bibinfo {author} {\bibfnamefont {A.}~\bibnamefont {Ziletti}}, \ and\
  \bibinfo {author} {\bibfnamefont {M.}~\bibnamefont {Scheffler}},\ }\bibfield
  {title} {\enquote {\bibinfo {title} {Crowd-sourcing materials-science
  challenges with the {NOMAD 2018 Kaggle} competition},}\ }\href {\doibase
  10.1038/s41524-019-0239-3} {\bibfield  {journal} {\bibinfo  {journal} {npj
  Computational Materials}\ }\textbf {\bibinfo {volume} {5}},\ \bibinfo {pages}
  {111} (\bibinfo {year} {2019})}\BibitemShut {NoStop}%
\bibitem [{\citenamefont {Garrido~Torres}\ \emph {et~al.}(2021)\citenamefont
  {Garrido~Torres}, \citenamefont {Gharakhanyan}, \citenamefont {Artrith},
  \citenamefont {Eegholm},\ and\ \citenamefont
  {Urban}}]{Temperature-machine-learning}%
  \BibitemOpen
  \bibfield  {author} {\bibinfo {author} {\bibfnamefont {J.~A.}\ \bibnamefont
  {Garrido~Torres}}, \bibinfo {author} {\bibfnamefont {V.}~\bibnamefont
  {Gharakhanyan}}, \bibinfo {author} {\bibfnamefont {N.}~\bibnamefont
  {Artrith}}, \bibinfo {author} {\bibfnamefont {T.~H.}\ \bibnamefont
  {Eegholm}}, \ and\ \bibinfo {author} {\bibfnamefont {A.}~\bibnamefont
  {Urban}},\ }\bibfield  {title} {\enquote {\bibinfo {title} {Augmenting
  {zero-Kelvin} quantum mechanics with machine learning for the prediction of
  chemical reactions at high temperatures},}\ }\href {\doibase
  10.1038/s41467-021-27154-2} {\bibfield  {journal} {\bibinfo  {journal}
  {Nature Communications}\ }\textbf {\bibinfo {volume} {12}},\ \bibinfo {pages}
  {7012} (\bibinfo {year} {2021})}\BibitemShut {NoStop}%
\bibitem [{\citenamefont {Weinreich}\ \emph {et~al.}(2022)\citenamefont
  {Weinreich}, \citenamefont {Lemm}, \citenamefont {von Rudorff},\ and\
  \citenamefont {von Lilienfeld}}]{Machine-learning-phase-space-averages}%
  \BibitemOpen
  \bibfield  {author} {\bibinfo {author} {\bibfnamefont {J.}~\bibnamefont
  {Weinreich}}, \bibinfo {author} {\bibfnamefont {D.}~\bibnamefont {Lemm}},
  \bibinfo {author} {\bibfnamefont {G.~F.}\ \bibnamefont {von Rudorff}}, \ and\
  \bibinfo {author} {\bibfnamefont {O.~A.}\ \bibnamefont {von Lilienfeld}},\
  }\bibfield  {title} {\enquote {\bibinfo {title} {{Ab initio machine learning
  of phase space averages}},}\ }\href {https://doi.org/10.1063/5.0095674}
  {\bibfield  {journal} {\bibinfo  {journal} {The Journal of Chemical Physics}\
  }\textbf {\bibinfo {volume} {157}} (\bibinfo {year} {2022})}\BibitemShut
  {NoStop}%
\bibitem [{\citenamefont {Tozer}, \citenamefont {Ingamells},\ and\
  \citenamefont {Handy}(1996)}]{Machine-learning-XC-potentials}%
  \BibitemOpen
  \bibfield  {author} {\bibinfo {author} {\bibfnamefont {D.~J.}\ \bibnamefont
  {Tozer}}, \bibinfo {author} {\bibfnamefont {V.~E.}\ \bibnamefont
  {Ingamells}}, \ and\ \bibinfo {author} {\bibfnamefont {N.~C.}\ \bibnamefont
  {Handy}},\ }\bibfield  {title} {\enquote {\bibinfo {title}
  {{Exchange‐correlation potentials}},}\ }\href {\doibase 10.1063/1.472753}
  {\bibfield  {journal} {\bibinfo  {journal} {The Journal of Chemical Physics}\
  }\textbf {\bibinfo {volume} {105}},\ \bibinfo {pages} {9200--9213} (\bibinfo
  {year} {1996})}\BibitemShut {NoStop}%
\bibitem [{\citenamefont {Snyder}\ \emph {et~al.}(2012)\citenamefont {Snyder},
  \citenamefont {Rupp}, \citenamefont {Hansen}, \citenamefont {M\"uller},\ and\
  \citenamefont {Burke}}]{Burke-ML-Density-Functionals}%
  \BibitemOpen
  \bibfield  {author} {\bibinfo {author} {\bibfnamefont {J.~C.}\ \bibnamefont
  {Snyder}}, \bibinfo {author} {\bibfnamefont {M.}~\bibnamefont {Rupp}},
  \bibinfo {author} {\bibfnamefont {K.}~\bibnamefont {Hansen}}, \bibinfo
  {author} {\bibfnamefont {K.-R.}\ \bibnamefont {M\"uller}}, \ and\ \bibinfo
  {author} {\bibfnamefont {K.}~\bibnamefont {Burke}},\ }\bibfield  {title}
  {\enquote {\bibinfo {title} {Finding density functionals with machine
  learning},}\ }\href {\doibase 10.1103/PhysRevLett.108.253002} {\bibfield
  {journal} {\bibinfo  {journal} {Phys. Rev. Lett.}\ }\textbf {\bibinfo
  {volume} {108}},\ \bibinfo {pages} {253002} (\bibinfo {year}
  {2012})}\BibitemShut {NoStop}%
\bibitem [{\citenamefont {Nagai}, \citenamefont {Akashi},\ and\ \citenamefont
  {Sugino}(2020)}]{ML-DFT-molecules}%
  \BibitemOpen
  \bibfield  {author} {\bibinfo {author} {\bibfnamefont {R.}~\bibnamefont
  {Nagai}}, \bibinfo {author} {\bibfnamefont {R.}~\bibnamefont {Akashi}}, \
  and\ \bibinfo {author} {\bibfnamefont {O.}~\bibnamefont {Sugino}},\
  }\bibfield  {title} {\enquote {\bibinfo {title} {Completing density
  functional theory by machine learning hidden messages from molecules},}\
  }\href {\doibase 10.1038/s41524-020-0310-0} {\bibfield  {journal} {\bibinfo
  {journal} {npj Computational Materials}\ }\textbf {\bibinfo {volume} {6}},\
  \bibinfo {pages} {43} (\bibinfo {year} {2020})}\BibitemShut {NoStop}%
\bibitem [{\citenamefont {Behler}\ and\ \citenamefont
  {Parrinello}(2007)}]{Behler-Parrinello-NNPs}%
  \BibitemOpen
  \bibfield  {author} {\bibinfo {author} {\bibfnamefont {J.}~\bibnamefont
  {Behler}}\ and\ \bibinfo {author} {\bibfnamefont {M.}~\bibnamefont
  {Parrinello}},\ }\bibfield  {title} {\enquote {\bibinfo {title} {Generalized
  neural-network representation of high-dimensional potential-energy
  surfaces},}\ }\href {\doibase 10.1103/PhysRevLett.98.146401} {\bibfield
  {journal} {\bibinfo  {journal} {Phys. Rev. Lett.}\ }\textbf {\bibinfo
  {volume} {98}},\ \bibinfo {pages} {146401} (\bibinfo {year}
  {2007})}\BibitemShut {NoStop}%
\bibitem [{\citenamefont {Bart\'ok}\ \emph {et~al.}(2010)\citenamefont
  {Bart\'ok}, \citenamefont {Payne}, \citenamefont {Kondor},\ and\
  \citenamefont {Cs\'anyi}}]{Gaussian-Approximation-Potentials}%
  \BibitemOpen
  \bibfield  {author} {\bibinfo {author} {\bibfnamefont {A.~P.}\ \bibnamefont
  {Bart\'ok}}, \bibinfo {author} {\bibfnamefont {M.~C.}\ \bibnamefont {Payne}},
  \bibinfo {author} {\bibfnamefont {R.}~\bibnamefont {Kondor}}, \ and\ \bibinfo
  {author} {\bibfnamefont {G.}~\bibnamefont {Cs\'anyi}},\ }\bibfield  {title}
  {\enquote {\bibinfo {title} {Gaussian approximation potentials: The accuracy
  of quantum mechanics, without the electrons},}\ }\href {\doibase
  10.1103/PhysRevLett.104.136403} {\bibfield  {journal} {\bibinfo  {journal}
  {Phys. Rev. Lett.}\ }\textbf {\bibinfo {volume} {104}},\ \bibinfo {pages}
  {136403} (\bibinfo {year} {2010})}\BibitemShut {NoStop}%
\bibitem [{\citenamefont {Deringer}, \citenamefont {Caro},\ and\ \citenamefont
  {Csányi}(2019)}]{Gabor-MLIP-review}%
  \BibitemOpen
  \bibfield  {author} {\bibinfo {author} {\bibfnamefont {V.~L.}\ \bibnamefont
  {Deringer}}, \bibinfo {author} {\bibfnamefont {M.~A.}\ \bibnamefont {Caro}},
  \ and\ \bibinfo {author} {\bibfnamefont {G.}~\bibnamefont {Csányi}},\
  }\bibfield  {title} {\enquote {\bibinfo {title} {Machine learning interatomic
  potentials as emerging tools for materials science},}\ }\href@noop {}
  {\bibfield  {journal} {\bibinfo  {journal} {Advanced Materials}\ }\textbf
  {\bibinfo {volume} {31}},\ \bibinfo {pages} {1902765} (\bibinfo {year}
  {2019})}\BibitemShut {NoStop}%
\bibitem [{\citenamefont {Hohenberg}\ and\ \citenamefont
  {Kohn}(1964)}]{DFT-Hohenberg-Kohn}%
  \BibitemOpen
  \bibfield  {author} {\bibinfo {author} {\bibfnamefont {P.}~\bibnamefont
  {Hohenberg}}\ and\ \bibinfo {author} {\bibfnamefont {W.}~\bibnamefont
  {Kohn}},\ }\bibfield  {title} {\enquote {\bibinfo {title} {Inhomogeneous
  electron gas},}\ }\href {\doibase 10.1103/PhysRev.136.B864} {\bibfield
  {journal} {\bibinfo  {journal} {Phys. Rev.}\ }\textbf {\bibinfo {volume}
  {136}},\ \bibinfo {pages} {B864--B871} (\bibinfo {year} {1964})}\BibitemShut
  {NoStop}%
\bibitem [{\citenamefont {Kohn}\ and\ \citenamefont
  {Sham}(1965)}]{DFT-Kohn-Sham}%
  \BibitemOpen
  \bibfield  {author} {\bibinfo {author} {\bibfnamefont {W.}~\bibnamefont
  {Kohn}}\ and\ \bibinfo {author} {\bibfnamefont {L.~J.}\ \bibnamefont
  {Sham}},\ }\bibfield  {title} {\enquote {\bibinfo {title} {Self-consistent
  equations including exchange and correlation effects},}\ }\href {\doibase
  10.1103/PhysRev.140.A1133} {\bibfield  {journal} {\bibinfo  {journal} {Phys.
  Rev.}\ }\textbf {\bibinfo {volume} {140}},\ \bibinfo {pages} {A1133--A1138}
  (\bibinfo {year} {1965})}\BibitemShut {NoStop}%
\bibitem [{\citenamefont {Bart\'ok}\ \emph {et~al.}(2018)\citenamefont
  {Bart\'ok}, \citenamefont {Kermode}, \citenamefont {Bernstein},\ and\
  \citenamefont {Cs\'anyi}}]{Gabor-Silicon-GAP}%
  \BibitemOpen
  \bibfield  {author} {\bibinfo {author} {\bibfnamefont {A.~P.}\ \bibnamefont
  {Bart\'ok}}, \bibinfo {author} {\bibfnamefont {J.}~\bibnamefont {Kermode}},
  \bibinfo {author} {\bibfnamefont {N.}~\bibnamefont {Bernstein}}, \ and\
  \bibinfo {author} {\bibfnamefont {G.}~\bibnamefont {Cs\'anyi}},\ }\bibfield
  {title} {\enquote {\bibinfo {title} {Machine learning a general-purpose
  interatomic potential for silicon},}\ }\href {\doibase
  10.1103/PhysRevX.8.041048} {\bibfield  {journal} {\bibinfo  {journal} {Phys.
  Rev. X}\ }\textbf {\bibinfo {volume} {8}},\ \bibinfo {pages} {041048}
  (\bibinfo {year} {2018})}\BibitemShut {NoStop}%
\bibitem [{\citenamefont {Cheng}\ \emph {et~al.}(2020)\citenamefont {Cheng},
  \citenamefont {Mazzola}, \citenamefont {Pickard},\ and\ \citenamefont
  {Ceriotti}}]{Bingqing-Chris-hydrogen-MLP}%
  \BibitemOpen
  \bibfield  {author} {\bibinfo {author} {\bibfnamefont {B.}~\bibnamefont
  {Cheng}}, \bibinfo {author} {\bibfnamefont {G.}~\bibnamefont {Mazzola}},
  \bibinfo {author} {\bibfnamefont {C.~J.}\ \bibnamefont {Pickard}}, \ and\
  \bibinfo {author} {\bibfnamefont {M.}~\bibnamefont {Ceriotti}},\ }\bibfield
  {title} {\enquote {\bibinfo {title} {Evidence for supercritical behaviour of
  high-pressure liquid hydrogen},}\ }\href {\doibase 10.1038/s41586-020-2677-y}
  {\bibfield  {journal} {\bibinfo  {journal} {Nature}\ }\textbf {\bibinfo
  {volume} {585}},\ \bibinfo {pages} {217--220} (\bibinfo {year}
  {2020})}\BibitemShut {NoStop}%
\bibitem [{\citenamefont {Cheng}\ \emph {et~al.}(2019)\citenamefont {Cheng},
  \citenamefont {Engel}, \citenamefont {Behler}, \citenamefont {Dellago},\ and\
  \citenamefont {Ceriotti}}]{Bingqing-water-MLP}%
  \BibitemOpen
  \bibfield  {author} {\bibinfo {author} {\bibfnamefont {B.}~\bibnamefont
  {Cheng}}, \bibinfo {author} {\bibfnamefont {E.~A.}\ \bibnamefont {Engel}},
  \bibinfo {author} {\bibfnamefont {J.}~\bibnamefont {Behler}}, \bibinfo
  {author} {\bibfnamefont {C.}~\bibnamefont {Dellago}}, \ and\ \bibinfo
  {author} {\bibfnamefont {M.}~\bibnamefont {Ceriotti}},\ }\bibfield  {title}
  {\enquote {\bibinfo {title} {Ab initio thermodynamics of liquid and solid
  water},}\ }\href {\doibase 10.1073/pnas.1815117116} {\bibfield  {journal}
  {\bibinfo  {journal} {Proceedings of the National Academy of Sciences}\
  }\textbf {\bibinfo {volume} {116}},\ \bibinfo {pages} {1110--1115} (\bibinfo
  {year} {2019})}\BibitemShut {NoStop}%
\bibitem [{\citenamefont {Sosso}\ \emph {et~al.}(2012)\citenamefont {Sosso},
  \citenamefont {Miceli}, \citenamefont {Caravati}, \citenamefont {Behler},\
  and\ \citenamefont {Bernasconi}}]{GeTe-MLP}%
  \BibitemOpen
  \bibfield  {author} {\bibinfo {author} {\bibfnamefont {G.~C.}\ \bibnamefont
  {Sosso}}, \bibinfo {author} {\bibfnamefont {G.}~\bibnamefont {Miceli}},
  \bibinfo {author} {\bibfnamefont {S.}~\bibnamefont {Caravati}}, \bibinfo
  {author} {\bibfnamefont {J.}~\bibnamefont {Behler}}, \ and\ \bibinfo {author}
  {\bibfnamefont {M.}~\bibnamefont {Bernasconi}},\ }\bibfield  {title}
  {\enquote {\bibinfo {title} {Neural network interatomic potential for the
  phase change material {GeTe}},}\ }\href {\doibase 10.1103/PhysRevB.85.174103}
  {\bibfield  {journal} {\bibinfo  {journal} {Phys. Rev. B}\ }\textbf {\bibinfo
  {volume} {85}},\ \bibinfo {pages} {174103} (\bibinfo {year}
  {2012})}\BibitemShut {NoStop}%
\bibitem [{\citenamefont {Mocanu}\ \emph {et~al.}(2018)\citenamefont {Mocanu},
  \citenamefont {Konstantinou}, \citenamefont {Lee}, \citenamefont {Bernstein},
  \citenamefont {Deringer}, \citenamefont {Cs{\'a}nyi},\ and\ \citenamefont
  {Elliott}}]{Gabor-GeSbTe-MLP}%
  \BibitemOpen
  \bibfield  {author} {\bibinfo {author} {\bibfnamefont {F.~C.}\ \bibnamefont
  {Mocanu}}, \bibinfo {author} {\bibfnamefont {K.}~\bibnamefont
  {Konstantinou}}, \bibinfo {author} {\bibfnamefont {T.~H.}\ \bibnamefont
  {Lee}}, \bibinfo {author} {\bibfnamefont {N.}~\bibnamefont {Bernstein}},
  \bibinfo {author} {\bibfnamefont {V.~L.}\ \bibnamefont {Deringer}}, \bibinfo
  {author} {\bibfnamefont {G.}~\bibnamefont {Cs{\'a}nyi}}, \ and\ \bibinfo
  {author} {\bibfnamefont {S.~R.}\ \bibnamefont {Elliott}},\ }\bibfield
  {title} {\enquote {\bibinfo {title} {Modeling the phase-change memory
  material, {Ge2Sb2Te5}, with a machine-learned interatomic potential},}\
  }\href {\doibase 10.1021/acs.jpcb.8b06476} {\bibfield  {journal} {\bibinfo
  {journal} {The Journal of Physical Chemistry B}\ }\textbf {\bibinfo {volume}
  {122}},\ \bibinfo {pages} {8998--9006} (\bibinfo {year} {2018})}\BibitemShut
  {NoStop}%
\bibitem [{\citenamefont {Thompson}\ \emph {et~al.}(2015)\citenamefont
  {Thompson}, \citenamefont {Swiler}, \citenamefont {Trott}, \citenamefont
  {Foiles},\ and\ \citenamefont {Tucker}}]{SNAP-MLP}%
  \BibitemOpen
  \bibfield  {author} {\bibinfo {author} {\bibfnamefont {A.}~\bibnamefont
  {Thompson}}, \bibinfo {author} {\bibfnamefont {L.}~\bibnamefont {Swiler}},
  \bibinfo {author} {\bibfnamefont {C.}~\bibnamefont {Trott}}, \bibinfo
  {author} {\bibfnamefont {S.}~\bibnamefont {Foiles}}, \ and\ \bibinfo {author}
  {\bibfnamefont {G.}~\bibnamefont {Tucker}},\ }\bibfield  {title} {\enquote
  {\bibinfo {title} {Spectral neighbor analysis method for automated generation
  of quantum-accurate interatomic potentials},}\ }\href {\doibase
  https://doi.org/10.1016/j.jcp.2014.12.018} {\bibfield  {journal} {\bibinfo
  {journal} {Journal of Computational Physics}\ }\textbf {\bibinfo {volume}
  {285}},\ \bibinfo {pages} {316--330} (\bibinfo {year} {2015})}\BibitemShut
  {NoStop}%
\bibitem [{\citenamefont {Zhang}\ \emph {et~al.}(2018)\citenamefont {Zhang},
  \citenamefont {Han}, \citenamefont {Wang}, \citenamefont {Car},\ and\
  \citenamefont {E}}]{DeepMD-MLP}%
  \BibitemOpen
  \bibfield  {author} {\bibinfo {author} {\bibfnamefont {L.}~\bibnamefont
  {Zhang}}, \bibinfo {author} {\bibfnamefont {J.}~\bibnamefont {Han}}, \bibinfo
  {author} {\bibfnamefont {H.}~\bibnamefont {Wang}}, \bibinfo {author}
  {\bibfnamefont {R.}~\bibnamefont {Car}}, \ and\ \bibinfo {author}
  {\bibfnamefont {W.}~\bibnamefont {E}},\ }\bibfield  {title} {\enquote
  {\bibinfo {title} {Deep potential molecular dynamics: A scalable model with
  the accuracy of quantum mechanics},}\ }\href {\doibase
  10.1103/PhysRevLett.120.143001} {\bibfield  {journal} {\bibinfo  {journal}
  {Phys. Rev. Lett.}\ }\textbf {\bibinfo {volume} {120}},\ \bibinfo {pages}
  {143001} (\bibinfo {year} {2018})}\BibitemShut {NoStop}%
\bibitem [{\citenamefont {Pickard}(2022)}]{Chris-EDDPs}%
  \BibitemOpen
  \bibfield  {author} {\bibinfo {author} {\bibfnamefont {C.~J.}\ \bibnamefont
  {Pickard}},\ }\bibfield  {title} {\enquote {\bibinfo {title} {Ephemeral data
  derived potentials for random structure search},}\ }\href {\doibase
  10.1103/PhysRevB.106.014102} {\bibfield  {journal} {\bibinfo  {journal}
  {Phys. Rev. B}\ }\textbf {\bibinfo {volume} {106}},\ \bibinfo {pages}
  {014102} (\bibinfo {year} {2022})}\BibitemShut {NoStop}%
\bibitem [{\citenamefont {Shapeev}(2016)}]{MTP}%
  \BibitemOpen
  \bibfield  {author} {\bibinfo {author} {\bibfnamefont {A.~V.}\ \bibnamefont
  {Shapeev}},\ }\bibfield  {title} {\enquote {\bibinfo {title} {Moment tensor
  potentials: A class of systematically improvable interatomic potentials},}\
  }\href {\doibase 10.1137/15M1054183} {\bibfield  {journal} {\bibinfo
  {journal} {Multiscale Modeling \& Simulation}\ }\textbf {\bibinfo {volume}
  {14}},\ \bibinfo {pages} {1153--1173} (\bibinfo {year} {2016})}\BibitemShut
  {NoStop}%
\bibitem [{\citenamefont {Drautz}(2020)}]{ACE-I}%
  \BibitemOpen
  \bibfield  {author} {\bibinfo {author} {\bibfnamefont {R.}~\bibnamefont
  {Drautz}},\ }\bibfield  {title} {\enquote {\bibinfo {title} {Atomic cluster
  expansion of scalar, vectorial, and tensorial properties including magnetism
  and charge transfer},}\ }\href {\doibase 10.1103/PhysRevB.102.024104}
  {\bibfield  {journal} {\bibinfo  {journal} {Phys. Rev. B}\ }\textbf {\bibinfo
  {volume} {102}},\ \bibinfo {pages} {024104} (\bibinfo {year}
  {2020})}\BibitemShut {NoStop}%
\bibitem [{\citenamefont {Dusson}\ \emph {et~al.}(2022)\citenamefont {Dusson},
  \citenamefont {Bachmayr}, \citenamefont {Cs{\'a}nyi}, \citenamefont {Drautz},
  \citenamefont {Etter}, \citenamefont {{van der Oord}},\ and\ \citenamefont
  {Ortner}}]{ACE-II}%
  \BibitemOpen
  \bibfield  {author} {\bibinfo {author} {\bibfnamefont {G.}~\bibnamefont
  {Dusson}}, \bibinfo {author} {\bibfnamefont {M.}~\bibnamefont {Bachmayr}},
  \bibinfo {author} {\bibfnamefont {G.}~\bibnamefont {Cs{\'a}nyi}}, \bibinfo
  {author} {\bibfnamefont {R.}~\bibnamefont {Drautz}}, \bibinfo {author}
  {\bibfnamefont {S.}~\bibnamefont {Etter}}, \bibinfo {author} {\bibfnamefont
  {C.}~\bibnamefont {{van der Oord}}}, \ and\ \bibinfo {author} {\bibfnamefont
  {C.}~\bibnamefont {Ortner}},\ }\bibfield  {title} {\enquote {\bibinfo {title}
  {Atomic cluster expansion: Completeness, efficiency and stability},}\ }\href
  {\doibase https://doi.org/10.1016/j.jcp.2022.110946} {\bibfield  {journal}
  {\bibinfo  {journal} {Journal of Computational Physics}\ }\textbf {\bibinfo
  {volume} {454}},\ \bibinfo {pages} {110946} (\bibinfo {year}
  {2022})}\BibitemShut {NoStop}%
\bibitem [{\citenamefont {Batatia}\ \emph {et~al.}(2022)\citenamefont
  {Batatia}, \citenamefont {Kovacs}, \citenamefont {Simm}, \citenamefont
  {Ortner},\ and\ \citenamefont {Csanyi}}]{MACE}%
  \BibitemOpen
  \bibfield  {author} {\bibinfo {author} {\bibfnamefont {I.}~\bibnamefont
  {Batatia}}, \bibinfo {author} {\bibfnamefont {D.~P.}\ \bibnamefont {Kovacs}},
  \bibinfo {author} {\bibfnamefont {G.}~\bibnamefont {Simm}}, \bibinfo {author}
  {\bibfnamefont {C.}~\bibnamefont {Ortner}}, \ and\ \bibinfo {author}
  {\bibfnamefont {G.}~\bibnamefont {Csanyi}},\ }\bibfield  {title} {\enquote
  {\bibinfo {title} {{MACE}: Higher order equivariant message passing neural
  networks for fast and accurate force fields},}\ }in\ \href
  {https://proceedings.neurips.cc/paper_files/paper/2022/file/4a36c3c51af11ed9f34615b81edb5bbc-Paper-Conference.pdf}
  {\emph {\bibinfo {booktitle} {Advances in Neural Information Processing
  Systems}}},\ Vol.~\bibinfo {volume} {35},\ \bibinfo {editor} {edited by\
  \bibinfo {editor} {\bibfnamefont {S.}~\bibnamefont {Koyejo}}, \bibinfo
  {editor} {\bibfnamefont {S.}~\bibnamefont {Mohamed}}, \bibinfo {editor}
  {\bibfnamefont {A.}~\bibnamefont {Agarwal}}, \bibinfo {editor} {\bibfnamefont
  {D.}~\bibnamefont {Belgrave}}, \bibinfo {editor} {\bibfnamefont
  {K.}~\bibnamefont {Cho}}, \ and\ \bibinfo {editor} {\bibfnamefont
  {A.}~\bibnamefont {Oh}}}\ (\bibinfo  {publisher} {Curran Associates, Inc.},\
  \bibinfo {year} {2022})\ pp.\ \bibinfo {pages} {11423--11436}\BibitemShut
  {NoStop}%
\bibitem [{\citenamefont {Musaelian}\ \emph {et~al.}(2023)\citenamefont
  {Musaelian}, \citenamefont {Batzner}, \citenamefont {Johansson},
  \citenamefont {Sun}, \citenamefont {Owen}, \citenamefont {Kornbluth},\ and\
  \citenamefont {Kozinsky}}]{Kozinsky-Allegro-NNP-systematic}%
  \BibitemOpen
  \bibfield  {author} {\bibinfo {author} {\bibfnamefont {A.}~\bibnamefont
  {Musaelian}}, \bibinfo {author} {\bibfnamefont {S.}~\bibnamefont {Batzner}},
  \bibinfo {author} {\bibfnamefont {A.}~\bibnamefont {Johansson}}, \bibinfo
  {author} {\bibfnamefont {L.}~\bibnamefont {Sun}}, \bibinfo {author}
  {\bibfnamefont {C.~J.}\ \bibnamefont {Owen}}, \bibinfo {author}
  {\bibfnamefont {M.}~\bibnamefont {Kornbluth}}, \ and\ \bibinfo {author}
  {\bibfnamefont {B.}~\bibnamefont {Kozinsky}},\ }\bibfield  {title} {\enquote
  {\bibinfo {title} {Learning local equivariant representations for large-scale
  atomistic dynamics},}\ }\href {\doibase 10.1038/s41467-023-36329-y}
  {\bibfield  {journal} {\bibinfo  {journal} {Nature Communications}\ }\textbf
  {\bibinfo {volume} {14}},\ \bibinfo {pages} {579} (\bibinfo {year}
  {2023})}\BibitemShut {NoStop}%
\bibitem [{\citenamefont {Behler}\ and\ \citenamefont
  {Cs{\'a}nyi}(2021)}]{Gabor-Behler-MLP-perspective}%
  \BibitemOpen
  \bibfield  {author} {\bibinfo {author} {\bibfnamefont {J.}~\bibnamefont
  {Behler}}\ and\ \bibinfo {author} {\bibfnamefont {G.}~\bibnamefont
  {Cs{\'a}nyi}},\ }\bibfield  {title} {\enquote {\bibinfo {title} {Machine
  learning potentials for extended systems: a perspective},}\ }\href {\doibase
  10.1140/epjb/s10051-021-00156-1} {\bibfield  {journal} {\bibinfo  {journal}
  {The European Physical Journal B}\ }\textbf {\bibinfo {volume} {94}},\
  \bibinfo {pages} {142} (\bibinfo {year} {2021})}\BibitemShut {NoStop}%
\bibitem [{\citenamefont {Behler}(2021)}]{Behler-MLP-Four-Generations-Review}%
  \BibitemOpen
  \bibfield  {author} {\bibinfo {author} {\bibfnamefont {J.}~\bibnamefont
  {Behler}},\ }\bibfield  {title} {\enquote {\bibinfo {title} {Four generations
  of high-dimensional neural network potentials},}\ }\href {\doibase
  10.1021/acs.chemrev.0c00868} {\bibfield  {journal} {\bibinfo  {journal}
  {Chemical Reviews}\ }\textbf {\bibinfo {volume} {121}},\ \bibinfo {pages}
  {10037--10072} (\bibinfo {year} {2021})}\BibitemShut {NoStop}%
\bibitem [{\citenamefont {Chen}\ and\ \citenamefont
  {Ong}(2022)}]{Universal-graph-IAP}%
  \BibitemOpen
  \bibfield  {author} {\bibinfo {author} {\bibfnamefont {C.}~\bibnamefont
  {Chen}}\ and\ \bibinfo {author} {\bibfnamefont {S.~P.}\ \bibnamefont {Ong}},\
  }\bibfield  {title} {\enquote {\bibinfo {title} {A universal graph deep
  learning interatomic potential for the periodic table},}\ }\href {\doibase
  10.1038/s43588-022-00349-3} {\bibfield  {journal} {\bibinfo  {journal}
  {Nature Computational Science}\ }\textbf {\bibinfo {volume} {2}},\ \bibinfo
  {pages} {718--728} (\bibinfo {year} {2022})}\BibitemShut {NoStop}%
\bibitem [{\citenamefont {Choudhary}\ \emph {et~al.}(2023)\citenamefont
  {Choudhary}, \citenamefont {DeCost}, \citenamefont {Major}, \citenamefont
  {Butler}, \citenamefont {Thiyagalingam},\ and\ \citenamefont
  {Tavazza}}]{Unified-graph-NN}%
  \BibitemOpen
  \bibfield  {author} {\bibinfo {author} {\bibfnamefont {K.}~\bibnamefont
  {Choudhary}}, \bibinfo {author} {\bibfnamefont {B.}~\bibnamefont {DeCost}},
  \bibinfo {author} {\bibfnamefont {L.}~\bibnamefont {Major}}, \bibinfo
  {author} {\bibfnamefont {K.}~\bibnamefont {Butler}}, \bibinfo {author}
  {\bibfnamefont {J.}~\bibnamefont {Thiyagalingam}}, \ and\ \bibinfo {author}
  {\bibfnamefont {F.}~\bibnamefont {Tavazza}},\ }\bibfield  {title} {\enquote
  {\bibinfo {title} {Unified graph neural network force-field for the periodic
  table: solid state applications},}\ }\href {\doibase 10.1039/D2DD00096B}
  {\bibfield  {journal} {\bibinfo  {journal} {Digital Discovery}\ }\textbf
  {\bibinfo {volume} {2}},\ \bibinfo {pages} {346--355} (\bibinfo {year}
  {2023})}\BibitemShut {NoStop}%
\bibitem [{\citenamefont {Nigam}\ \emph {et~al.}(2022)\citenamefont {Nigam},
  \citenamefont {Pozdnyakov}, \citenamefont {Fraux},\ and\ \citenamefont
  {Ceriotti}}]{Atom-centred-message-passing-unified}%
  \BibitemOpen
  \bibfield  {author} {\bibinfo {author} {\bibfnamefont {J.}~\bibnamefont
  {Nigam}}, \bibinfo {author} {\bibfnamefont {S.}~\bibnamefont {Pozdnyakov}},
  \bibinfo {author} {\bibfnamefont {G.}~\bibnamefont {Fraux}}, \ and\ \bibinfo
  {author} {\bibfnamefont {M.}~\bibnamefont {Ceriotti}},\ }\bibfield  {title}
  {\enquote {\bibinfo {title} {{Unified theory of atom-centered representations
  and message-passing machine-learning schemes}},}\ }\href
  {https://doi.org/10.1063/5.0087042} {\bibfield  {journal} {\bibinfo
  {journal} {The Journal of Chemical Physics}\ }\textbf {\bibinfo {volume}
  {156}} (\bibinfo {year} {2022})}\BibitemShut {NoStop}%
\bibitem [{\citenamefont {Behler}(2011)}]{Behler-descriptors}%
  \BibitemOpen
  \bibfield  {author} {\bibinfo {author} {\bibfnamefont {J.}~\bibnamefont
  {Behler}},\ }\bibfield  {title} {\enquote {\bibinfo {title} {{Atom-centered
  symmetry functions for constructing high-dimensional neural network
  potentials}},}\ }\href {https://doi.org/10.1063/1.3553717} {\bibfield
  {journal} {\bibinfo  {journal} {The Journal of Chemical Physics}\ }\textbf
  {\bibinfo {volume} {134}} (\bibinfo {year} {2011})},\ \bibinfo {note}
  {074106}\BibitemShut {NoStop}%
\bibitem [{\citenamefont {Bart\'ok}, \citenamefont {Kondor},\ and\
  \citenamefont {Cs\'anyi}(2013)}]{Gabor-SOAP}%
  \BibitemOpen
  \bibfield  {author} {\bibinfo {author} {\bibfnamefont {A.~P.}\ \bibnamefont
  {Bart\'ok}}, \bibinfo {author} {\bibfnamefont {R.}~\bibnamefont {Kondor}}, \
  and\ \bibinfo {author} {\bibfnamefont {G.}~\bibnamefont {Cs\'anyi}},\
  }\bibfield  {title} {\enquote {\bibinfo {title} {On representing chemical
  environments},}\ }\href {\doibase 10.1103/PhysRevB.87.184115} {\bibfield
  {journal} {\bibinfo  {journal} {Phys. Rev. B}\ }\textbf {\bibinfo {volume}
  {87}},\ \bibinfo {pages} {184115} (\bibinfo {year} {2013})}\BibitemShut
  {NoStop}%
\bibitem [{\citenamefont {Willatt}, \citenamefont {Musil},\ and\ \citenamefont
  {Ceriotti}(2019)}]{Atomic-density-representations}%
  \BibitemOpen
  \bibfield  {author} {\bibinfo {author} {\bibfnamefont {M.~J.}\ \bibnamefont
  {Willatt}}, \bibinfo {author} {\bibfnamefont {F.}~\bibnamefont {Musil}}, \
  and\ \bibinfo {author} {\bibfnamefont {M.}~\bibnamefont {Ceriotti}},\
  }\bibfield  {title} {\enquote {\bibinfo {title} {{Atom-density
  representations for machine learning}},}\ }\href@noop {} {\bibfield
  {journal} {\bibinfo  {journal} {The Journal of Chemical Physics}\ }\textbf
  {\bibinfo {volume} {150}} (\bibinfo {year} {2019})},\ \bibinfo {note}
  {154110}\BibitemShut {NoStop}%
\bibitem [{\citenamefont {Musil}\ \emph {et~al.}(2021)\citenamefont {Musil},
  \citenamefont {Grisafi}, \citenamefont {Bart{\'o}k}, \citenamefont {Ortner},
  \citenamefont {Cs{\'a}nyi},\ and\ \citenamefont
  {Ceriotti}}]{Gabor-descriptors}%
  \BibitemOpen
  \bibfield  {author} {\bibinfo {author} {\bibfnamefont {F.}~\bibnamefont
  {Musil}}, \bibinfo {author} {\bibfnamefont {A.}~\bibnamefont {Grisafi}},
  \bibinfo {author} {\bibfnamefont {A.~P.}\ \bibnamefont {Bart{\'o}k}},
  \bibinfo {author} {\bibfnamefont {C.}~\bibnamefont {Ortner}}, \bibinfo
  {author} {\bibfnamefont {G.}~\bibnamefont {Cs{\'a}nyi}}, \ and\ \bibinfo
  {author} {\bibfnamefont {M.}~\bibnamefont {Ceriotti}},\ }\bibfield  {title}
  {\enquote {\bibinfo {title} {Physics-inspired structural representations for
  molecules and materials},}\ }\href {\doibase 10.1021/acs.chemrev.1c00021}
  {\bibfield  {journal} {\bibinfo  {journal} {Chemical Reviews}\ }\textbf
  {\bibinfo {volume} {121}},\ \bibinfo {pages} {9759--9815} (\bibinfo {year}
  {2021})}\BibitemShut {NoStop}%
\bibitem [{\citenamefont {Smith}\ \emph {et~al.}(2018)\citenamefont {Smith},
  \citenamefont {Nebgen}, \citenamefont {Lubbers}, \citenamefont {Isayev},\
  and\ \citenamefont {Roitberg}}]{Efficient-database-generation}%
  \BibitemOpen
  \bibfield  {author} {\bibinfo {author} {\bibfnamefont {J.~S.}\ \bibnamefont
  {Smith}}, \bibinfo {author} {\bibfnamefont {B.}~\bibnamefont {Nebgen}},
  \bibinfo {author} {\bibfnamefont {N.}~\bibnamefont {Lubbers}}, \bibinfo
  {author} {\bibfnamefont {O.}~\bibnamefont {Isayev}}, \ and\ \bibinfo {author}
  {\bibfnamefont {A.~E.}\ \bibnamefont {Roitberg}},\ }\bibfield  {title}
  {\enquote {\bibinfo {title} {{Less is more: Sampling chemical space with
  active learning}},}\ }\href {https://doi.org/10.1063/1.5023802} {\bibfield
  {journal} {\bibinfo  {journal} {The Journal of Chemical Physics}\ }\textbf
  {\bibinfo {volume} {148}} (\bibinfo {year} {2018})}\BibitemShut {NoStop}%
\bibitem [{\citenamefont {Novikov}\ \emph {et~al.}(2020)\citenamefont
  {Novikov}, \citenamefont {Gubaev}, \citenamefont {Podryabinkin},\ and\
  \citenamefont {Shapeev}}]{Shapeev-MLIP}%
  \BibitemOpen
  \bibfield  {author} {\bibinfo {author} {\bibfnamefont {I.~S.}\ \bibnamefont
  {Novikov}}, \bibinfo {author} {\bibfnamefont {K.}~\bibnamefont {Gubaev}},
  \bibinfo {author} {\bibfnamefont {E.~V.}\ \bibnamefont {Podryabinkin}}, \
  and\ \bibinfo {author} {\bibfnamefont {A.~V.}\ \bibnamefont {Shapeev}},\
  }\bibfield  {title} {\enquote {\bibinfo {title} {The {MLIP} package: moment
  tensor potentials with {MPI} and active learning},}\ }\href {\doibase
  10.1088/2632-2153/abc9fe} {\bibfield  {journal} {\bibinfo  {journal} {Machine
  Learning: Science and Technology}\ }\textbf {\bibinfo {volume} {2}},\
  \bibinfo {pages} {025002} (\bibinfo {year} {2020})}\BibitemShut {NoStop}%
\bibitem [{\citenamefont {Miksch}\ \emph {et~al.}(2021)\citenamefont {Miksch},
  \citenamefont {Morawietz}, \citenamefont {K{\"a}stner}, \citenamefont
  {Urban},\ and\ \citenamefont {Artrith}}]{MLP-construction-strategies}%
  \BibitemOpen
  \bibfield  {author} {\bibinfo {author} {\bibfnamefont {A.~M.}\ \bibnamefont
  {Miksch}}, \bibinfo {author} {\bibfnamefont {T.}~\bibnamefont {Morawietz}},
  \bibinfo {author} {\bibfnamefont {J.}~\bibnamefont {K{\"a}stner}}, \bibinfo
  {author} {\bibfnamefont {A.}~\bibnamefont {Urban}}, \ and\ \bibinfo {author}
  {\bibfnamefont {N.}~\bibnamefont {Artrith}},\ }\bibfield  {title} {\enquote
  {\bibinfo {title} {Strategies for the construction of machine-learning
  potentials for accurate and efficient atomic-scale simulations},}\ }\href
  {\doibase 10.1088/2632-2153/abfd96} {\bibfield  {journal} {\bibinfo
  {journal} {Machine Learning: Science and Technology}\ }\textbf {\bibinfo
  {volume} {2}},\ \bibinfo {pages} {031001} (\bibinfo {year}
  {2021})}\BibitemShut {NoStop}%
\bibitem [{\citenamefont {Podryabinkin}\ and\ \citenamefont
  {Shapeev}(2017)}]{Active-Learning-Shapeev}%
  \BibitemOpen
  \bibfield  {author} {\bibinfo {author} {\bibfnamefont {E.~V.}\ \bibnamefont
  {Podryabinkin}}\ and\ \bibinfo {author} {\bibfnamefont {A.~V.}\ \bibnamefont
  {Shapeev}},\ }\bibfield  {title} {\enquote {\bibinfo {title} {Active learning
  of linearly parametrized interatomic potentials},}\ }\href {\doibase
  https://doi.org/10.1016/j.commatsci.2017.08.031} {\bibfield  {journal}
  {\bibinfo  {journal} {Computational Materials Science}\ }\textbf {\bibinfo
  {volume} {140}},\ \bibinfo {pages} {171--180} (\bibinfo {year}
  {2017})}\BibitemShut {NoStop}%
\bibitem [{\citenamefont {Zhang}\ \emph {et~al.}(2019)\citenamefont {Zhang},
  \citenamefont {Lin}, \citenamefont {Wang}, \citenamefont {Car},\ and\
  \citenamefont {E}}]{Active-Learning-Car}%
  \BibitemOpen
  \bibfield  {author} {\bibinfo {author} {\bibfnamefont {L.}~\bibnamefont
  {Zhang}}, \bibinfo {author} {\bibfnamefont {D.-Y.}\ \bibnamefont {Lin}},
  \bibinfo {author} {\bibfnamefont {H.}~\bibnamefont {Wang}}, \bibinfo {author}
  {\bibfnamefont {R.}~\bibnamefont {Car}}, \ and\ \bibinfo {author}
  {\bibfnamefont {W.}~\bibnamefont {E}},\ }\bibfield  {title} {\enquote
  {\bibinfo {title} {Active learning of uniformly accurate interatomic
  potentials for materials simulation},}\ }\href {\doibase
  10.1103/PhysRevMaterials.3.023804} {\bibfield  {journal} {\bibinfo  {journal}
  {Phys. Rev. Mater.}\ }\textbf {\bibinfo {volume} {3}},\ \bibinfo {pages}
  {023804} (\bibinfo {year} {2019})}\BibitemShut {NoStop}%
\bibitem [{\citenamefont {van~der Oord}\ \emph {et~al.}(2023)\citenamefont
  {van~der Oord}, \citenamefont {Sachs}, \citenamefont {Kov{\'a}cs},
  \citenamefont {Ortner},\ and\ \citenamefont
  {Cs{\'a}nyi}}]{Hyperactive-learning}%
  \BibitemOpen
  \bibfield  {author} {\bibinfo {author} {\bibfnamefont {C.}~\bibnamefont
  {van~der Oord}}, \bibinfo {author} {\bibfnamefont {M.}~\bibnamefont {Sachs}},
  \bibinfo {author} {\bibfnamefont {D.~P.}\ \bibnamefont {Kov{\'a}cs}},
  \bibinfo {author} {\bibfnamefont {C.}~\bibnamefont {Ortner}}, \ and\ \bibinfo
  {author} {\bibfnamefont {G.}~\bibnamefont {Cs{\'a}nyi}},\ }\href {\doibase
  10.1038/s41524-023-01104-6} {\enquote {\bibinfo {title} {Hyperactive learning
  for data-driven interatomic potentials},}\ } (\bibinfo {year}
  {2023})\BibitemShut {NoStop}%
\bibitem [{\citenamefont {Kulichenko}\ \emph {et~al.}(2023)\citenamefont
  {Kulichenko}, \citenamefont {Barros}, \citenamefont {Lubbers}, \citenamefont
  {Li}, \citenamefont {Messerly}, \citenamefont {Tretiak}, \citenamefont
  {Smith},\ and\ \citenamefont {Nebgen}}]{Uncertainty-driven-active-learning}%
  \BibitemOpen
  \bibfield  {author} {\bibinfo {author} {\bibfnamefont {M.}~\bibnamefont
  {Kulichenko}}, \bibinfo {author} {\bibfnamefont {K.}~\bibnamefont {Barros}},
  \bibinfo {author} {\bibfnamefont {N.}~\bibnamefont {Lubbers}}, \bibinfo
  {author} {\bibfnamefont {Y.~W.}\ \bibnamefont {Li}}, \bibinfo {author}
  {\bibfnamefont {R.}~\bibnamefont {Messerly}}, \bibinfo {author}
  {\bibfnamefont {S.}~\bibnamefont {Tretiak}}, \bibinfo {author} {\bibfnamefont
  {J.~S.}\ \bibnamefont {Smith}}, \ and\ \bibinfo {author} {\bibfnamefont
  {B.}~\bibnamefont {Nebgen}},\ }\bibfield  {title} {\enquote {\bibinfo {title}
  {Uncertainty-driven dynamics for active learning of interatomic
  potentials},}\ }\href {\doibase 10.1038/s43588-023-00406-5} {\bibfield
  {journal} {\bibinfo  {journal} {Nature Computational Science}\ }\textbf
  {\bibinfo {volume} {3}},\ \bibinfo {pages} {230--239} (\bibinfo {year}
  {2023})}\BibitemShut {NoStop}%
\bibitem [{\citenamefont {Dolgirev}, \citenamefont {Kruglov},\ and\
  \citenamefont {Oganov}(2016)}]{USPEX-MLP}%
  \BibitemOpen
  \bibfield  {author} {\bibinfo {author} {\bibfnamefont {P.~E.}\ \bibnamefont
  {Dolgirev}}, \bibinfo {author} {\bibfnamefont {I.~A.}\ \bibnamefont
  {Kruglov}}, \ and\ \bibinfo {author} {\bibfnamefont {A.~R.}\ \bibnamefont
  {Oganov}},\ }\bibfield  {title} {\enquote {\bibinfo {title} {Machine learning
  scheme for fast extraction of chemically interpretable interatomic
  potentials},}\ }\href {\doibase 10.1063/1.4961886} {\bibfield  {journal}
  {\bibinfo  {journal} {AIP Advances}\ }\textbf {\bibinfo {volume} {6}},\
  \bibinfo {pages} {085318} (\bibinfo {year} {2016})}\BibitemShut {NoStop}%
\bibitem [{\citenamefont {Hajinazar}, \citenamefont {Shao},\ and\ \citenamefont
  {Kolmogorov}(2017)}]{Maise-MLP-I}%
  \BibitemOpen
  \bibfield  {author} {\bibinfo {author} {\bibfnamefont {S.}~\bibnamefont
  {Hajinazar}}, \bibinfo {author} {\bibfnamefont {J.}~\bibnamefont {Shao}}, \
  and\ \bibinfo {author} {\bibfnamefont {A.~N.}\ \bibnamefont {Kolmogorov}},\
  }\bibfield  {title} {\enquote {\bibinfo {title} {Stratified construction of
  neural network based interatomic models for multicomponent materials},}\
  }\href {\doibase 10.1103/PhysRevB.95.014114} {\bibfield  {journal} {\bibinfo
  {journal} {Phys. Rev. B}\ }\textbf {\bibinfo {volume} {95}},\ \bibinfo
  {pages} {014114} (\bibinfo {year} {2017})}\BibitemShut {NoStop}%
\bibitem [{\citenamefont {Hajinazar}\ \emph {et~al.}(2021)\citenamefont
  {Hajinazar}, \citenamefont {Thorn}, \citenamefont {Sandoval}, \citenamefont
  {Kharabadze},\ and\ \citenamefont {Kolmogorov}}]{Maise-MLP-II}%
  \BibitemOpen
  \bibfield  {author} {\bibinfo {author} {\bibfnamefont {S.}~\bibnamefont
  {Hajinazar}}, \bibinfo {author} {\bibfnamefont {A.}~\bibnamefont {Thorn}},
  \bibinfo {author} {\bibfnamefont {E.~D.}\ \bibnamefont {Sandoval}}, \bibinfo
  {author} {\bibfnamefont {S.}~\bibnamefont {Kharabadze}}, \ and\ \bibinfo
  {author} {\bibfnamefont {A.~N.}\ \bibnamefont {Kolmogorov}},\ }\bibfield
  {title} {\enquote {\bibinfo {title} {{MAISE}: Construction of neural network
  interatomic models and evolutionary structure optimization},}\ }\href
  {\doibase https://doi.org/10.1016/j.cpc.2020.107679} {\bibfield  {journal}
  {\bibinfo  {journal} {Computer Physics Communications}\ }\textbf {\bibinfo
  {volume} {259}},\ \bibinfo {pages} {107679} (\bibinfo {year}
  {2021})}\BibitemShut {NoStop}%
\bibitem [{\citenamefont {Deringer}, \citenamefont {Pickard},\ and\
  \citenamefont {Cs\'anyi}(2018)}]{GAP-RSS-boron}%
  \BibitemOpen
  \bibfield  {author} {\bibinfo {author} {\bibfnamefont {V.~L.}\ \bibnamefont
  {Deringer}}, \bibinfo {author} {\bibfnamefont {C.~J.}\ \bibnamefont
  {Pickard}}, \ and\ \bibinfo {author} {\bibfnamefont {G.}~\bibnamefont
  {Cs\'anyi}},\ }\bibfield  {title} {\enquote {\bibinfo {title} {Data-driven
  learning of total and local energies in elemental boron},}\ }\href {\doibase
  10.1103/PhysRevLett.120.156001} {\bibfield  {journal} {\bibinfo  {journal}
  {Phys. Rev. Lett.}\ }\textbf {\bibinfo {volume} {120}},\ \bibinfo {pages}
  {156001} (\bibinfo {year} {2018})}\BibitemShut {NoStop}%
\bibitem [{\citenamefont {Deringer}\ \emph {et~al.}(2018)\citenamefont
  {Deringer}, \citenamefont {Proserpio}, \citenamefont {Cs{\'a}nyi},\ and\
  \citenamefont {Pickard}}]{GAP-RSS-phosphorus}%
  \BibitemOpen
  \bibfield  {author} {\bibinfo {author} {\bibfnamefont {V.~L.}\ \bibnamefont
  {Deringer}}, \bibinfo {author} {\bibfnamefont {D.~M.}\ \bibnamefont
  {Proserpio}}, \bibinfo {author} {\bibfnamefont {G.}~\bibnamefont
  {Cs{\'a}nyi}}, \ and\ \bibinfo {author} {\bibfnamefont {C.~J.}\ \bibnamefont
  {Pickard}},\ }\bibfield  {title} {\enquote {\bibinfo {title} {Data-driven
  learning and prediction of inorganic crystal structures},}\ }\href {\doibase
  10.1039/C8FD00034D} {\bibfield  {journal} {\bibinfo  {journal} {Faraday
  Discuss.}\ }\textbf {\bibinfo {volume} {211}},\ \bibinfo {pages} {45--59}
  (\bibinfo {year} {2018})}\BibitemShut {NoStop}%
\bibitem [{\citenamefont {Oganov}\ \emph {et~al.}(2019)\citenamefont {Oganov},
  \citenamefont {Pickard}, \citenamefont {Zhu},\ and\ \citenamefont
  {Needs}}]{Structure-prediction-review}%
  \BibitemOpen
  \bibfield  {author} {\bibinfo {author} {\bibfnamefont {A.~R.}\ \bibnamefont
  {Oganov}}, \bibinfo {author} {\bibfnamefont {C.~J.}\ \bibnamefont {Pickard}},
  \bibinfo {author} {\bibfnamefont {Q.}~\bibnamefont {Zhu}}, \ and\ \bibinfo
  {author} {\bibfnamefont {R.~J.}\ \bibnamefont {Needs}},\ }\bibfield  {title}
  {\enquote {\bibinfo {title} {Structure prediction drives materials
  discovery},}\ }\href {\doibase 10.1038/s41578-019-0101-8} {\bibfield
  {journal} {\bibinfo  {journal} {Nature Reviews Materials}\ }\textbf {\bibinfo
  {volume} {4}},\ \bibinfo {pages} {331--348} (\bibinfo {year}
  {2019})}\BibitemShut {NoStop}%
\bibitem [{\citenamefont {Conway}, \citenamefont {Pickard},\ and\ \citenamefont
  {Hermann}(2023)}]{Lewis-Chris-CSP-review}%
  \BibitemOpen
  \bibfield  {author} {\bibinfo {author} {\bibfnamefont {L.~J.}\ \bibnamefont
  {Conway}}, \bibinfo {author} {\bibfnamefont {C.~J.}\ \bibnamefont {Pickard}},
  \ and\ \bibinfo {author} {\bibfnamefont {A.}~\bibnamefont {Hermann}},\
  }\bibfield  {title} {\enquote {\bibinfo {title} {3.12 - first principles
  crystal structure prediction},}\ }in\ \href {\doibase
  https://doi.org/10.1016/B978-0-12-823144-9.00173-4} {\emph {\bibinfo
  {booktitle} {Comprehensive Inorganic Chemistry III (Third Edition)}}},\
  \bibinfo {editor} {edited by\ \bibinfo {editor} {\bibfnamefont
  {J.}~\bibnamefont {Reedijk}}\ and\ \bibinfo {editor} {\bibfnamefont {K.~R.}\
  \bibnamefont {Poeppelmeier}}}\ (\bibinfo  {publisher} {Elsevier},\ \bibinfo
  {address} {Oxford},\ \bibinfo {year} {2023})\ \bibinfo {edition} {3rd}\ ed.,\
  pp.\ \bibinfo {pages} {393--420}\BibitemShut {NoStop}%
\bibitem [{\citenamefont {Pickard}\ and\ \citenamefont
  {Needs}(2006)}]{Chris-Silane}%
  \BibitemOpen
  \bibfield  {author} {\bibinfo {author} {\bibfnamefont {C.~J.}\ \bibnamefont
  {Pickard}}\ and\ \bibinfo {author} {\bibfnamefont {R.~J.}\ \bibnamefont
  {Needs}},\ }\bibfield  {title} {\enquote {\bibinfo {title} {High-pressure
  phases of silane},}\ }\href {\doibase 10.1103/PhysRevLett.97.045504}
  {\bibfield  {journal} {\bibinfo  {journal} {Phys. Rev. Lett.}\ }\textbf
  {\bibinfo {volume} {97}},\ \bibinfo {pages} {045504} (\bibinfo {year}
  {2006})}\BibitemShut {NoStop}%
\bibitem [{\citenamefont {Pickard}\ and\ \citenamefont {Needs}(2011)}]{AIRSS}%
  \BibitemOpen
  \bibfield  {author} {\bibinfo {author} {\bibfnamefont {C.~J.}\ \bibnamefont
  {Pickard}}\ and\ \bibinfo {author} {\bibfnamefont {R.~J.}\ \bibnamefont
  {Needs}},\ }\bibfield  {title} {\enquote {\bibinfo {title} {Ab initio random
  structure searching},}\ }\href {\doibase 10.1088/0953-8984/23/5/053201}
  {\bibfield  {journal} {\bibinfo  {journal} {Journal of Physics: Condensed
  Matter}\ }\textbf {\bibinfo {volume} {23}},\ \bibinfo {pages} {053201}
  (\bibinfo {year} {2011})}\BibitemShut {NoStop}%
\bibitem [{\citenamefont {Glass}, \citenamefont {Oganov},\ and\ \citenamefont
  {Hansen}(2006)}]{USPEX}%
  \BibitemOpen
  \bibfield  {author} {\bibinfo {author} {\bibfnamefont {C.~W.}\ \bibnamefont
  {Glass}}, \bibinfo {author} {\bibfnamefont {A.~R.}\ \bibnamefont {Oganov}}, \
  and\ \bibinfo {author} {\bibfnamefont {N.}~\bibnamefont {Hansen}},\
  }\bibfield  {title} {\enquote {\bibinfo {title} {{USPEX}---evolutionary
  crystal structure prediction},}\ }\href {\doibase
  https://doi.org/10.1016/j.cpc.2006.07.020} {\bibfield  {journal} {\bibinfo
  {journal} {Computer Physics Communications}\ }\textbf {\bibinfo {volume}
  {175}},\ \bibinfo {pages} {713--720} (\bibinfo {year} {2006})}\BibitemShut
  {NoStop}%
\bibitem [{\citenamefont {Lonie}\ and\ \citenamefont
  {Zurek}(2011)}]{Eva-Zurek-XtalOpt}%
  \BibitemOpen
  \bibfield  {author} {\bibinfo {author} {\bibfnamefont {D.~C.}\ \bibnamefont
  {Lonie}}\ and\ \bibinfo {author} {\bibfnamefont {E.}~\bibnamefont {Zurek}},\
  }\bibfield  {title} {\enquote {\bibinfo {title} {{XtalOpt}: An open-source
  evolutionary algorithm for crystal structure prediction},}\ }\href {\doibase
  https://doi.org/10.1016/j.cpc.2010.07.048} {\bibfield  {journal} {\bibinfo
  {journal} {Computer Physics Communications}\ }\textbf {\bibinfo {volume}
  {182}},\ \bibinfo {pages} {372--387} (\bibinfo {year} {2011})}\BibitemShut
  {NoStop}%
\bibitem [{\citenamefont {Wang}\ \emph {et~al.}(2010)\citenamefont {Wang},
  \citenamefont {Lv}, \citenamefont {Zhu},\ and\ \citenamefont {Ma}}]{Calypso}%
  \BibitemOpen
  \bibfield  {author} {\bibinfo {author} {\bibfnamefont {Y.}~\bibnamefont
  {Wang}}, \bibinfo {author} {\bibfnamefont {J.}~\bibnamefont {Lv}}, \bibinfo
  {author} {\bibfnamefont {L.}~\bibnamefont {Zhu}}, \ and\ \bibinfo {author}
  {\bibfnamefont {Y.}~\bibnamefont {Ma}},\ }\bibfield  {title} {\enquote
  {\bibinfo {title} {Crystal structure prediction via particle-swarm
  optimization},}\ }\href@noop {} {\bibfield  {journal} {\bibinfo  {journal}
  {Phys. Rev. B}\ }\textbf {\bibinfo {volume} {82}},\ \bibinfo {pages} {094116}
  (\bibinfo {year} {2010})}\BibitemShut {NoStop}%
\bibitem [{\citenamefont {Wales}\ and\ \citenamefont
  {Doye}(1997)}]{Basin-hopping}%
  \BibitemOpen
  \bibfield  {author} {\bibinfo {author} {\bibfnamefont {D.~J.}\ \bibnamefont
  {Wales}}\ and\ \bibinfo {author} {\bibfnamefont {J.~P.~K.}\ \bibnamefont
  {Doye}},\ }\bibfield  {title} {\enquote {\bibinfo {title} {Global
  optimization by basin-hopping and the lowest energy structures of
  {Lennard-Jones} clusters containing up to 110 atoms},}\ }\href
  {https://doi.org/10.1021/jp970984n} {\bibfield  {journal} {\bibinfo
  {journal} {The Journal of Physical Chemistry A}\ }\textbf {\bibinfo {volume}
  {101}},\ \bibinfo {pages} {5111--5116} (\bibinfo {year} {1997})}\BibitemShut
  {NoStop}%
\bibitem [{\citenamefont {Pickard}\ and\ \citenamefont
  {Needs}(2007{\natexlab{a}})}]{Chris-hydrogen-phase-III}%
  \BibitemOpen
  \bibfield  {author} {\bibinfo {author} {\bibfnamefont {C.~J.}\ \bibnamefont
  {Pickard}}\ and\ \bibinfo {author} {\bibfnamefont {R.~J.}\ \bibnamefont
  {Needs}},\ }\bibfield  {title} {\enquote {\bibinfo {title} {Structure of
  phase {III} of solid hydrogen},}\ }\href {\doibase 10.1038/nphys625}
  {\bibfield  {journal} {\bibinfo  {journal} {Nature Physics}\ }\textbf
  {\bibinfo {volume} {3}},\ \bibinfo {pages} {473--476} (\bibinfo {year}
  {2007}{\natexlab{a}})}\BibitemShut {NoStop}%
\bibitem [{\citenamefont {Pickard}\ and\ \citenamefont
  {Needs}(2010)}]{Chris-HG-aluminium}%
  \BibitemOpen
  \bibfield  {author} {\bibinfo {author} {\bibfnamefont {C.~J.}\ \bibnamefont
  {Pickard}}\ and\ \bibinfo {author} {\bibfnamefont {R.~J.}\ \bibnamefont
  {Needs}},\ }\bibfield  {title} {\enquote {\bibinfo {title} {Aluminium at
  terapascal pressures},}\ }\href {\doibase 10.1038/nmat2796} {\bibfield
  {journal} {\bibinfo  {journal} {Nature Materials}\ }\textbf {\bibinfo
  {volume} {9}},\ \bibinfo {pages} {624--627} (\bibinfo {year}
  {2010})}\BibitemShut {NoStop}%
\bibitem [{\citenamefont {Liu}\ \emph {et~al.}(2020)\citenamefont {Liu},
  \citenamefont {Gao}, \citenamefont {Hermann}, \citenamefont {Wang},
  \citenamefont {Miao}, \citenamefont {Pickard}, \citenamefont {Needs},
  \citenamefont {Wang}, \citenamefont {Xing},\ and\ \citenamefont
  {Sun}}]{Chris-helium-ammonia}%
  \BibitemOpen
  \bibfield  {author} {\bibinfo {author} {\bibfnamefont {C.}~\bibnamefont
  {Liu}}, \bibinfo {author} {\bibfnamefont {H.}~\bibnamefont {Gao}}, \bibinfo
  {author} {\bibfnamefont {A.}~\bibnamefont {Hermann}}, \bibinfo {author}
  {\bibfnamefont {Y.}~\bibnamefont {Wang}}, \bibinfo {author} {\bibfnamefont
  {M.}~\bibnamefont {Miao}}, \bibinfo {author} {\bibfnamefont {C.~J.}\
  \bibnamefont {Pickard}}, \bibinfo {author} {\bibfnamefont {R.~J.}\
  \bibnamefont {Needs}}, \bibinfo {author} {\bibfnamefont {H.-T.}\ \bibnamefont
  {Wang}}, \bibinfo {author} {\bibfnamefont {D.}~\bibnamefont {Xing}}, \ and\
  \bibinfo {author} {\bibfnamefont {J.}~\bibnamefont {Sun}},\ }\bibfield
  {title} {\enquote {\bibinfo {title} {Plastic and superionic helium ammonia
  compounds under high pressure and high temperature},}\ }\href {\doibase
  10.1103/PhysRevX.10.021007} {\bibfield  {journal} {\bibinfo  {journal} {Phys.
  Rev. X}\ }\textbf {\bibinfo {volume} {10}},\ \bibinfo {pages} {021007}
  (\bibinfo {year} {2020})}\BibitemShut {NoStop}%
\bibitem [{\citenamefont {Duan}\ \emph {et~al.}(2014)\citenamefont {Duan},
  \citenamefont {Liu}, \citenamefont {Tian}, \citenamefont {Li}, \citenamefont
  {Huang}, \citenamefont {Zhao}, \citenamefont {Yu}, \citenamefont {Liu},
  \citenamefont {Tian},\ and\ \citenamefont
  {Cui}}]{H3S-theoretical-prediction}%
  \BibitemOpen
  \bibfield  {author} {\bibinfo {author} {\bibfnamefont {D.}~\bibnamefont
  {Duan}}, \bibinfo {author} {\bibfnamefont {Y.}~\bibnamefont {Liu}}, \bibinfo
  {author} {\bibfnamefont {F.}~\bibnamefont {Tian}}, \bibinfo {author}
  {\bibfnamefont {D.}~\bibnamefont {Li}}, \bibinfo {author} {\bibfnamefont
  {X.}~\bibnamefont {Huang}}, \bibinfo {author} {\bibfnamefont
  {Z.}~\bibnamefont {Zhao}}, \bibinfo {author} {\bibfnamefont {H.}~\bibnamefont
  {Yu}}, \bibinfo {author} {\bibfnamefont {B.}~\bibnamefont {Liu}}, \bibinfo
  {author} {\bibfnamefont {W.}~\bibnamefont {Tian}}, \ and\ \bibinfo {author}
  {\bibfnamefont {T.}~\bibnamefont {Cui}},\ }\bibfield  {title} {\enquote
  {\bibinfo {title} {Pressure-induced metallization of dense {(H2S)2H2 with
  high-Tc} superconductivity},}\ }\href {\doibase 10.1038/srep06968} {\bibfield
   {journal} {\bibinfo  {journal} {Scientific Reports}\ }\textbf {\bibinfo
  {volume} {4}},\ \bibinfo {pages} {6968} (\bibinfo {year} {2014})}\BibitemShut
  {NoStop}%
\bibitem [{\citenamefont {Peng}\ \emph {et~al.}(2017)\citenamefont {Peng},
  \citenamefont {Sun}, \citenamefont {Pickard}, \citenamefont {Needs},
  \citenamefont {Wu},\ and\ \citenamefont {Ma}}]{Chris-clathrate-hydrides}%
  \BibitemOpen
  \bibfield  {author} {\bibinfo {author} {\bibfnamefont {F.}~\bibnamefont
  {Peng}}, \bibinfo {author} {\bibfnamefont {Y.}~\bibnamefont {Sun}}, \bibinfo
  {author} {\bibfnamefont {C.~J.}\ \bibnamefont {Pickard}}, \bibinfo {author}
  {\bibfnamefont {R.~J.}\ \bibnamefont {Needs}}, \bibinfo {author}
  {\bibfnamefont {Q.}~\bibnamefont {Wu}}, \ and\ \bibinfo {author}
  {\bibfnamefont {Y.}~\bibnamefont {Ma}},\ }\bibfield  {title} {\enquote
  {\bibinfo {title} {Hydrogen clathrate structures in rare earth hydrides at
  high pressures: Possible route to room-temperature superconductivity},}\
  }\href {\doibase 10.1103/PhysRevLett.119.107001} {\bibfield  {journal}
  {\bibinfo  {journal} {Phys. Rev. Lett.}\ }\textbf {\bibinfo {volume} {119}},\
  \bibinfo {pages} {107001} (\bibinfo {year} {2017})}\BibitemShut {NoStop}%
\bibitem [{\citenamefont {Liu}\ \emph {et~al.}(2017)\citenamefont {Liu},
  \citenamefont {Naumov}, \citenamefont {Hoffmann}, \citenamefont {Ashcroft},\
  and\ \citenamefont {Hemley}}]{Hemley-LaH10-prediction}%
  \BibitemOpen
  \bibfield  {author} {\bibinfo {author} {\bibfnamefont {H.}~\bibnamefont
  {Liu}}, \bibinfo {author} {\bibfnamefont {I.~I.}\ \bibnamefont {Naumov}},
  \bibinfo {author} {\bibfnamefont {R.}~\bibnamefont {Hoffmann}}, \bibinfo
  {author} {\bibfnamefont {N.~W.}\ \bibnamefont {Ashcroft}}, \ and\ \bibinfo
  {author} {\bibfnamefont {R.~J.}\ \bibnamefont {Hemley}},\ }\bibfield  {title}
  {\enquote {\bibinfo {title} {Potential high-{T}c superconducting lanthanum
  and yttrium hydrides at high pressure},}\ }\href {\doibase
  10.1073/pnas.1704505114} {\bibfield  {journal} {\bibinfo  {journal}
  {Proceedings of the National Academy of Sciences}\ }\textbf {\bibinfo
  {volume} {114}},\ \bibinfo {pages} {6990--6995} (\bibinfo {year}
  {2017})}\BibitemShut {NoStop}%
\bibitem [{\citenamefont {Ouyang}, \citenamefont {Xie},\ and\ \citenamefont
  {Jiang}(2015)}]{Au-cluster-NNP}%
  \BibitemOpen
  \bibfield  {author} {\bibinfo {author} {\bibfnamefont {R.}~\bibnamefont
  {Ouyang}}, \bibinfo {author} {\bibfnamefont {Y.}~\bibnamefont {Xie}}, \ and\
  \bibinfo {author} {\bibfnamefont {D.-e.}\ \bibnamefont {Jiang}},\ }\bibfield
  {title} {\enquote {\bibinfo {title} {Global minimization of gold clusters by
  combining neural network potentials and the basin-hopping method},}\ }\href
  {\doibase 10.1039/C5NR03903G} {\bibfield  {journal} {\bibinfo  {journal}
  {Nanoscale}\ }\textbf {\bibinfo {volume} {7}},\ \bibinfo {pages}
  {14817--14821} (\bibinfo {year} {2015})}\BibitemShut {NoStop}%
\bibitem [{\citenamefont {Deringer}, \citenamefont {Csányi},\ and\
  \citenamefont {Proserpio}(2017)}]{Gabor-Carbon-GAP-searching}%
  \BibitemOpen
  \bibfield  {author} {\bibinfo {author} {\bibfnamefont {V.~L.}\ \bibnamefont
  {Deringer}}, \bibinfo {author} {\bibfnamefont {G.}~\bibnamefont {Csányi}}, \
  and\ \bibinfo {author} {\bibfnamefont {D.~M.}\ \bibnamefont {Proserpio}},\
  }\bibfield  {title} {\enquote {\bibinfo {title} {Extracting crystal chemistry
  from amorphous carbon structures},}\ }\href {\doibase
  https://doi.org/10.1002/cphc.201700151} {\bibfield  {journal} {\bibinfo
  {journal} {ChemPhysChem}\ }\textbf {\bibinfo {volume} {18}},\ \bibinfo
  {pages} {873--877} (\bibinfo {year} {2017})}\BibitemShut {NoStop}%
\bibitem [{\citenamefont {Patra}\ \emph {et~al.}(2017)\citenamefont {Patra},
  \citenamefont {Meenakshisundaram}, \citenamefont {Hung},\ and\ \citenamefont
  {Simmons}}]{Evolutionary-algorithms-learn}%
  \BibitemOpen
  \bibfield  {author} {\bibinfo {author} {\bibfnamefont {T.~K.}\ \bibnamefont
  {Patra}}, \bibinfo {author} {\bibfnamefont {V.}~\bibnamefont
  {Meenakshisundaram}}, \bibinfo {author} {\bibfnamefont {J.-H.}\ \bibnamefont
  {Hung}}, \ and\ \bibinfo {author} {\bibfnamefont {D.~S.}\ \bibnamefont
  {Simmons}},\ }\bibfield  {title} {\enquote {\bibinfo {title}
  {Neural-network-biased genetic algorithms for materials design: Evolutionary
  algorithms that learn},}\ }\href {\doibase 10.1021/acscombsci.6b00136}
  {\bibfield  {journal} {\bibinfo  {journal} {ACS Combinatorial Science}\
  }\textbf {\bibinfo {volume} {19}},\ \bibinfo {pages} {96--107} (\bibinfo
  {year} {2017})}\BibitemShut {NoStop}%
\bibitem [{\citenamefont {Tong}\ \emph {et~al.}(2018)\citenamefont {Tong},
  \citenamefont {Xue}, \citenamefont {Lv}, \citenamefont {Wang},\ and\
  \citenamefont {Ma}}]{Calypso-NNP}%
  \BibitemOpen
  \bibfield  {author} {\bibinfo {author} {\bibfnamefont {Q.}~\bibnamefont
  {Tong}}, \bibinfo {author} {\bibfnamefont {L.}~\bibnamefont {Xue}}, \bibinfo
  {author} {\bibfnamefont {J.}~\bibnamefont {Lv}}, \bibinfo {author}
  {\bibfnamefont {Y.}~\bibnamefont {Wang}}, \ and\ \bibinfo {author}
  {\bibfnamefont {Y.}~\bibnamefont {Ma}},\ }\bibfield  {title} {\enquote
  {\bibinfo {title} {Accelerating {CALYPSO} structure prediction by data-driven
  learning of a potential energy surface},}\ }\href {\doibase
  10.1039/C8FD00055G} {\bibfield  {journal} {\bibinfo  {journal} {Faraday
  Discuss.}\ }\textbf {\bibinfo {volume} {211}},\ \bibinfo {pages} {31--43}
  (\bibinfo {year} {2018})}\BibitemShut {NoStop}%
\bibitem [{\citenamefont {Podryabinkin}\ \emph {et~al.}(2019)\citenamefont
  {Podryabinkin}, \citenamefont {Tikhonov}, \citenamefont {Shapeev},\ and\
  \citenamefont {Oganov}}]{USPEX-MTP}%
  \BibitemOpen
  \bibfield  {author} {\bibinfo {author} {\bibfnamefont {E.~V.}\ \bibnamefont
  {Podryabinkin}}, \bibinfo {author} {\bibfnamefont {E.~V.}\ \bibnamefont
  {Tikhonov}}, \bibinfo {author} {\bibfnamefont {A.~V.}\ \bibnamefont
  {Shapeev}}, \ and\ \bibinfo {author} {\bibfnamefont {A.~R.}\ \bibnamefont
  {Oganov}},\ }\bibfield  {title} {\enquote {\bibinfo {title} {Accelerating
  crystal structure prediction by machine-learning interatomic potentials with
  active learning},}\ }\href {\doibase 10.1103/PhysRevB.99.064114} {\bibfield
  {journal} {\bibinfo  {journal} {Phys. Rev. B}\ }\textbf {\bibinfo {volume}
  {99}},\ \bibinfo {pages} {064114} (\bibinfo {year} {2019})}\BibitemShut
  {NoStop}%
\bibitem [{\citenamefont {Gubaev}\ \emph {et~al.}(2019)\citenamefont {Gubaev},
  \citenamefont {Podryabinkin}, \citenamefont {Hart},\ and\ \citenamefont
  {Shapeev}}]{Shapeev-MLP-searches}%
  \BibitemOpen
  \bibfield  {author} {\bibinfo {author} {\bibfnamefont {K.}~\bibnamefont
  {Gubaev}}, \bibinfo {author} {\bibfnamefont {E.~V.}\ \bibnamefont
  {Podryabinkin}}, \bibinfo {author} {\bibfnamefont {G.~L.}\ \bibnamefont
  {Hart}}, \ and\ \bibinfo {author} {\bibfnamefont {A.~V.}\ \bibnamefont
  {Shapeev}},\ }\bibfield  {title} {\enquote {\bibinfo {title} {Accelerating
  high-throughput searches for new alloys with active learning of interatomic
  potentials},}\ }\href {\doibase
  https://doi.org/10.1016/j.commatsci.2018.09.031} {\bibfield  {journal}
  {\bibinfo  {journal} {Computational Materials Science}\ }\textbf {\bibinfo
  {volume} {156}},\ \bibinfo {pages} {148--156} (\bibinfo {year}
  {2019})}\BibitemShut {NoStop}%
\bibitem [{\citenamefont {Thorn}\ \emph {et~al.}(2019)\citenamefont {Thorn},
  \citenamefont {Rojas-Nunez}, \citenamefont {Hajinazar}, \citenamefont
  {Baltazar},\ and\ \citenamefont
  {Kolmogorov}}]{Kolmogorov-Au-cluster-searching}%
  \BibitemOpen
  \bibfield  {author} {\bibinfo {author} {\bibfnamefont {A.}~\bibnamefont
  {Thorn}}, \bibinfo {author} {\bibfnamefont {J.}~\bibnamefont {Rojas-Nunez}},
  \bibinfo {author} {\bibfnamefont {S.}~\bibnamefont {Hajinazar}}, \bibinfo
  {author} {\bibfnamefont {S.~E.}\ \bibnamefont {Baltazar}}, \ and\ \bibinfo
  {author} {\bibfnamefont {A.~N.}\ \bibnamefont {Kolmogorov}},\ }\bibfield
  {title} {\enquote {\bibinfo {title} {Toward ab initio ground states of gold
  clusters via neural network modeling},}\ }\href {\doibase
  10.1021/acs.jpcc.9b08517} {\bibfield  {journal} {\bibinfo  {journal} {The
  Journal of Physical Chemistry C}\ }\textbf {\bibinfo {volume} {123}},\
  \bibinfo {pages} {30088--30098} (\bibinfo {year} {2019})}\BibitemShut
  {NoStop}%
\bibitem [{\citenamefont {Bisbo}\ and\ \citenamefont
  {Hammer}(2020)}]{Surrogate-potential-Hammer}%
  \BibitemOpen
  \bibfield  {author} {\bibinfo {author} {\bibfnamefont {M.~K.}\ \bibnamefont
  {Bisbo}}\ and\ \bibinfo {author} {\bibfnamefont {B.}~\bibnamefont {Hammer}},\
  }\bibfield  {title} {\enquote {\bibinfo {title} {Efficient global structure
  optimization with a machine-learned surrogate model},}\ }\href {\doibase
  10.1103/PhysRevLett.124.086102} {\bibfield  {journal} {\bibinfo  {journal}
  {Phys. Rev. Lett.}\ }\textbf {\bibinfo {volume} {124}},\ \bibinfo {pages}
  {086102} (\bibinfo {year} {2020})}\BibitemShut {NoStop}%
\bibitem [{\citenamefont {Kaappa}, \citenamefont {del R\'{\i}o},\ and\
  \citenamefont {Jacobsen}(2021)}]{Surrogate-potential-Jacobsen}%
  \BibitemOpen
  \bibfield  {author} {\bibinfo {author} {\bibfnamefont {S.}~\bibnamefont
  {Kaappa}}, \bibinfo {author} {\bibfnamefont {E.~G.}\ \bibnamefont {del
  R\'{\i}o}}, \ and\ \bibinfo {author} {\bibfnamefont {K.~W.}\ \bibnamefont
  {Jacobsen}},\ }\bibfield  {title} {\enquote {\bibinfo {title} {Global
  optimization of atomic structures with gradient-enhanced gaussian process
  regression},}\ }\href {\doibase 10.1103/PhysRevB.103.174114} {\bibfield
  {journal} {\bibinfo  {journal} {Phys. Rev. B}\ }\textbf {\bibinfo {volume}
  {103}},\ \bibinfo {pages} {174114} (\bibinfo {year} {2021})}\BibitemShut
  {NoStop}%
\bibitem [{\citenamefont {{Dasenbrock-Gammon}}\ \emph
  {et~al.}(2023)\citenamefont {{Dasenbrock-Gammon}}, \citenamefont {Snider},
  \citenamefont {McBride}, \citenamefont {Pasan}, \citenamefont {Durkee},
  \citenamefont {{Khalvashi-Sutter}}, \citenamefont {Munasinghe}, \citenamefont
  {Dissanayake}, \citenamefont {Lawler}, \citenamefont {Salamat},\ and\
  \citenamefont {Dias}}]{Dasenbrock-Gammon2023}%
  \BibitemOpen
  \bibfield  {author} {\bibinfo {author} {\bibfnamefont {N.}~\bibnamefont
  {{Dasenbrock-Gammon}}}, \bibinfo {author} {\bibfnamefont {E.}~\bibnamefont
  {Snider}}, \bibinfo {author} {\bibfnamefont {R.}~\bibnamefont {McBride}},
  \bibinfo {author} {\bibfnamefont {H.}~\bibnamefont {Pasan}}, \bibinfo
  {author} {\bibfnamefont {D.}~\bibnamefont {Durkee}}, \bibinfo {author}
  {\bibfnamefont {N.}~\bibnamefont {{Khalvashi-Sutter}}}, \bibinfo {author}
  {\bibfnamefont {S.}~\bibnamefont {Munasinghe}}, \bibinfo {author}
  {\bibfnamefont {S.~E.}\ \bibnamefont {Dissanayake}}, \bibinfo {author}
  {\bibfnamefont {K.~V.}\ \bibnamefont {Lawler}}, \bibinfo {author}
  {\bibfnamefont {A.}~\bibnamefont {Salamat}}, \ and\ \bibinfo {author}
  {\bibfnamefont {R.~P.}\ \bibnamefont {Dias}},\ }\bibfield  {title} {\enquote
  {\bibinfo {title} {Evidence of near-ambient superconductivity in a
  {{N-doped}} lutetium hydride},}\ }\href {\doibase 10.1038/s41586-023-05742-0}
  {\bibfield  {journal} {\bibinfo  {journal} {Nature}\ }\textbf {\bibinfo
  {volume} {615}},\ \bibinfo {pages} {244--250} (\bibinfo {year}
  {2023})}\BibitemShut {NoStop}%
\bibitem [{\citenamefont {Xie}\ \emph {et~al.}(2023)\citenamefont {Xie},
  \citenamefont {Lu}, \citenamefont {Yu}, \citenamefont {Wang}, \citenamefont
  {Wang}, \citenamefont {Meng},\ and\ \citenamefont {Liu}}]{Xie2023}%
  \BibitemOpen
  \bibfield  {author} {\bibinfo {author} {\bibfnamefont {F.}~\bibnamefont
  {Xie}}, \bibinfo {author} {\bibfnamefont {T.}~\bibnamefont {Lu}}, \bibinfo
  {author} {\bibfnamefont {Z.}~\bibnamefont {Yu}}, \bibinfo {author}
  {\bibfnamefont {Y.}~\bibnamefont {Wang}}, \bibinfo {author} {\bibfnamefont
  {Z.}~\bibnamefont {Wang}}, \bibinfo {author} {\bibfnamefont {S.}~\bibnamefont
  {Meng}}, \ and\ \bibinfo {author} {\bibfnamefont {M.}~\bibnamefont {Liu}},\
  }\bibfield  {title} {\enquote {\bibinfo {title}
  {Lu\textendash{{H}}\textendash{{N Phase Diagram}} from {{First-Principles
  Calculations}}},}\ }\href {\doibase 10.1088/0256-307X/40/5/057401} {\bibfield
   {journal} {\bibinfo  {journal} {Chinese Physics Letters}\ }\textbf {\bibinfo
  {volume} {40}},\ \bibinfo {pages} {057401} (\bibinfo {year}
  {2023})}\BibitemShut {NoStop}%
\bibitem [{\citenamefont {Huo}\ \emph {et~al.}(2023)\citenamefont {Huo},
  \citenamefont {Duan}, \citenamefont {Ma}, \citenamefont {Zhang},
  \citenamefont {Jiang}, \citenamefont {An}, \citenamefont {Song},
  \citenamefont {Tian},\ and\ \citenamefont {Cui}}]{Huo2023}%
  \BibitemOpen
  \bibfield  {author} {\bibinfo {author} {\bibfnamefont {Z.}~\bibnamefont
  {Huo}}, \bibinfo {author} {\bibfnamefont {D.}~\bibnamefont {Duan}}, \bibinfo
  {author} {\bibfnamefont {T.}~\bibnamefont {Ma}}, \bibinfo {author}
  {\bibfnamefont {Z.}~\bibnamefont {Zhang}}, \bibinfo {author} {\bibfnamefont
  {Q.}~\bibnamefont {Jiang}}, \bibinfo {author} {\bibfnamefont
  {D.}~\bibnamefont {An}}, \bibinfo {author} {\bibfnamefont {H.}~\bibnamefont
  {Song}}, \bibinfo {author} {\bibfnamefont {F.}~\bibnamefont {Tian}}, \ and\
  \bibinfo {author} {\bibfnamefont {T.}~\bibnamefont {Cui}},\ }\bibfield
  {title} {\enquote {\bibinfo {title} {{First-principles study on the
  conventional superconductivity of {N-doped fcc-LuH3}}},}\ }\href {\doibase
  10.1063/5.0151844} {\bibfield  {journal} {\bibinfo  {journal} {Matter and
  Radiation at Extremes}\ }\textbf {\bibinfo {volume} {8}},\ \bibinfo {pages}
  {038402} (\bibinfo {year} {2023})},\ \Eprint
  {http://arxiv.org/abs/https://pubs.aip.org/aip/mre/article-pdf/doi/10.1063/5.0151844/18116117/038402\_1\_5.0151844.pdf}
  {https://pubs.aip.org/aip/mre/article-pdf/doi/10.1063/5.0151844/18116117/038402\_1\_5.0151844.pdf}
  \BibitemShut {NoStop}%
\bibitem [{\citenamefont {Hilleke}\ \emph {et~al.}(2023)\citenamefont
  {Hilleke}, \citenamefont {Wang}, \citenamefont {Luo}, \citenamefont {Geng},
  \citenamefont {Wang}, \citenamefont {Belli},\ and\ \citenamefont
  {Zurek}}]{Hilleke2023}%
  \BibitemOpen
  \bibfield  {author} {\bibinfo {author} {\bibfnamefont {K.~P.}\ \bibnamefont
  {Hilleke}}, \bibinfo {author} {\bibfnamefont {X.}~\bibnamefont {Wang}},
  \bibinfo {author} {\bibfnamefont {D.}~\bibnamefont {Luo}}, \bibinfo {author}
  {\bibfnamefont {N.}~\bibnamefont {Geng}}, \bibinfo {author} {\bibfnamefont
  {B.}~\bibnamefont {Wang}}, \bibinfo {author} {\bibfnamefont {F.}~\bibnamefont
  {Belli}}, \ and\ \bibinfo {author} {\bibfnamefont {E.}~\bibnamefont
  {Zurek}},\ }\bibfield  {title} {\enquote {\bibinfo {title} {Structure,
  stability, and superconductivity of {N-doped} lutetium hydrides at kbar
  pressures},}\ }\href {\doibase 10.1103/PhysRevB.108.014511} {\bibfield
  {journal} {\bibinfo  {journal} {Phys. Rev. B}\ }\textbf {\bibinfo {volume}
  {108}},\ \bibinfo {pages} {014511} (\bibinfo {year} {2023})}\BibitemShut
  {NoStop}%
\bibitem [{\citenamefont {Ferreira}\ \emph {et~al.}(2023)\citenamefont
  {Ferreira}, \citenamefont {Conway}, \citenamefont {Cucciari}, \citenamefont
  {Di~Cataldo}, \citenamefont {Giannessi}, \citenamefont {Kogler},
  \citenamefont {Eleno}, \citenamefont {Pickard}, \citenamefont {Heil},\ and\
  \citenamefont {Boeri}}]{Lewis-Chris-LuNH-search}%
  \BibitemOpen
  \bibfield  {author} {\bibinfo {author} {\bibfnamefont {P.~P.}\ \bibnamefont
  {Ferreira}}, \bibinfo {author} {\bibfnamefont {L.~J.}\ \bibnamefont
  {Conway}}, \bibinfo {author} {\bibfnamefont {A.}~\bibnamefont {Cucciari}},
  \bibinfo {author} {\bibfnamefont {S.}~\bibnamefont {Di~Cataldo}}, \bibinfo
  {author} {\bibfnamefont {F.}~\bibnamefont {Giannessi}}, \bibinfo {author}
  {\bibfnamefont {E.}~\bibnamefont {Kogler}}, \bibinfo {author} {\bibfnamefont
  {L.~T.~F.}\ \bibnamefont {Eleno}}, \bibinfo {author} {\bibfnamefont {C.~J.}\
  \bibnamefont {Pickard}}, \bibinfo {author} {\bibfnamefont {C.}~\bibnamefont
  {Heil}}, \ and\ \bibinfo {author} {\bibfnamefont {L.}~\bibnamefont {Boeri}},\
  }\bibfield  {title} {\enquote {\bibinfo {title} {Search for ambient
  superconductivity in the {Lu-N-H} system},}\ }\href {\doibase
  10.1038/s41467-023-41005-2} {\bibfield  {journal} {\bibinfo  {journal}
  {Nature Communications}\ }\textbf {\bibinfo {volume} {14}},\ \bibinfo {pages}
  {5367} (\bibinfo {year} {2023})}\BibitemShut {NoStop}%
\bibitem [{\citenamefont {Kim}\ \emph {et~al.}(2023)\citenamefont {Kim},
  \citenamefont {Conway}, \citenamefont {Pickard}, \citenamefont {Pascut},\
  and\ \citenamefont {Monserrat}}]{LuH-colour-theory}%
  \BibitemOpen
  \bibfield  {author} {\bibinfo {author} {\bibfnamefont {S.-W.}\ \bibnamefont
  {Kim}}, \bibinfo {author} {\bibfnamefont {L.~J.}\ \bibnamefont {Conway}},
  \bibinfo {author} {\bibfnamefont {C.~J.}\ \bibnamefont {Pickard}}, \bibinfo
  {author} {\bibfnamefont {G.~L.}\ \bibnamefont {Pascut}}, \ and\ \bibinfo
  {author} {\bibfnamefont {B.}~\bibnamefont {Monserrat}},\ }\href@noop {}
  {\enquote {\bibinfo {title} {Microscopic theory of colour in lutetium
  hydride},}\ } (\bibinfo {year} {2023}),\ \Eprint
  {http://arxiv.org/abs/2304.07326} {arXiv:2304.07326 [cond-mat.supr-con]}
  \BibitemShut {NoStop}%
\bibitem [{\citenamefont {Clark}\ \emph {et~al.}(2005)\citenamefont {Clark},
  \citenamefont {Segall}, \citenamefont {Pickard}, \citenamefont {Hasnip},
  \citenamefont {Probert}, \citenamefont {Refson},\ and\ \citenamefont
  {Payne}}]{CASTEP}%
  \BibitemOpen
  \bibfield  {author} {\bibinfo {author} {\bibfnamefont {S.~J.}\ \bibnamefont
  {Clark}}, \bibinfo {author} {\bibfnamefont {M.~D.}\ \bibnamefont {Segall}},
  \bibinfo {author} {\bibfnamefont {C.~J.}\ \bibnamefont {Pickard}}, \bibinfo
  {author} {\bibfnamefont {P.~J.}\ \bibnamefont {Hasnip}}, \bibinfo {author}
  {\bibfnamefont {M.~I.~J.}\ \bibnamefont {Probert}}, \bibinfo {author}
  {\bibfnamefont {K.}~\bibnamefont {Refson}}, \ and\ \bibinfo {author}
  {\bibfnamefont {M.~C.}\ \bibnamefont {Payne}},\ }\bibfield  {title} {\enquote
  {\bibinfo {title} {First principles methods using {CASTEP}},}\ }\href
  {\doibase https://doi.org/10.1524/zkri.220.5.567.65075} {\bibfield  {journal}
  {\bibinfo  {journal} {Zeitschrift f{\"u}r Kristallographie - Crystalline
  Materials}\ }\textbf {\bibinfo {volume} {220}},\ \bibinfo {pages} {567 --
  570} (\bibinfo {year} {2005})}\BibitemShut {NoStop}%
\bibitem [{\citenamefont {Jones}\ and\ \citenamefont
  {Chapman}(1924)}]{Lennard-Jones-I}%
  \BibitemOpen
  \bibfield  {author} {\bibinfo {author} {\bibfnamefont {J.~E.}\ \bibnamefont
  {Jones}}\ and\ \bibinfo {author} {\bibfnamefont {S.}~\bibnamefont
  {Chapman}},\ }\bibfield  {title} {\enquote {\bibinfo {title} {On the
  determination of molecular fields.—{I}. from the variation of the viscosity
  of a gas with temperature},}\ }\href@noop {} {\bibfield  {journal} {\bibinfo
  {journal} {Proceedings of the Royal Society of London. Series A, Containing
  Papers of a Mathematical and Physical Character}\ }\textbf {\bibinfo {volume}
  {106}},\ \bibinfo {pages} {441--462} (\bibinfo {year} {1924})}\BibitemShut
  {NoStop}%
\bibitem [{\citenamefont {Lennard-Jones}(1931)}]{Lennard-Jones-II}%
  \BibitemOpen
  \bibfield  {author} {\bibinfo {author} {\bibfnamefont {J.~E.}\ \bibnamefont
  {Lennard-Jones}},\ }\bibfield  {title} {\enquote {\bibinfo {title}
  {Cohesion},}\ }\href@noop {} {\bibfield  {journal} {\bibinfo  {journal}
  {Proceedings of the Physical Society}\ }\textbf {\bibinfo {volume} {43}},\
  \bibinfo {pages} {461} (\bibinfo {year} {1931})}\BibitemShut {NoStop}%
\bibitem [{\citenamefont {Born}\ and\ \citenamefont
  {Misra}(1940)}]{Extended-Lennard-Jones}%
  \BibitemOpen
  \bibfield  {author} {\bibinfo {author} {\bibfnamefont {M.}~\bibnamefont
  {Born}}\ and\ \bibinfo {author} {\bibfnamefont {R.~D.}\ \bibnamefont
  {Misra}},\ }\bibfield  {title} {\enquote {\bibinfo {title} {On the stability
  of crystal lattices. {IV}},}\ }\href {\doibase 10.1017/S0305004100017515}
  {\bibfield  {journal} {\bibinfo  {journal} {Mathematical Proceedings of the
  Cambridge Philosophical Society}\ }\textbf {\bibinfo {volume} {36}},\
  \bibinfo {pages} {466–478} (\bibinfo {year} {1940})}\BibitemShut {NoStop}%
\bibitem [{\citenamefont {Wang}\ \emph {et~al.}(2020)\citenamefont {Wang},
  \citenamefont {Ram{\'\i}rez-Hinestrosa}, \citenamefont {Dobnikar},\ and\
  \citenamefont {Frenkel}}]{Cutoff-LJ}%
  \BibitemOpen
  \bibfield  {author} {\bibinfo {author} {\bibfnamefont {X.}~\bibnamefont
  {Wang}}, \bibinfo {author} {\bibfnamefont {S.}~\bibnamefont
  {Ram{\'\i}rez-Hinestrosa}}, \bibinfo {author} {\bibfnamefont
  {J.}~\bibnamefont {Dobnikar}}, \ and\ \bibinfo {author} {\bibfnamefont
  {D.}~\bibnamefont {Frenkel}},\ }\bibfield  {title} {\enquote {\bibinfo
  {title} {The {Lennard-Jones} potential: when (not) to use it},}\ }\href
  {\doibase 10.1039/C9CP05445F} {\bibfield  {journal} {\bibinfo  {journal}
  {Phys. Chem. Chem. Phys.}\ }\textbf {\bibinfo {volume} {22}},\ \bibinfo
  {pages} {10624--10633} (\bibinfo {year} {2020})}\BibitemShut {NoStop}%
\bibitem [{\citenamefont {Schran}, \citenamefont {Brezina},\ and\ \citenamefont
  {Marsalek}(2020)}]{Committee-NNs}%
  \BibitemOpen
  \bibfield  {author} {\bibinfo {author} {\bibfnamefont {C.}~\bibnamefont
  {Schran}}, \bibinfo {author} {\bibfnamefont {K.}~\bibnamefont {Brezina}}, \
  and\ \bibinfo {author} {\bibfnamefont {O.}~\bibnamefont {Marsalek}},\
  }\bibfield  {title} {\enquote {\bibinfo {title} {Committee neural network
  potentials control generalization errors and enable active learning},}\
  }\href {\doibase 10.1063/5.0016004} {\bibfield  {journal} {\bibinfo
  {journal} {The Journal of Chemical Physics}\ }\textbf {\bibinfo {volume}
  {153}},\ \bibinfo {pages} {104105} (\bibinfo {year} {2020})}\BibitemShut
  {NoStop}%
\bibitem [{\citenamefont {Hansen}\ and\ \citenamefont
  {Salamon}(1990)}]{NN-ensembles}%
  \BibitemOpen
  \bibfield  {author} {\bibinfo {author} {\bibfnamefont {L.}~\bibnamefont
  {Hansen}}\ and\ \bibinfo {author} {\bibfnamefont {P.}~\bibnamefont
  {Salamon}},\ }\bibfield  {title} {\enquote {\bibinfo {title} {Neural network
  ensembles},}\ }\href {\doibase 10.1109/34.58871} {\bibfield  {journal}
  {\bibinfo  {journal} {IEEE Transactions on Pattern Analysis and Machine
  Intelligence}\ }\textbf {\bibinfo {volume} {12}},\ \bibinfo {pages}
  {993--1001} (\bibinfo {year} {1990})}\BibitemShut {NoStop}%
\bibitem [{\citenamefont {Poul}\ \emph {et~al.}(2023)\citenamefont {Poul},
  \citenamefont {Huber}, \citenamefont {Bitzek},\ and\ \citenamefont
  {Neugebauer}}]{Magnesium-small-cell-training}%
  \BibitemOpen
  \bibfield  {author} {\bibinfo {author} {\bibfnamefont {M.}~\bibnamefont
  {Poul}}, \bibinfo {author} {\bibfnamefont {L.}~\bibnamefont {Huber}},
  \bibinfo {author} {\bibfnamefont {E.}~\bibnamefont {Bitzek}}, \ and\ \bibinfo
  {author} {\bibfnamefont {J.}~\bibnamefont {Neugebauer}},\ }\bibfield  {title}
  {\enquote {\bibinfo {title} {Systematic atomic structure datasets for machine
  learning potentials: Application to defects in magnesium},}\ }\href {\doibase
  10.1103/PhysRevB.107.104103} {\bibfield  {journal} {\bibinfo  {journal}
  {Phys. Rev. B}\ }\textbf {\bibinfo {volume} {107}},\ \bibinfo {pages}
  {104103} (\bibinfo {year} {2023})}\BibitemShut {NoStop}%
\bibitem [{\citenamefont {Meziere}\ \emph {et~al.}(2023)\citenamefont
  {Meziere}, \citenamefont {Luo}, \citenamefont {Zia}, \citenamefont {Beland},
  \citenamefont {Daymond},\ and\ \citenamefont
  {Hart}}]{Hart-Zirconium-small-cell-training}%
  \BibitemOpen
  \bibfield  {author} {\bibinfo {author} {\bibfnamefont {J.~A.}\ \bibnamefont
  {Meziere}}, \bibinfo {author} {\bibfnamefont {Y.}~\bibnamefont {Luo}},
  \bibinfo {author} {\bibfnamefont {Y.}~\bibnamefont {Zia}}, \bibinfo {author}
  {\bibfnamefont {L.}~\bibnamefont {Beland}}, \bibinfo {author} {\bibfnamefont
  {M.}~\bibnamefont {Daymond}}, \ and\ \bibinfo {author} {\bibfnamefont
  {G.~L.~W.}\ \bibnamefont {Hart}},\ }\href@noop {} {\enquote {\bibinfo {title}
  {Accelerating training of {MLIPs} through small-cell training},}\ } (\bibinfo
  {year} {2023}),\ \Eprint {http://arxiv.org/abs/2304.01314} {arXiv:2304.01314
  [cond-mat.mtrl-sci]} \BibitemShut {NoStop}%
\bibitem [{\citenamefont {Levenberg}(1944)}]{Levenberg-Marquardt-I}%
  \BibitemOpen
  \bibfield  {author} {\bibinfo {author} {\bibfnamefont {K.}~\bibnamefont
  {Levenberg}},\ }\bibfield  {title} {\enquote {\bibinfo {title} {A method for
  the solution of certain non-linear problems in least squares},}\ }\href@noop
  {} {\bibfield  {journal} {\bibinfo  {journal} {Quart. Appl. Math.}\ }\textbf
  {\bibinfo {volume} {2}},\ \bibinfo {pages} {164--168} (\bibinfo {year}
  {1944})}\BibitemShut {NoStop}%
\bibitem [{\citenamefont {Mor{\'e}}(1978)}]{Levenberg-Marquardt-II}%
  \BibitemOpen
  \bibfield  {author} {\bibinfo {author} {\bibfnamefont {J.~J.}\ \bibnamefont
  {Mor{\'e}}},\ }\bibfield  {title} {\enquote {\bibinfo {title} {The
  {Levenberg-Marquardt} algorithm: Implementation and theory},}\ }in\
  \href@noop {} {\emph {\bibinfo {booktitle} {Numerical Analysis}}},\ \bibinfo
  {editor} {edited by\ \bibinfo {editor} {\bibfnamefont {G.~A.}\ \bibnamefont
  {Watson}}}\ (\bibinfo  {publisher} {Springer Berlin Heidelberg},\ \bibinfo
  {address} {Berlin, Heidelberg},\ \bibinfo {year} {1978})\ pp.\ \bibinfo
  {pages} {105--116}\BibitemShut {NoStop}%
\bibitem [{\citenamefont {Prechelt}(1998)}]{Early-stopping}%
  \BibitemOpen
  \bibfield  {author} {\bibinfo {author} {\bibfnamefont {L.}~\bibnamefont
  {Prechelt}},\ }\enquote {\bibinfo {title} {Early stopping - but when?}}\ in\
  \href {\doibase 10.1007/3-540-49430-8_3} {\emph {\bibinfo {booktitle} {Neural
  Networks: Tricks of the Trade}}},\ \bibinfo {editor} {edited by\ \bibinfo
  {editor} {\bibfnamefont {G.~B.}\ \bibnamefont {Orr}}\ and\ \bibinfo {editor}
  {\bibfnamefont {K.-R.}\ \bibnamefont {M{\"u}ller}}}\ (\bibinfo  {publisher}
  {Springer Berlin Heidelberg},\ \bibinfo {address} {Berlin, Heidelberg},\
  \bibinfo {year} {1998})\ pp.\ \bibinfo {pages} {55--69}\BibitemShut {NoStop}%
\bibitem [{\citenamefont {Chen}\ and\ \citenamefont {Plemmons}(2009)}]{NNLS}%
  \BibitemOpen
  \bibfield  {author} {\bibinfo {author} {\bibfnamefont {D.}~\bibnamefont
  {Chen}}\ and\ \bibinfo {author} {\bibfnamefont {R.}~\bibnamefont
  {Plemmons}},\ }\enquote {\bibinfo {title} {The birth of numerical
  analysis},}\ \ (\bibinfo  {publisher} {World Scientific},\ \bibinfo {year}
  {2009})\ Chap.~\bibinfo {chapter} {8},\ \bibinfo {edition} {1st}\
  ed.\BibitemShut {Stop}%
\bibitem [{\citenamefont {Neal}\ \emph {et~al.}(2019)\citenamefont {Neal},
  \citenamefont {Mittal}, \citenamefont {Baratin}, \citenamefont {Tantia},
  \citenamefont {Scicluna}, \citenamefont {Lacoste-Julien},\ and\ \citenamefont
  {Mitliagkas}}]{Bias-variance-tradeoff}%
  \BibitemOpen
  \bibfield  {author} {\bibinfo {author} {\bibfnamefont {B.}~\bibnamefont
  {Neal}}, \bibinfo {author} {\bibfnamefont {S.}~\bibnamefont {Mittal}},
  \bibinfo {author} {\bibfnamefont {A.}~\bibnamefont {Baratin}}, \bibinfo
  {author} {\bibfnamefont {V.}~\bibnamefont {Tantia}}, \bibinfo {author}
  {\bibfnamefont {M.}~\bibnamefont {Scicluna}}, \bibinfo {author}
  {\bibfnamefont {S.}~\bibnamefont {Lacoste-Julien}}, \ and\ \bibinfo {author}
  {\bibfnamefont {I.}~\bibnamefont {Mitliagkas}},\ }\href@noop {} {\enquote
  {\bibinfo {title} {A modern take on the bias-variance tradeoff in neural
  networks},}\ } (\bibinfo {year} {2019}),\ \Eprint
  {http://arxiv.org/abs/1810.08591} {arXiv:1810.08591 [cs.LG]} \BibitemShut
  {NoStop}%
\bibitem [{EDD()}]{EDDP-website}%
  \BibitemOpen
  \href@noop {} {}\bibinfo {howpublished}
  {\url{https://www.mtg.msm.cam.ac.uk/Codes/EDDP}}\BibitemShut {NoStop}%
\bibitem [{AIR()}]{AIRSS-website}%
  \BibitemOpen
  \href@noop {} {}\bibinfo {howpublished}
  {\url{https://www.mtg.msm.cam.ac.uk/Codes/AIRSS}}\BibitemShut {NoStop}%
\bibitem [{get()}]{getting-castep}%
  \BibitemOpen
  \href@noop {} {}\bibinfo {howpublished}
  {\url{http://www.castep.org/CASTEP/GettingCASTEP}}\BibitemShut {NoStop}%
\bibitem [{edd()}]{eddp-jl-github}%
  \BibitemOpen
  \href@noop {} {}\bibinfo {howpublished}
  {\url{https://github.com/zhubonan/EDDP.jl}}\BibitemShut {NoStop}%
\bibitem [{\citenamefont {Larsen}\ \emph {et~al.}(2017)\citenamefont {Larsen},
  \citenamefont {Mortensen}, \citenamefont {Blomqvist}, \citenamefont
  {Castelli}, \citenamefont {Christensen}, \citenamefont {Du{\l}ak},
  \citenamefont {Friis}, \citenamefont {Groves}, \citenamefont {Hammer},
  \citenamefont {Hargus}, \citenamefont {Hermes}, \citenamefont {Jennings},
  \citenamefont {Jensen}, \citenamefont {Kermode}, \citenamefont {Kitchin},
  \citenamefont {Kolsbjerg}, \citenamefont {Kubal}, \citenamefont {Kaasbjerg},
  \citenamefont {Lysgaard}, \citenamefont {Maronsson}, \citenamefont {Maxson},
  \citenamefont {Olsen}, \citenamefont {Pastewka}, \citenamefont {Peterson},
  \citenamefont {Rostgaard}, \citenamefont {Schi{\o}tz}, \citenamefont
  {Sch{\"u}tt}, \citenamefont {Strange}, \citenamefont {Thygesen},
  \citenamefont {Vegge}, \citenamefont {Vilhelmsen}, \citenamefont {Walter},
  \citenamefont {Zeng},\ and\ \citenamefont {Jacobsen}}]{ase-paper}%
  \BibitemOpen
  \bibfield  {author} {\bibinfo {author} {\bibfnamefont {A.~H.}\ \bibnamefont
  {Larsen}}, \bibinfo {author} {\bibfnamefont {J.~J.}\ \bibnamefont
  {Mortensen}}, \bibinfo {author} {\bibfnamefont {J.}~\bibnamefont
  {Blomqvist}}, \bibinfo {author} {\bibfnamefont {I.~E.}\ \bibnamefont
  {Castelli}}, \bibinfo {author} {\bibfnamefont {R.}~\bibnamefont
  {Christensen}}, \bibinfo {author} {\bibfnamefont {M.}~\bibnamefont
  {Du{\l}ak}}, \bibinfo {author} {\bibfnamefont {J.}~\bibnamefont {Friis}},
  \bibinfo {author} {\bibfnamefont {M.~N.}\ \bibnamefont {Groves}}, \bibinfo
  {author} {\bibfnamefont {B.}~\bibnamefont {Hammer}}, \bibinfo {author}
  {\bibfnamefont {C.}~\bibnamefont {Hargus}}, \bibinfo {author} {\bibfnamefont
  {E.~D.}\ \bibnamefont {Hermes}}, \bibinfo {author} {\bibfnamefont {P.~C.}\
  \bibnamefont {Jennings}}, \bibinfo {author} {\bibfnamefont {P.~B.}\
  \bibnamefont {Jensen}}, \bibinfo {author} {\bibfnamefont {J.}~\bibnamefont
  {Kermode}}, \bibinfo {author} {\bibfnamefont {J.~R.}\ \bibnamefont
  {Kitchin}}, \bibinfo {author} {\bibfnamefont {E.~L.}\ \bibnamefont
  {Kolsbjerg}}, \bibinfo {author} {\bibfnamefont {J.}~\bibnamefont {Kubal}},
  \bibinfo {author} {\bibfnamefont {K.}~\bibnamefont {Kaasbjerg}}, \bibinfo
  {author} {\bibfnamefont {S.}~\bibnamefont {Lysgaard}}, \bibinfo {author}
  {\bibfnamefont {J.~B.}\ \bibnamefont {Maronsson}}, \bibinfo {author}
  {\bibfnamefont {T.}~\bibnamefont {Maxson}}, \bibinfo {author} {\bibfnamefont
  {T.}~\bibnamefont {Olsen}}, \bibinfo {author} {\bibfnamefont
  {L.}~\bibnamefont {Pastewka}}, \bibinfo {author} {\bibfnamefont
  {A.}~\bibnamefont {Peterson}}, \bibinfo {author} {\bibfnamefont
  {C.}~\bibnamefont {Rostgaard}}, \bibinfo {author} {\bibfnamefont
  {J.}~\bibnamefont {Schi{\o}tz}}, \bibinfo {author} {\bibfnamefont
  {O.}~\bibnamefont {Sch{\"u}tt}}, \bibinfo {author} {\bibfnamefont
  {M.}~\bibnamefont {Strange}}, \bibinfo {author} {\bibfnamefont {K.~S.}\
  \bibnamefont {Thygesen}}, \bibinfo {author} {\bibfnamefont {T.}~\bibnamefont
  {Vegge}}, \bibinfo {author} {\bibfnamefont {L.}~\bibnamefont {Vilhelmsen}},
  \bibinfo {author} {\bibfnamefont {M.}~\bibnamefont {Walter}}, \bibinfo
  {author} {\bibfnamefont {Z.}~\bibnamefont {Zeng}}, \ and\ \bibinfo {author}
  {\bibfnamefont {K.~W.}\ \bibnamefont {Jacobsen}},\ }\bibfield  {title}
  {\enquote {\bibinfo {title} {The atomic simulation environment---a {Python}
  library for working with atoms},}\ }\href
  {http://stacks.iop.org/0953-8984/29/i=27/a=273002} {\bibfield  {journal}
  {\bibinfo  {journal} {Journal of Physics: Condensed Matter}\ }\textbf
  {\bibinfo {volume} {29}},\ \bibinfo {pages} {273002} (\bibinfo {year}
  {2017})}\BibitemShut {NoStop}%
\bibitem [{\citenamefont {Togo}(2023)}]{Phonopy}%
  \BibitemOpen
  \bibfield  {author} {\bibinfo {author} {\bibfnamefont {A.}~\bibnamefont
  {Togo}},\ }\bibfield  {title} {\enquote {\bibinfo {title} {First-principles
  phonon calculations with phonopy and phono3py},}\ }\href {\doibase
  10.7566/JPSJ.92.012001} {\bibfield  {journal} {\bibinfo  {journal} {J. Phys.
  Soc. Jpn.}\ }\textbf {\bibinfo {volume} {92}},\ \bibinfo {pages} {012001}
  (\bibinfo {year} {2023})}\BibitemShut {NoStop}%
\bibitem [{ddp()}]{ddp-scripts-github}%
  \BibitemOpen
  \href@noop {} {}\bibinfo {howpublished}
  {\url{https://github.com/sehunjoo/ddp-batch}}\BibitemShut {NoStop}%
\bibitem [{\citenamefont {Monserrat}(2018)}]{Tomeu-finite-differences}%
  \BibitemOpen
  \bibfield  {author} {\bibinfo {author} {\bibfnamefont {B.}~\bibnamefont
  {Monserrat}},\ }\bibfield  {title} {\enquote {\bibinfo {title}
  {Electron--phonon coupling from finite differences},}\ }\href {\doibase
  10.1088/1361-648X/aaa737} {\bibfield  {journal} {\bibinfo  {journal} {Journal
  of Physics: Condensed Matter}\ }\textbf {\bibinfo {volume} {30}},\ \bibinfo
  {pages} {083001} (\bibinfo {year} {2018})}\BibitemShut {NoStop}%
\bibitem [{\citenamefont {Thompson}\ \emph {et~al.}(2022)\citenamefont
  {Thompson}, \citenamefont {Aktulga}, \citenamefont {Berger}, \citenamefont
  {Bolintineanu}, \citenamefont {Brown}, \citenamefont {Crozier}, \citenamefont
  {in~'t Veld}, \citenamefont {Kohlmeyer}, \citenamefont {Moore}, \citenamefont
  {Nguyen}, \citenamefont {Shan}, \citenamefont {Stevens}, \citenamefont
  {Tranchida}, \citenamefont {Trott},\ and\ \citenamefont {Plimpton}}]{Lammps}%
  \BibitemOpen
  \bibfield  {author} {\bibinfo {author} {\bibfnamefont {A.~P.}\ \bibnamefont
  {Thompson}}, \bibinfo {author} {\bibfnamefont {H.~M.}\ \bibnamefont
  {Aktulga}}, \bibinfo {author} {\bibfnamefont {R.}~\bibnamefont {Berger}},
  \bibinfo {author} {\bibfnamefont {D.~S.}\ \bibnamefont {Bolintineanu}},
  \bibinfo {author} {\bibfnamefont {W.~M.}\ \bibnamefont {Brown}}, \bibinfo
  {author} {\bibfnamefont {P.~S.}\ \bibnamefont {Crozier}}, \bibinfo {author}
  {\bibfnamefont {P.~J.}\ \bibnamefont {in~'t Veld}}, \bibinfo {author}
  {\bibfnamefont {A.}~\bibnamefont {Kohlmeyer}}, \bibinfo {author}
  {\bibfnamefont {S.~G.}\ \bibnamefont {Moore}}, \bibinfo {author}
  {\bibfnamefont {T.~D.}\ \bibnamefont {Nguyen}}, \bibinfo {author}
  {\bibfnamefont {R.}~\bibnamefont {Shan}}, \bibinfo {author} {\bibfnamefont
  {M.~J.}\ \bibnamefont {Stevens}}, \bibinfo {author} {\bibfnamefont
  {J.}~\bibnamefont {Tranchida}}, \bibinfo {author} {\bibfnamefont
  {C.}~\bibnamefont {Trott}}, \ and\ \bibinfo {author} {\bibfnamefont {S.~J.}\
  \bibnamefont {Plimpton}},\ }\bibfield  {title} {\enquote {\bibinfo {title}
  {{LAMMPS} - a flexible simulation tool for particle-based materials modeling
  at the atomic, meso, and continuum scales},}\ }\href {\doibase
  10.1016/j.cpc.2021.108171} {\bibfield  {journal} {\bibinfo  {journal} {Comp.
  Phys. Comm.}\ }\textbf {\bibinfo {volume} {271}},\ \bibinfo {pages} {108171}
  (\bibinfo {year} {2022})}\BibitemShut {NoStop}%
\bibitem [{\citenamefont {Erhart}\ \emph {et~al.}(2015)\citenamefont {Erhart},
  \citenamefont {Sadigh}, \citenamefont {Schleife},\ and\ \citenamefont
  {{\AA}berg}}]{Erhart2015}%
  \BibitemOpen
  \bibfield  {author} {\bibinfo {author} {\bibfnamefont {P.}~\bibnamefont
  {Erhart}}, \bibinfo {author} {\bibfnamefont {B.}~\bibnamefont {Sadigh}},
  \bibinfo {author} {\bibfnamefont {A.}~\bibnamefont {Schleife}}, \ and\
  \bibinfo {author} {\bibfnamefont {D.}~\bibnamefont {{\AA}berg}},\ }\bibfield
  {title} {\enquote {\bibinfo {title} {First-principles study of codoping in
  lanthanum bromide},}\ }\href {\doibase 10.1103/PhysRevB.91.165206} {\bibfield
   {journal} {\bibinfo  {journal} {Physical Review B}\ }\textbf {\bibinfo
  {volume} {91}},\ \bibinfo {pages} {165206} (\bibinfo {year}
  {2015})}\BibitemShut {NoStop}%
\bibitem [{\citenamefont {Qamar}\ \emph {et~al.}(2022)\citenamefont {Qamar},
  \citenamefont {Mrovec}, \citenamefont {Lysogorskiy}, \citenamefont
  {Bochkarev},\ and\ \citenamefont {Drautz}}]{Ralf-Drautz-carbon-ace}%
  \BibitemOpen
  \bibfield  {author} {\bibinfo {author} {\bibfnamefont {M.}~\bibnamefont
  {Qamar}}, \bibinfo {author} {\bibfnamefont {M.}~\bibnamefont {Mrovec}},
  \bibinfo {author} {\bibfnamefont {Y.}~\bibnamefont {Lysogorskiy}}, \bibinfo
  {author} {\bibfnamefont {A.}~\bibnamefont {Bochkarev}}, \ and\ \bibinfo
  {author} {\bibfnamefont {R.}~\bibnamefont {Drautz}},\ }\href@noop {}
  {\enquote {\bibinfo {title} {Atomic cluster expansion for quantum-accurate
  large-scale simulations of carbon},}\ } (\bibinfo {year} {2022}),\ \Eprint
  {http://arxiv.org/abs/2210.09161} {arXiv:2210.09161 [cond-mat.mtrl-sci]}
  \BibitemShut {NoStop}%
\bibitem [{EDD()}]{EDDP-benchmark}%
  \BibitemOpen
  \href@noop {} {}\bibinfo {howpublished}
  {\url{https://github.com/PascalSalzbrenner/EDDP_benchmarks}}\BibitemShut
  {NoStop}%
\bibitem [{\citenamefont {Perdew}, \citenamefont {Burke},\ and\ \citenamefont
  {Ernzerhof}(1996)}]{PBE}%
  \BibitemOpen
  \bibfield  {author} {\bibinfo {author} {\bibfnamefont {J.~P.}\ \bibnamefont
  {Perdew}}, \bibinfo {author} {\bibfnamefont {K.}~\bibnamefont {Burke}}, \
  and\ \bibinfo {author} {\bibfnamefont {M.}~\bibnamefont {Ernzerhof}},\
  }\bibfield  {title} {\enquote {\bibinfo {title} {Generalized gradient
  approximation made simple},}\ }\href {\doibase 10.1103/PhysRevLett.77.3865}
  {\bibfield  {journal} {\bibinfo  {journal} {Phys. Rev. Lett.}\ }\textbf
  {\bibinfo {volume} {77}},\ \bibinfo {pages} {3865--3868} (\bibinfo {year}
  {1996})}\BibitemShut {NoStop}%
\bibitem [{\citenamefont {Tkatchenko}\ and\ \citenamefont
  {Scheffler}(2009)}]{Tkatchenko2009}%
  \BibitemOpen
  \bibfield  {author} {\bibinfo {author} {\bibfnamefont {A.}~\bibnamefont
  {Tkatchenko}}\ and\ \bibinfo {author} {\bibfnamefont {M.}~\bibnamefont
  {Scheffler}},\ }\bibfield  {title} {\enquote {\bibinfo {title} {Accurate
  {{Molecular Van Der Waals Interactions}} from {{Ground-State Electron
  Density}} and {{Free-Atom Reference Data}}},}\ }\href {\doibase
  10.1103/PhysRevLett.102.073005} {\bibfield  {journal} {\bibinfo  {journal}
  {Phys. Rev. Lett.}\ }\textbf {\bibinfo {volume} {102}},\ \bibinfo {pages}
  {073005} (\bibinfo {year} {2009})}\BibitemShut {NoStop}%
\bibitem [{\citenamefont {Perdew}\ \emph {et~al.}(2008)\citenamefont {Perdew},
  \citenamefont {Ruzsinszky}, \citenamefont {Csonka}, \citenamefont {Vydrov},
  \citenamefont {Scuseria}, \citenamefont {Constantin}, \citenamefont {Zhou},\
  and\ \citenamefont {Burke}}]{PBEsol}%
  \BibitemOpen
  \bibfield  {author} {\bibinfo {author} {\bibfnamefont {J.~P.}\ \bibnamefont
  {Perdew}}, \bibinfo {author} {\bibfnamefont {A.}~\bibnamefont {Ruzsinszky}},
  \bibinfo {author} {\bibfnamefont {G.~I.}\ \bibnamefont {Csonka}}, \bibinfo
  {author} {\bibfnamefont {O.~A.}\ \bibnamefont {Vydrov}}, \bibinfo {author}
  {\bibfnamefont {G.~E.}\ \bibnamefont {Scuseria}}, \bibinfo {author}
  {\bibfnamefont {L.~A.}\ \bibnamefont {Constantin}}, \bibinfo {author}
  {\bibfnamefont {X.}~\bibnamefont {Zhou}}, \ and\ \bibinfo {author}
  {\bibfnamefont {K.}~\bibnamefont {Burke}},\ }\bibfield  {title} {\enquote
  {\bibinfo {title} {Restoring the density-gradient expansion for exchange in
  solids and surfaces},}\ }\href {\doibase 10.1103/PhysRevLett.100.136406}
  {\bibfield  {journal} {\bibinfo  {journal} {Phys. Rev. Lett.}\ }\textbf
  {\bibinfo {volume} {100}},\ \bibinfo {pages} {136406} (\bibinfo {year}
  {2008})}\BibitemShut {NoStop}%
\bibitem [{\citenamefont {Salzbrenner}\ \emph {et~al.}(2023)\citenamefont
  {Salzbrenner}, \citenamefont {Joo}, \citenamefont {Conway}, \citenamefont
  {Cooke}, \citenamefont {Zhu}, \citenamefont {Matraszek}, \citenamefont
  {Witt},\ and\ \citenamefont {Pickard}}]{Materials-cloud-data}%
  \BibitemOpen
  \bibfield  {author} {\bibinfo {author} {\bibfnamefont {P.~T.}\ \bibnamefont
  {Salzbrenner}}, \bibinfo {author} {\bibfnamefont {S.~H.}\ \bibnamefont
  {Joo}}, \bibinfo {author} {\bibfnamefont {L.~J.}\ \bibnamefont {Conway}},
  \bibinfo {author} {\bibfnamefont {P.~I.~C.}\ \bibnamefont {Cooke}}, \bibinfo
  {author} {\bibfnamefont {B.}~\bibnamefont {Zhu}}, \bibinfo {author}
  {\bibfnamefont {M.~P.}\ \bibnamefont {Matraszek}}, \bibinfo {author}
  {\bibfnamefont {W.~C.}\ \bibnamefont {Witt}}, \ and\ \bibinfo {author}
  {\bibfnamefont {C.~J.}\ \bibnamefont {Pickard}},\ }\href {\doibase
  10.24435/materialscloud:44-c5} {\enquote {\bibinfo {title} {Developments and
  further applications of ephemeral data derived potentials},}\ }\bibinfo
  {howpublished} {\url{https://doi.org/10.24435/materialscloud:44-c5}}
  (\bibinfo {year} {2023}),\ \Eprint {http://arxiv.org/abs/2023.150} {Materials
  Cloud Archive:2023.150} \BibitemShut {NoStop}%
\bibitem [{\citenamefont {Emsley}(2011)}]{Nature-building-blocks}%
  \BibitemOpen
  \bibfield  {author} {\bibinfo {author} {\bibfnamefont {J.}~\bibnamefont
  {Emsley}},\ }\href@noop {} {\emph {\bibinfo {title} {Nature's Building
  Blocks: An A-Z Guide to the Elements}}},\ \bibinfo {edition} {2nd}\ ed.\
  (\bibinfo  {publisher} {Oxford University Press},\ \bibinfo {year}
  {2011})\BibitemShut {NoStop}%
\bibitem [{\citenamefont {Hoffmann}\ \emph {et~al.}(2016)\citenamefont
  {Hoffmann}, \citenamefont {Kabanov}, \citenamefont {Golov},\ and\
  \citenamefont {Proserpio}}]{SACADA-Carbon-database}%
  \BibitemOpen
  \bibfield  {author} {\bibinfo {author} {\bibfnamefont {R.}~\bibnamefont
  {Hoffmann}}, \bibinfo {author} {\bibfnamefont {A.~A.}\ \bibnamefont
  {Kabanov}}, \bibinfo {author} {\bibfnamefont {A.~A.}\ \bibnamefont {Golov}},
  \ and\ \bibinfo {author} {\bibfnamefont {D.~M.}\ \bibnamefont {Proserpio}},\
  }\bibfield  {title} {\enquote {\bibinfo {title} {Homo citans and carbon
  allotropes: For an ethics of citation},}\ }\href@noop {} {\bibfield
  {journal} {\bibinfo  {journal} {Angewandte Chemie International Edition}\
  }\textbf {\bibinfo {volume} {55}},\ \bibinfo {pages} {10962--10976} (\bibinfo
  {year} {2016})}\BibitemShut {NoStop}%
\bibitem [{\citenamefont {Shi}\ \emph {et~al.}(2018)\citenamefont {Shi},
  \citenamefont {He}, \citenamefont {Pickard}, \citenamefont {Tang},\ and\
  \citenamefont {Zhong}}]{Chris-Carbon-group-graph-theory}%
  \BibitemOpen
  \bibfield  {author} {\bibinfo {author} {\bibfnamefont {X.}~\bibnamefont
  {Shi}}, \bibinfo {author} {\bibfnamefont {C.}~\bibnamefont {He}}, \bibinfo
  {author} {\bibfnamefont {C.~J.}\ \bibnamefont {Pickard}}, \bibinfo {author}
  {\bibfnamefont {C.}~\bibnamefont {Tang}}, \ and\ \bibinfo {author}
  {\bibfnamefont {J.}~\bibnamefont {Zhong}},\ }\bibfield  {title} {\enquote
  {\bibinfo {title} {Stochastic generation of complex crystal structures
  combining group and graph theory with application to carbon},}\ }\href
  {\doibase 10.1103/PhysRevB.97.014104} {\bibfield  {journal} {\bibinfo
  {journal} {Phys. Rev. B}\ }\textbf {\bibinfo {volume} {97}},\ \bibinfo
  {pages} {014104} (\bibinfo {year} {2018})}\BibitemShut {NoStop}%
\bibitem [{\citenamefont {Rowe}\ \emph {et~al.}(2020)\citenamefont {Rowe},
  \citenamefont {Deringer}, \citenamefont {Gasparotto}, \citenamefont
  {Cs{\'a}nyi},\ and\ \citenamefont {Michaelides}}]{Carbon-GAP-20}%
  \BibitemOpen
  \bibfield  {author} {\bibinfo {author} {\bibfnamefont {P.}~\bibnamefont
  {Rowe}}, \bibinfo {author} {\bibfnamefont {V.~L.}\ \bibnamefont {Deringer}},
  \bibinfo {author} {\bibfnamefont {P.}~\bibnamefont {Gasparotto}}, \bibinfo
  {author} {\bibfnamefont {G.}~\bibnamefont {Cs{\'a}nyi}}, \ and\ \bibinfo
  {author} {\bibfnamefont {A.}~\bibnamefont {Michaelides}},\ }\bibfield
  {title} {\enquote {\bibinfo {title} {{An accurate and transferable machine
  learning potential for carbon}},}\ }\href {https://doi.org/10.1063/5.0005084}
  {\bibfield  {journal} {\bibinfo  {journal} {The Journal of Chemical Physics}\
  }\textbf {\bibinfo {volume} {153}} (\bibinfo {year} {2020})},\ \bibinfo
  {note} {034702}\BibitemShut {NoStop}%
\bibitem [{\citenamefont {Shaidu}\ \emph {et~al.}(2021)\citenamefont {Shaidu},
  \citenamefont {K{\"u}{\c c}{\"u}kbenli}, \citenamefont {Lot}, \citenamefont
  {Pellegrini}, \citenamefont {Kaxiras},\ and\ \citenamefont
  {de~Gironcoli}}]{Carbon-PANNA}%
  \BibitemOpen
  \bibfield  {author} {\bibinfo {author} {\bibfnamefont {Y.}~\bibnamefont
  {Shaidu}}, \bibinfo {author} {\bibfnamefont {E.}~\bibnamefont {K{\"u}{\c
  c}{\"u}kbenli}}, \bibinfo {author} {\bibfnamefont {R.}~\bibnamefont {Lot}},
  \bibinfo {author} {\bibfnamefont {F.}~\bibnamefont {Pellegrini}}, \bibinfo
  {author} {\bibfnamefont {E.}~\bibnamefont {Kaxiras}}, \ and\ \bibinfo
  {author} {\bibfnamefont {S.}~\bibnamefont {de~Gironcoli}},\ }\bibfield
  {title} {\enquote {\bibinfo {title} {A systematic approach to generating
  accurate neural network potentials: the case of carbon},}\ }\href {\doibase
  10.1038/s41524-021-00508-6} {\bibfield  {journal} {\bibinfo  {journal} {npj
  Computational Materials}\ }\textbf {\bibinfo {volume} {7}},\ \bibinfo {pages}
  {52} (\bibinfo {year} {2021})}\BibitemShut {NoStop}%
\bibitem [{\citenamefont {Lot}\ \emph {et~al.}(2020)\citenamefont {Lot},
  \citenamefont {Pellegrini}, \citenamefont {Shaidu},\ and\ \citenamefont
  {K{\"u}{\c c}{\"u}kbenli}}]{PANNA-potential}%
  \BibitemOpen
  \bibfield  {author} {\bibinfo {author} {\bibfnamefont {R.}~\bibnamefont
  {Lot}}, \bibinfo {author} {\bibfnamefont {F.}~\bibnamefont {Pellegrini}},
  \bibinfo {author} {\bibfnamefont {Y.}~\bibnamefont {Shaidu}}, \ and\ \bibinfo
  {author} {\bibfnamefont {E.}~\bibnamefont {K{\"u}{\c c}{\"u}kbenli}},\
  }\bibfield  {title} {\enquote {\bibinfo {title} {Panna: Properties from
  artificial neural network architectures},}\ }\href {\doibase
  https://doi.org/10.1016/j.cpc.2020.107402} {\bibfield  {journal} {\bibinfo
  {journal} {Computer Physics Communications}\ }\textbf {\bibinfo {volume}
  {256}},\ \bibinfo {pages} {107402} (\bibinfo {year} {2020})}\BibitemShut
  {NoStop}%
\bibitem [{\citenamefont {Kónya}\ and\ \citenamefont
  {Nagy}(2018)}]{Radioactive-decay}%
  \BibitemOpen
  \bibfield  {author} {\bibinfo {author} {\bibfnamefont {J.}~\bibnamefont
  {Kónya}}\ and\ \bibinfo {author} {\bibfnamefont {N.~M.}\ \bibnamefont
  {Nagy}},\ }\href@noop {} {\emph {\bibinfo {title} {Nuclear and
  Radiochemistry}}},\ \bibinfo {edition} {2nd}\ ed.\ (\bibinfo  {publisher}
  {Elsevier},\ \bibinfo {year} {2018})\BibitemShut {NoStop}%
\bibitem [{\citenamefont {McMahon}\ and\ \citenamefont
  {Nelmes}(2006)}]{McMahon-Nelmes-metal-structures}%
  \BibitemOpen
  \bibfield  {author} {\bibinfo {author} {\bibfnamefont {M.~I.}\ \bibnamefont
  {McMahon}}\ and\ \bibinfo {author} {\bibfnamefont {R.~J.}\ \bibnamefont
  {Nelmes}},\ }\bibfield  {title} {\enquote {\bibinfo {title} {High-pressure
  structures and phase transformations in elemental metals},}\ }\href {\doibase
  10.1039/B517777B} {\bibfield  {journal} {\bibinfo  {journal} {Chem. Soc.
  Rev.}\ }\textbf {\bibinfo {volume} {35}},\ \bibinfo {pages} {943--963}
  (\bibinfo {year} {2006})}\BibitemShut {NoStop}%
\bibitem [{\citenamefont {Godwal}\ \emph {et~al.}(1990)\citenamefont {Godwal},
  \citenamefont {Meade}, \citenamefont {Jeanloz}, \citenamefont {Garcia},
  \citenamefont {Liu},\ and\ \citenamefont {Cohen}}]{Lead-melting-DAC}%
  \BibitemOpen
  \bibfield  {author} {\bibinfo {author} {\bibfnamefont {B.~K.}\ \bibnamefont
  {Godwal}}, \bibinfo {author} {\bibfnamefont {C.}~\bibnamefont {Meade}},
  \bibinfo {author} {\bibfnamefont {R.}~\bibnamefont {Jeanloz}}, \bibinfo
  {author} {\bibfnamefont {A.}~\bibnamefont {Garcia}}, \bibinfo {author}
  {\bibfnamefont {A.~Y.}\ \bibnamefont {Liu}}, \ and\ \bibinfo {author}
  {\bibfnamefont {M.~L.}\ \bibnamefont {Cohen}},\ }\bibfield  {title} {\enquote
  {\bibinfo {title} {Ultrahigh-pressure melting of lead: A multidisciplinary
  study},}\ }\href {\doibase 10.1126/science.248.4954.462} {\bibfield
  {journal} {\bibinfo  {journal} {Science}\ }\textbf {\bibinfo {volume}
  {248}},\ \bibinfo {pages} {462--465} (\bibinfo {year} {1990})}\BibitemShut
  {NoStop}%
\bibitem [{\citenamefont {Partouche-Sebban}\ \emph {et~al.}(2005)\citenamefont
  {Partouche-Sebban}, \citenamefont {P{\'e}lissier}, \citenamefont {Abeyta},
  \citenamefont {Anderson}, \citenamefont {Byers}, \citenamefont
  {Dennis-Koller}, \citenamefont {Esparza}, \citenamefont {Hixson},
  \citenamefont {Holtkamp}, \citenamefont {Jensen}, \citenamefont {King},
  \citenamefont {Rigg}, \citenamefont {Rodriguez}, \citenamefont {Shampine},
  \citenamefont {Stone}, \citenamefont {Westley}, \citenamefont {Borror},\ and\
  \citenamefont {Kruschwitz}}]{Lead-melting-shockwave}%
  \BibitemOpen
  \bibfield  {author} {\bibinfo {author} {\bibfnamefont {D.}~\bibnamefont
  {Partouche-Sebban}}, \bibinfo {author} {\bibfnamefont {J.~L.}\ \bibnamefont
  {P{\'e}lissier}}, \bibinfo {author} {\bibfnamefont {F.~G.}\ \bibnamefont
  {Abeyta}}, \bibinfo {author} {\bibfnamefont {W.~W.}\ \bibnamefont
  {Anderson}}, \bibinfo {author} {\bibfnamefont {M.~E.}\ \bibnamefont {Byers}},
  \bibinfo {author} {\bibfnamefont {D.}~\bibnamefont {Dennis-Koller}}, \bibinfo
  {author} {\bibfnamefont {J.~S.}\ \bibnamefont {Esparza}}, \bibinfo {author}
  {\bibfnamefont {R.~S.}\ \bibnamefont {Hixson}}, \bibinfo {author}
  {\bibfnamefont {D.~B.}\ \bibnamefont {Holtkamp}}, \bibinfo {author}
  {\bibfnamefont {B.~J.}\ \bibnamefont {Jensen}}, \bibinfo {author}
  {\bibfnamefont {J.~C.}\ \bibnamefont {King}}, \bibinfo {author}
  {\bibfnamefont {P.~A.}\ \bibnamefont {Rigg}}, \bibinfo {author}
  {\bibfnamefont {P.}~\bibnamefont {Rodriguez}}, \bibinfo {author}
  {\bibfnamefont {D.~L.}\ \bibnamefont {Shampine}}, \bibinfo {author}
  {\bibfnamefont {J.~B.}\ \bibnamefont {Stone}}, \bibinfo {author}
  {\bibfnamefont {D.~T.}\ \bibnamefont {Westley}}, \bibinfo {author}
  {\bibfnamefont {S.~D.}\ \bibnamefont {Borror}}, \ and\ \bibinfo {author}
  {\bibfnamefont {C.~A.}\ \bibnamefont {Kruschwitz}},\ }\bibfield  {title}
  {\enquote {\bibinfo {title} {Measurement of the shock-heated melt curve of
  lead using pyrometry and reflectometry},}\ }\href {\doibase
  10.1063/1.1849436} {\bibfield  {journal} {\bibinfo  {journal} {Journal of
  Applied Physics}\ }\textbf {\bibinfo {volume} {97}},\ \bibinfo {pages}
  {043521} (\bibinfo {year} {2005})}\BibitemShut {NoStop}%
\bibitem [{\citenamefont {Errandonea}(2010)}]{Lead-melting-Bridgman}%
  \BibitemOpen
  \bibfield  {author} {\bibinfo {author} {\bibfnamefont {D.}~\bibnamefont
  {Errandonea}},\ }\bibfield  {title} {\enquote {\bibinfo {title} {The melting
  curve of ten metals up to 12 {GPa} and 1600 {K}},}\ }\href {\doibase
  10.1063/1.3468149} {\bibfield  {journal} {\bibinfo  {journal} {Journal of
  Applied Physics}\ }\textbf {\bibinfo {volume} {108}},\ \bibinfo {pages}
  {033517} (\bibinfo {year} {2010})}\BibitemShut {NoStop}%
\bibitem [{\citenamefont {Cricchio}\ \emph {et~al.}(2006)\citenamefont
  {Cricchio}, \citenamefont {Belonoshko}, \citenamefont {Burakovsky},
  \citenamefont {Preston},\ and\ \citenamefont {Ahuja}}]{Lead-melting-DFT}%
  \BibitemOpen
  \bibfield  {author} {\bibinfo {author} {\bibfnamefont {F.}~\bibnamefont
  {Cricchio}}, \bibinfo {author} {\bibfnamefont {A.~B.}\ \bibnamefont
  {Belonoshko}}, \bibinfo {author} {\bibfnamefont {L.}~\bibnamefont
  {Burakovsky}}, \bibinfo {author} {\bibfnamefont {D.~L.}\ \bibnamefont
  {Preston}}, \ and\ \bibinfo {author} {\bibfnamefont {R.}~\bibnamefont
  {Ahuja}},\ }\bibfield  {title} {\enquote {\bibinfo {title} {High-pressure
  melting of lead},}\ }\href@noop {} {\bibfield  {journal} {\bibinfo  {journal}
  {Phys. Rev. B}\ }\textbf {\bibinfo {volume} {73}},\ \bibinfo {pages} {140103}
  (\bibinfo {year} {2006})}\BibitemShut {NoStop}%
\bibitem [{\citenamefont {Strange}(1998)}]{Relativistic-QM}%
  \BibitemOpen
  \bibfield  {author} {\bibinfo {author} {\bibfnamefont {P.}~\bibnamefont
  {Strange}},\ }\href@noop {} {\emph {\bibinfo {title} {Relativistic Quantum
  Mechanics: With Applications in Condensed Matter and Atomic Physics}}},\
  \bibinfo {edition} {1st}\ ed.\ (\bibinfo  {publisher} {Cambridge University
  Press},\ \bibinfo {year} {1998})\BibitemShut {NoStop}%
\bibitem [{\citenamefont {Smirnov}(2018)}]{Pb-phases-SOC}%
  \BibitemOpen
  \bibfield  {author} {\bibinfo {author} {\bibfnamefont {N.~A.}\ \bibnamefont
  {Smirnov}},\ }\bibfield  {title} {\enquote {\bibinfo {title} {Effect of
  spin-orbit interactions on the structural stability, thermodynamic
  properties, and transport properties of lead under pressure},}\ }\href
  {\doibase 10.1103/PhysRevB.97.094114} {\bibfield  {journal} {\bibinfo
  {journal} {Phys. Rev. B}\ }\textbf {\bibinfo {volume} {97}},\ \bibinfo
  {pages} {094114} (\bibinfo {year} {2018})}\BibitemShut {NoStop}%
\bibitem [{\citenamefont {Verstraete}\ \emph {et~al.}(2008)\citenamefont
  {Verstraete}, \citenamefont {Torrent}, \citenamefont {Jollet}, \citenamefont
  {Z\'erah},\ and\ \citenamefont {Gonze}}]{Pb-SOC-phonons}%
  \BibitemOpen
  \bibfield  {author} {\bibinfo {author} {\bibfnamefont {M.~J.}\ \bibnamefont
  {Verstraete}}, \bibinfo {author} {\bibfnamefont {M.}~\bibnamefont {Torrent}},
  \bibinfo {author} {\bibfnamefont {F.}~\bibnamefont {Jollet}}, \bibinfo
  {author} {\bibfnamefont {G.}~\bibnamefont {Z\'erah}}, \ and\ \bibinfo
  {author} {\bibfnamefont {X.}~\bibnamefont {Gonze}},\ }\bibfield  {title}
  {\enquote {\bibinfo {title} {Density functional perturbation theory with
  spin-orbit coupling: Phonon band structure of lead},}\ }\href {\doibase
  10.1103/PhysRevB.78.045119} {\bibfield  {journal} {\bibinfo  {journal} {Phys.
  Rev. B}\ }\textbf {\bibinfo {volume} {78}},\ \bibinfo {pages} {045119}
  (\bibinfo {year} {2008})}\BibitemShut {NoStop}%
\bibitem [{\citenamefont {Heid}\ \emph {et~al.}(2010)\citenamefont {Heid},
  \citenamefont {Bohnen}, \citenamefont {Sklyadneva},\ and\ \citenamefont
  {Chulkov}}]{Pb-SOC-superconductivity}%
  \BibitemOpen
  \bibfield  {author} {\bibinfo {author} {\bibfnamefont {R.}~\bibnamefont
  {Heid}}, \bibinfo {author} {\bibfnamefont {K.-P.}\ \bibnamefont {Bohnen}},
  \bibinfo {author} {\bibfnamefont {I.~Y.}\ \bibnamefont {Sklyadneva}}, \ and\
  \bibinfo {author} {\bibfnamefont {E.~V.}\ \bibnamefont {Chulkov}},\
  }\bibfield  {title} {\enquote {\bibinfo {title} {Effect of spin-orbit
  coupling on the electron-phonon interaction of the superconductors {Pb and
  Tl}},}\ }\href {\doibase 10.1103/PhysRevB.81.174527} {\bibfield  {journal}
  {\bibinfo  {journal} {Phys. Rev. B}\ }\textbf {\bibinfo {volume} {81}},\
  \bibinfo {pages} {174527} (\bibinfo {year} {2010})}\BibitemShut {NoStop}%
\bibitem [{\citenamefont {Giustino}(2014)}]{Giustino-DFT}%
  \BibitemOpen
  \bibfield  {author} {\bibinfo {author} {\bibfnamefont {F.}~\bibnamefont
  {Giustino}},\ }\href@noop {} {\emph {\bibinfo {title} {Materials Modelling
  using Density Functional Theory: Properties and Predictions}}},\ \bibinfo
  {edition} {1st}\ ed.\ (\bibinfo  {publisher} {Oxford University Press},\
  \bibinfo {year} {2014})\BibitemShut {NoStop}%
\bibitem [{\citenamefont {Lloyd-Williams}\ and\ \citenamefont
  {Monserrat}(2015)}]{Tomeu-nondiagonal-supercells}%
  \BibitemOpen
  \bibfield  {author} {\bibinfo {author} {\bibfnamefont {J.~H.}\ \bibnamefont
  {Lloyd-Williams}}\ and\ \bibinfo {author} {\bibfnamefont {B.}~\bibnamefont
  {Monserrat}},\ }\bibfield  {title} {\enquote {\bibinfo {title} {Lattice
  dynamics and electron-phonon coupling calculations using nondiagonal
  supercells},}\ }\href {\doibase 10.1103/PhysRevB.92.184301} {\bibfield
  {journal} {\bibinfo  {journal} {Phys. Rev. B}\ }\textbf {\bibinfo {volume}
  {92}},\ \bibinfo {pages} {184301} (\bibinfo {year} {2015})}\BibitemShut
  {NoStop}%
\bibitem [{\citenamefont {Kohn}(1959)}]{Kohn-anomaly}%
  \BibitemOpen
  \bibfield  {author} {\bibinfo {author} {\bibfnamefont {W.}~\bibnamefont
  {Kohn}},\ }\bibfield  {title} {\enquote {\bibinfo {title} {Image of the
  {Fermi} surface in the vibration spectrum of a metal},}\ }\href {\doibase
  10.1103/PhysRevLett.2.393} {\bibfield  {journal} {\bibinfo  {journal} {Phys.
  Rev. Lett.}\ }\textbf {\bibinfo {volume} {2}},\ \bibinfo {pages} {393--394}
  (\bibinfo {year} {1959})}\BibitemShut {NoStop}%
\bibitem [{\citenamefont {Wang}\ \emph {et~al.}(2022)\citenamefont {Wang},
  \citenamefont {Pan}, \citenamefont {Wang}, \citenamefont {Chen},
  \citenamefont {Wang},\ and\ \citenamefont {Geng}}]{Uranium-phonons-MTP}%
  \BibitemOpen
  \bibfield  {author} {\bibinfo {author} {\bibfnamefont {H.}~\bibnamefont
  {Wang}}, \bibinfo {author} {\bibfnamefont {X.-L.}\ \bibnamefont {Pan}},
  \bibinfo {author} {\bibfnamefont {Y.-F.}\ \bibnamefont {Wang}}, \bibinfo
  {author} {\bibfnamefont {X.-R.}\ \bibnamefont {Chen}}, \bibinfo {author}
  {\bibfnamefont {Y.-X.}\ \bibnamefont {Wang}}, \ and\ \bibinfo {author}
  {\bibfnamefont {H.-Y.}\ \bibnamefont {Geng}},\ }\bibfield  {title} {\enquote
  {\bibinfo {title} {Lattice dynamics and elastic properties of $\alpha$-{U} at
  high-temperature and high-pressure by machine learning potential
  simulations},}\ }\href {\doibase
  https://doi.org/10.1016/j.jnucmat.2022.154029} {\bibfield  {journal}
  {\bibinfo  {journal} {Journal of Nuclear Materials}\ }\textbf {\bibinfo
  {volume} {572}},\ \bibinfo {pages} {154029} (\bibinfo {year}
  {2022})}\BibitemShut {NoStop}%
\bibitem [{\citenamefont {Brockhouse}\ \emph {et~al.}(1962)\citenamefont
  {Brockhouse}, \citenamefont {Arase}, \citenamefont {Caglioti}, \citenamefont
  {Rao},\ and\ \citenamefont {Woods}}]{Lead-phonons-experiment}%
  \BibitemOpen
  \bibfield  {author} {\bibinfo {author} {\bibfnamefont {B.~N.}\ \bibnamefont
  {Brockhouse}}, \bibinfo {author} {\bibfnamefont {T.}~\bibnamefont {Arase}},
  \bibinfo {author} {\bibfnamefont {G.}~\bibnamefont {Caglioti}}, \bibinfo
  {author} {\bibfnamefont {K.~R.}\ \bibnamefont {Rao}}, \ and\ \bibinfo
  {author} {\bibfnamefont {A.~D.~B.}\ \bibnamefont {Woods}},\ }\bibfield
  {title} {\enquote {\bibinfo {title} {Crystal dynamics of lead. {I.}
  {Dispersion} curves at 100 {K}},}\ }\href {\doibase 10.1103/PhysRev.128.1099}
  {\bibfield  {journal} {\bibinfo  {journal} {Phys. Rev.}\ }\textbf {\bibinfo
  {volume} {128}},\ \bibinfo {pages} {1099--1111} (\bibinfo {year}
  {1962})}\BibitemShut {NoStop}%
\bibitem [{\citenamefont {Kuznetsov}\ \emph {et~al.}(2002)\citenamefont
  {Kuznetsov}, \citenamefont {Dmitriev}, \citenamefont {Dubrovinsky},
  \citenamefont {Prakapenka},\ and\ \citenamefont
  {Weber}}]{Lead-FCC-HCP-experiment}%
  \BibitemOpen
  \bibfield  {author} {\bibinfo {author} {\bibfnamefont {A.}~\bibnamefont
  {Kuznetsov}}, \bibinfo {author} {\bibfnamefont {V.}~\bibnamefont {Dmitriev}},
  \bibinfo {author} {\bibfnamefont {L.}~\bibnamefont {Dubrovinsky}}, \bibinfo
  {author} {\bibfnamefont {V.}~\bibnamefont {Prakapenka}}, \ and\ \bibinfo
  {author} {\bibfnamefont {H.-P.}\ \bibnamefont {Weber}},\ }\bibfield  {title}
  {\enquote {\bibinfo {title} {{FCC}--{HCP} phase boundary in lead},}\ }\href
  {\doibase https://doi.org/10.1016/S0038-1098(02)00112-6} {\bibfield
  {journal} {\bibinfo  {journal} {Solid State Communications}\ }\textbf
  {\bibinfo {volume} {122}},\ \bibinfo {pages} {125--127} (\bibinfo {year}
  {2002})}\BibitemShut {NoStop}%
\bibitem [{\citenamefont {Dove}(1993)}]{Dove-Lattice-Dynamics}%
  \BibitemOpen
  \bibfield  {author} {\bibinfo {author} {\bibfnamefont {M.~T.}\ \bibnamefont
  {Dove}},\ }\href {\doibase 10.1017/CBO9780511619885} {\emph {\bibinfo {title}
  {Introduction to Lattice Dynamics}}}\ (\bibinfo  {publisher} {Cambridge
  University Press},\ \bibinfo {year} {1993})\BibitemShut {NoStop}%
\bibitem [{\citenamefont {Boyer}(1979)}]{Quasi-harmonic-approximation}%
  \BibitemOpen
  \bibfield  {author} {\bibinfo {author} {\bibfnamefont {L.~L.}\ \bibnamefont
  {Boyer}},\ }\bibfield  {title} {\enquote {\bibinfo {title} {Calculation of
  thermal expansion, compressiblity, an melting in alkali halides: {NaCl} and
  {KCl}},}\ }\href {\doibase 10.1103/PhysRevLett.42.584} {\bibfield  {journal}
  {\bibinfo  {journal} {Phys. Rev. Lett.}\ }\textbf {\bibinfo {volume} {42}},\
  \bibinfo {pages} {584--587} (\bibinfo {year} {1979})}\BibitemShut {NoStop}%
\bibitem [{\citenamefont {Mei}\ and\ \citenamefont
  {Davenport}(1992)}]{Coexistence-melting-I}%
  \BibitemOpen
  \bibfield  {author} {\bibinfo {author} {\bibfnamefont {J.}~\bibnamefont
  {Mei}}\ and\ \bibinfo {author} {\bibfnamefont {J.~W.}\ \bibnamefont
  {Davenport}},\ }\bibfield  {title} {\enquote {\bibinfo {title} {Free-energy
  calculations and the melting point of {Al}},}\ }\href {\doibase
  10.1103/PhysRevB.46.21} {\bibfield  {journal} {\bibinfo  {journal} {Phys.
  Rev. B}\ }\textbf {\bibinfo {volume} {46}},\ \bibinfo {pages} {21--25}
  (\bibinfo {year} {1992})}\BibitemShut {NoStop}%
\bibitem [{\citenamefont {Morris}\ \emph {et~al.}(1994)\citenamefont {Morris},
  \citenamefont {Wang}, \citenamefont {Ho},\ and\ \citenamefont
  {Chan}}]{Coexistence-melting-II}%
  \BibitemOpen
  \bibfield  {author} {\bibinfo {author} {\bibfnamefont {J.~R.}\ \bibnamefont
  {Morris}}, \bibinfo {author} {\bibfnamefont {C.~Z.}\ \bibnamefont {Wang}},
  \bibinfo {author} {\bibfnamefont {K.~M.}\ \bibnamefont {Ho}}, \ and\ \bibinfo
  {author} {\bibfnamefont {C.~T.}\ \bibnamefont {Chan}},\ }\bibfield  {title}
  {\enquote {\bibinfo {title} {Melting line of aluminum from simulations of
  coexisting phases},}\ }\href {\doibase 10.1103/PhysRevB.49.3109} {\bibfield
  {journal} {\bibinfo  {journal} {Phys. Rev. B}\ }\textbf {\bibinfo {volume}
  {49}},\ \bibinfo {pages} {3109--3115} (\bibinfo {year} {1994})}\BibitemShut
  {NoStop}%
\bibitem [{\citenamefont {Belonoshko}(1994)}]{Coexistence-melting-III}%
  \BibitemOpen
  \bibfield  {author} {\bibinfo {author} {\bibfnamefont {A.~B.}\ \bibnamefont
  {Belonoshko}},\ }\bibfield  {title} {\enquote {\bibinfo {title} {Molecular
  dynamics of {MgSiO3} perovskite at high pressures: Equation of state,
  structure, and melting transition},}\ }\href {\doibase
  https://doi.org/10.1016/0016-7037(94)90265-8} {\bibfield  {journal} {\bibinfo
   {journal} {Geochimica et Cosmochimica Acta}\ }\textbf {\bibinfo {volume}
  {58}},\ \bibinfo {pages} {4039--4047} (\bibinfo {year} {1994})}\BibitemShut
  {NoStop}%
\bibitem [{\citenamefont {Ogitsu}\ \emph {et~al.}(2003)\citenamefont {Ogitsu},
  \citenamefont {Schwegler}, \citenamefont {Gygi},\ and\ \citenamefont
  {Galli}}]{Coexistence-melting-IV}%
  \BibitemOpen
  \bibfield  {author} {\bibinfo {author} {\bibfnamefont {T.}~\bibnamefont
  {Ogitsu}}, \bibinfo {author} {\bibfnamefont {E.}~\bibnamefont {Schwegler}},
  \bibinfo {author} {\bibfnamefont {F.}~\bibnamefont {Gygi}}, \ and\ \bibinfo
  {author} {\bibfnamefont {G.}~\bibnamefont {Galli}},\ }\bibfield  {title}
  {\enquote {\bibinfo {title} {Melting of lithium hydride under pressure},}\
  }\href {\doibase 10.1103/PhysRevLett.91.175502} {\bibfield  {journal}
  {\bibinfo  {journal} {Phys. Rev. Lett.}\ }\textbf {\bibinfo {volume} {91}},\
  \bibinfo {pages} {175502} (\bibinfo {year} {2003})}\BibitemShut {NoStop}%
\bibitem [{\citenamefont {Lindemann}(1910)}]{Lindemann-criterion}%
  \BibitemOpen
  \bibfield  {author} {\bibinfo {author} {\bibfnamefont {F.~A.}\ \bibnamefont
  {Lindemann}},\ }\bibfield  {title} {\enquote {\bibinfo {title} {{\"U}ber die
  {Berechnung} molekularer {Eigenfrequenzen}},}\ }\href@noop {} {\bibfield
  {journal} {\bibinfo  {journal} {Physikalische Zeitschrift}\ }\textbf
  {\bibinfo {volume} {11}},\ \bibinfo {pages} {609--612} (\bibinfo {year}
  {1910})}\BibitemShut {NoStop}%
\bibitem [{\citenamefont {Gilvarry}(1956)}]{Lindemann-Gruneisen-Laws}%
  \BibitemOpen
  \bibfield  {author} {\bibinfo {author} {\bibfnamefont {J.~J.}\ \bibnamefont
  {Gilvarry}},\ }\bibfield  {title} {\enquote {\bibinfo {title} {The
  {Lindemann} and {G}r\"uneisen laws},}\ }\href {\doibase
  10.1103/PhysRev.102.308} {\bibfield  {journal} {\bibinfo  {journal} {Phys.
  Rev.}\ }\textbf {\bibinfo {volume} {102}},\ \bibinfo {pages} {308--316}
  (\bibinfo {year} {1956})}\BibitemShut {NoStop}%
\bibitem [{\citenamefont {Pickard}, \citenamefont {Errea},\ and\ \citenamefont
  {Eremets}(2020)}]{Chris-superconducting-hydride-review}%
  \BibitemOpen
  \bibfield  {author} {\bibinfo {author} {\bibfnamefont {C.~J.}\ \bibnamefont
  {Pickard}}, \bibinfo {author} {\bibfnamefont {I.}~\bibnamefont {Errea}}, \
  and\ \bibinfo {author} {\bibfnamefont {M.~I.}\ \bibnamefont {Eremets}},\
  }\bibfield  {title} {\enquote {\bibinfo {title} {Superconducting hydrides
  under pressure},}\ }\href {\doibase 10.1146/annurev-conmatphys-031218-013413}
  {\bibfield  {journal} {\bibinfo  {journal} {Annual Review of Condensed Matter
  Physics}\ }\textbf {\bibinfo {volume} {11}},\ \bibinfo {pages} {57--76}
  (\bibinfo {year} {2020})}\BibitemShut {NoStop}%
\bibitem [{\citenamefont {Drozdov}\ \emph {et~al.}(2019)\citenamefont
  {Drozdov}, \citenamefont {Kong}, \citenamefont {Minkov}, \citenamefont
  {Besedin}, \citenamefont {Kuzovnikov}, \citenamefont {Mozaffari},
  \citenamefont {Balicas}, \citenamefont {Balakirev}, \citenamefont {Graf},
  \citenamefont {Prakapenka}, \citenamefont {Greenberg}, \citenamefont
  {Knyazev}, \citenamefont {Tkacz},\ and\ \citenamefont
  {Eremets}}]{Eremets-LaH10}%
  \BibitemOpen
  \bibfield  {author} {\bibinfo {author} {\bibfnamefont {A.~P.}\ \bibnamefont
  {Drozdov}}, \bibinfo {author} {\bibfnamefont {P.~P.}\ \bibnamefont {Kong}},
  \bibinfo {author} {\bibfnamefont {V.~S.}\ \bibnamefont {Minkov}}, \bibinfo
  {author} {\bibfnamefont {S.~P.}\ \bibnamefont {Besedin}}, \bibinfo {author}
  {\bibfnamefont {M.~A.}\ \bibnamefont {Kuzovnikov}}, \bibinfo {author}
  {\bibfnamefont {S.}~\bibnamefont {Mozaffari}}, \bibinfo {author}
  {\bibfnamefont {L.}~\bibnamefont {Balicas}}, \bibinfo {author} {\bibfnamefont
  {F.~F.}\ \bibnamefont {Balakirev}}, \bibinfo {author} {\bibfnamefont {D.~E.}\
  \bibnamefont {Graf}}, \bibinfo {author} {\bibfnamefont {V.~B.}\ \bibnamefont
  {Prakapenka}}, \bibinfo {author} {\bibfnamefont {E.}~\bibnamefont
  {Greenberg}}, \bibinfo {author} {\bibfnamefont {D.~A.}\ \bibnamefont
  {Knyazev}}, \bibinfo {author} {\bibfnamefont {M.}~\bibnamefont {Tkacz}}, \
  and\ \bibinfo {author} {\bibfnamefont {M.~I.}\ \bibnamefont {Eremets}},\
  }\bibfield  {title} {\enquote {\bibinfo {title} {Superconductivity at 250 {K}
  in lanthanum hydride under high pressures},}\ }\href {\doibase
  10.1038/s41586-019-1201-8} {\bibfield  {journal} {\bibinfo  {journal}
  {Nature}\ }\textbf {\bibinfo {volume} {569}},\ \bibinfo {pages} {528--531}
  (\bibinfo {year} {2019})}\BibitemShut {NoStop}%
\bibitem [{\citenamefont {Somayazulu}\ \emph {et~al.}(2019)\citenamefont
  {Somayazulu}, \citenamefont {Ahart}, \citenamefont {Mishra}, \citenamefont
  {Geballe}, \citenamefont {Baldini}, \citenamefont {Meng}, \citenamefont
  {Struzhkin},\ and\ \citenamefont {Hemley}}]{Hemley-LaH10-experiment}%
  \BibitemOpen
  \bibfield  {author} {\bibinfo {author} {\bibfnamefont {M.}~\bibnamefont
  {Somayazulu}}, \bibinfo {author} {\bibfnamefont {M.}~\bibnamefont {Ahart}},
  \bibinfo {author} {\bibfnamefont {A.~K.}\ \bibnamefont {Mishra}}, \bibinfo
  {author} {\bibfnamefont {Z.~M.}\ \bibnamefont {Geballe}}, \bibinfo {author}
  {\bibfnamefont {M.}~\bibnamefont {Baldini}}, \bibinfo {author} {\bibfnamefont
  {Y.}~\bibnamefont {Meng}}, \bibinfo {author} {\bibfnamefont {V.~V.}\
  \bibnamefont {Struzhkin}}, \ and\ \bibinfo {author} {\bibfnamefont {R.~J.}\
  \bibnamefont {Hemley}},\ }\bibfield  {title} {\enquote {\bibinfo {title}
  {Evidence for superconductivity above 260 {K} in lanthanum superhydride at
  megabar pressures},}\ }\href {\doibase 10.1103/PhysRevLett.122.027001}
  {\bibfield  {journal} {\bibinfo  {journal} {Phys. Rev. Lett.}\ }\textbf
  {\bibinfo {volume} {122}},\ \bibinfo {pages} {027001} (\bibinfo {year}
  {2019})}\BibitemShut {NoStop}%
\bibitem [{\citenamefont {Liu}\ \emph {et~al.}(2018)\citenamefont {Liu},
  \citenamefont {Naumov}, \citenamefont {Geballe}, \citenamefont {Somayazulu},
  \citenamefont {Tse},\ and\ \citenamefont {Hemley}}]{LaH10-Russell}%
  \BibitemOpen
  \bibfield  {author} {\bibinfo {author} {\bibfnamefont {H.}~\bibnamefont
  {Liu}}, \bibinfo {author} {\bibfnamefont {I.~I.}\ \bibnamefont {Naumov}},
  \bibinfo {author} {\bibfnamefont {Z.~M.}\ \bibnamefont {Geballe}}, \bibinfo
  {author} {\bibfnamefont {M.}~\bibnamefont {Somayazulu}}, \bibinfo {author}
  {\bibfnamefont {J.~S.}\ \bibnamefont {Tse}}, \ and\ \bibinfo {author}
  {\bibfnamefont {R.~J.}\ \bibnamefont {Hemley}},\ }\bibfield  {title}
  {\enquote {\bibinfo {title} {Dynamics and superconductivity in compressed
  lanthanum superhydride},}\ }\href {\doibase 10.1103/PhysRevB.98.100102}
  {\bibfield  {journal} {\bibinfo  {journal} {Phys. Rev. B}\ }\textbf {\bibinfo
  {volume} {98}},\ \bibinfo {pages} {100102} (\bibinfo {year}
  {2018})}\BibitemShut {NoStop}%
\bibitem [{\citenamefont {Causs\'e}, \citenamefont {Geneste},\ and\
  \citenamefont {Loubeyre}(2023)}]{LaH10-Paul}%
  \BibitemOpen
  \bibfield  {author} {\bibinfo {author} {\bibfnamefont {M.}~\bibnamefont
  {Causs\'e}}, \bibinfo {author} {\bibfnamefont {G.}~\bibnamefont {Geneste}}, \
  and\ \bibinfo {author} {\bibfnamefont {P.}~\bibnamefont {Loubeyre}},\
  }\bibfield  {title} {\enquote {\bibinfo {title} {Superionicity of
  {H}$^{\ensuremath{\delta}\ensuremath{-}}$ in {LaH}$_{10}$ superhydride},}\
  }\href {\doibase 10.1103/PhysRevB.107.L060301} {\bibfield  {journal}
  {\bibinfo  {journal} {Phys. Rev. B}\ }\textbf {\bibinfo {volume} {107}},\
  \bibinfo {pages} {L060301} (\bibinfo {year} {2023})}\BibitemShut {NoStop}%
\bibitem [{\citenamefont {Pickard}\ and\ \citenamefont
  {Needs}(2007{\natexlab{b}})}]{Pickard2007-aluminium-hydride}%
  \BibitemOpen
  \bibfield  {author} {\bibinfo {author} {\bibfnamefont {C.~J.}\ \bibnamefont
  {Pickard}}\ and\ \bibinfo {author} {\bibfnamefont {R.~J.}\ \bibnamefont
  {Needs}},\ }\bibfield  {title} {\enquote {\bibinfo {title} {Metallization of
  aluminum hydride at high pressures: A first-principles study},}\ }\href
  {\doibase 10.1103/PhysRevB.76.144114} {\bibfield  {journal} {\bibinfo
  {journal} {Phys. Rev. B}\ }\textbf {\bibinfo {volume} {76}},\ \bibinfo
  {pages} {144114} (\bibinfo {year} {2007}{\natexlab{b}})}\BibitemShut
  {NoStop}%
\bibitem [{\citenamefont {Ye}\ \emph {et~al.}(2018)\citenamefont {Ye},
  \citenamefont {Zarifi}, \citenamefont {Zurek}, \citenamefont {Hoffmann},\
  and\ \citenamefont {Ashcroft}}]{ScH-Ashcroft}%
  \BibitemOpen
  \bibfield  {author} {\bibinfo {author} {\bibfnamefont {X.}~\bibnamefont
  {Ye}}, \bibinfo {author} {\bibfnamefont {N.}~\bibnamefont {Zarifi}}, \bibinfo
  {author} {\bibfnamefont {E.}~\bibnamefont {Zurek}}, \bibinfo {author}
  {\bibfnamefont {R.}~\bibnamefont {Hoffmann}}, \ and\ \bibinfo {author}
  {\bibfnamefont {N.~W.}\ \bibnamefont {Ashcroft}},\ }\bibfield  {title}
  {\enquote {\bibinfo {title} {High hydrides of scandium under pressure:
  Potential superconductors},}\ }\href@noop {} {\bibfield  {journal} {\bibinfo
  {journal} {The Journal of Physical Chemistry C}\ }\textbf {\bibinfo {volume}
  {122}},\ \bibinfo {pages} {6298--6309} (\bibinfo {year} {2018})}\BibitemShut
  {NoStop}%
\bibitem [{\citenamefont {Wang}\ \emph {et~al.}(2021)\citenamefont {Wang},
  \citenamefont {Ding}, \citenamefont {Delaire},\ and\ \citenamefont
  {Arya}}]{Arya2021-Non-Arrhenius-AgCrSe2}%
  \BibitemOpen
  \bibfield  {author} {\bibinfo {author} {\bibfnamefont {J.}~\bibnamefont
  {Wang}}, \bibinfo {author} {\bibfnamefont {J.}~\bibnamefont {Ding}}, \bibinfo
  {author} {\bibfnamefont {O.}~\bibnamefont {Delaire}}, \ and\ \bibinfo
  {author} {\bibfnamefont {G.}~\bibnamefont {Arya}},\ }\bibfield  {title}
  {\enquote {\bibinfo {title} {Atomistic mechanisms underlying non-{Arrhenius}
  ion transport in superionic conductor {AgCrSe2}},}\ }\href {\doibase
  10.1021/acsaem.1c01237} {\bibfield  {journal} {\bibinfo  {journal} {ACS
  Applied Energy Materials}\ }\textbf {\bibinfo {volume} {4}},\ \bibinfo
  {pages} {7157--7167} (\bibinfo {year} {2021})}\BibitemShut {NoStop}%
\bibitem [{\citenamefont {Kincs}\ and\ \citenamefont
  {Martin}(1996)}]{Martin1996-Non-Arrhenius}%
  \BibitemOpen
  \bibfield  {author} {\bibinfo {author} {\bibfnamefont {J.}~\bibnamefont
  {Kincs}}\ and\ \bibinfo {author} {\bibfnamefont {S.~W.}\ \bibnamefont
  {Martin}},\ }\bibfield  {title} {\enquote {\bibinfo {title} {Non-{Arrhenius}
  conductivity in glass: Mobility and conductivity saturation effects},}\
  }\href {\doibase 10.1103/PhysRevLett.76.70} {\bibfield  {journal} {\bibinfo
  {journal} {Phys. Rev. Lett.}\ }\textbf {\bibinfo {volume} {76}},\ \bibinfo
  {pages} {70--73} (\bibinfo {year} {1996})}\BibitemShut {NoStop}%
\bibitem [{\citenamefont {Qi}\ \emph {et~al.}(2021)\citenamefont {Qi},
  \citenamefont {Banerjee}, \citenamefont {Zuo}, \citenamefont {Chen},
  \citenamefont {Zhu}, \citenamefont {{Holekevi Chandrappa}}, \citenamefont
  {Li},\ and\ \citenamefont {Ong}}]{Ong2021-Non-Arrhenius}%
  \BibitemOpen
  \bibfield  {author} {\bibinfo {author} {\bibfnamefont {J.}~\bibnamefont
  {Qi}}, \bibinfo {author} {\bibfnamefont {S.}~\bibnamefont {Banerjee}},
  \bibinfo {author} {\bibfnamefont {Y.}~\bibnamefont {Zuo}}, \bibinfo {author}
  {\bibfnamefont {C.}~\bibnamefont {Chen}}, \bibinfo {author} {\bibfnamefont
  {Z.}~\bibnamefont {Zhu}}, \bibinfo {author} {\bibfnamefont {M.}~\bibnamefont
  {{Holekevi Chandrappa}}}, \bibinfo {author} {\bibfnamefont {X.}~\bibnamefont
  {Li}}, \ and\ \bibinfo {author} {\bibfnamefont {S.}~\bibnamefont {Ong}},\
  }\bibfield  {title} {\enquote {\bibinfo {title} {Bridging the gap between
  simulated and experimental ionic conductivities in lithium superionic
  conductors},}\ }\href {\doibase https://doi.org/10.1016/j.mtphys.2021.100463}
  {\bibfield  {journal} {\bibinfo  {journal} {Materials Today Physics}\
  }\textbf {\bibinfo {volume} {21}},\ \bibinfo {pages} {100463} (\bibinfo
  {year} {2021})}\BibitemShut {NoStop}%
\bibitem [{\citenamefont {Williams}\ \emph {et~al.}(1997)\citenamefont
  {Williams}, \citenamefont {Partin}, \citenamefont {Lincoln}, \citenamefont
  {Kouvetakis},\ and\ \citenamefont {O'Keeffe}}]{Williams1997}%
  \BibitemOpen
  \bibfield  {author} {\bibinfo {author} {\bibfnamefont {D.}~\bibnamefont
  {Williams}}, \bibinfo {author} {\bibfnamefont {D.}~\bibnamefont {Partin}},
  \bibinfo {author} {\bibfnamefont {F.}~\bibnamefont {Lincoln}}, \bibinfo
  {author} {\bibfnamefont {J.}~\bibnamefont {Kouvetakis}}, \ and\ \bibinfo
  {author} {\bibfnamefont {M.}~\bibnamefont {O'Keeffe}},\ }\bibfield  {title}
  {\enquote {\bibinfo {title} {The {{Disordered Crystal Structures}} of
  {{Zn}}({{CN}})2and {{Ga}}({{CN}})3},}\ }\href {\doibase
  10.1006/jssc.1997.7571} {\bibfield  {journal} {\bibinfo  {journal} {Journal
  of Solid State Chemistry}\ }\textbf {\bibinfo {volume} {134}},\ \bibinfo
  {pages} {164--169} (\bibinfo {year} {1997})}\BibitemShut {NoStop}%
\bibitem [{\citenamefont {Chapman}\ and\ \citenamefont
  {Chupas}(2007)}]{Chapman2007}%
  \BibitemOpen
  \bibfield  {author} {\bibinfo {author} {\bibfnamefont {K.~W.}\ \bibnamefont
  {Chapman}}\ and\ \bibinfo {author} {\bibfnamefont {P.~J.}\ \bibnamefont
  {Chupas}},\ }\bibfield  {title} {\enquote {\bibinfo {title} {Pressure
  {{Enhancement}} of {{Negative Thermal Expansion Behavior}} and {{Induced
  Framework Softening}} in {{Zinc Cyanide}}},}\ }\href {\doibase
  10.1021/ja073791e} {\bibfield  {journal} {\bibinfo  {journal} {Journal of the
  American Chemical Society}\ }\textbf {\bibinfo {volume} {129}},\ \bibinfo
  {pages} {10090--10091} (\bibinfo {year} {2007})}\BibitemShut {NoStop}%
\bibitem [{\citenamefont {Goodwin}\ and\ \citenamefont
  {Kepert}(2005)}]{Goodwin2005}%
  \BibitemOpen
  \bibfield  {author} {\bibinfo {author} {\bibfnamefont {A.~L.}\ \bibnamefont
  {Goodwin}}\ and\ \bibinfo {author} {\bibfnamefont {C.~J.}\ \bibnamefont
  {Kepert}},\ }\bibfield  {title} {\enquote {\bibinfo {title} {Negative thermal
  expansion and low-frequency modes in cyanide-bridged framework materials},}\
  }\href {\doibase 10.1103/PhysRevB.71.140301} {\bibfield  {journal} {\bibinfo
  {journal} {Physical Review B}\ }\textbf {\bibinfo {volume} {71}},\ \bibinfo
  {pages} {140301} (\bibinfo {year} {2005})}\BibitemShut {NoStop}%
\bibitem [{\citenamefont {Lapidus}\ \emph {et~al.}(2013)\citenamefont
  {Lapidus}, \citenamefont {Halder}, \citenamefont {Chupas},\ and\
  \citenamefont {Chapman}}]{Lapidus2013}%
  \BibitemOpen
  \bibfield  {author} {\bibinfo {author} {\bibfnamefont {S.~H.}\ \bibnamefont
  {Lapidus}}, \bibinfo {author} {\bibfnamefont {G.~J.}\ \bibnamefont {Halder}},
  \bibinfo {author} {\bibfnamefont {P.~J.}\ \bibnamefont {Chupas}}, \ and\
  \bibinfo {author} {\bibfnamefont {K.~W.}\ \bibnamefont {Chapman}},\
  }\bibfield  {title} {\enquote {\bibinfo {title} {Exploiting {{High
  Pressures}} to {{Generate Porosity}}, {{Polymorphism}}, {{And Lattice
  Expansion}} in the {{Nonporous Molecular Framework Zn}}({{CN}})
  {\textsubscript{2}}},}\ }\href {\doibase 10.1021/ja4012707} {\bibfield
  {journal} {\bibinfo  {journal} {Journal of the American Chemical Society}\
  }\textbf {\bibinfo {volume} {135}},\ \bibinfo {pages} {7621--7628} (\bibinfo
  {year} {2013})}\BibitemShut {NoStop}%
\bibitem [{\citenamefont {Fang}\ \emph {et~al.}(2013)\citenamefont {Fang},
  \citenamefont {Dove}, \citenamefont {Rimmer},\ and\ \citenamefont
  {Misquitta}}]{Fang2013}%
  \BibitemOpen
  \bibfield  {author} {\bibinfo {author} {\bibfnamefont {H.}~\bibnamefont
  {Fang}}, \bibinfo {author} {\bibfnamefont {M.~T.}\ \bibnamefont {Dove}},
  \bibinfo {author} {\bibfnamefont {L.~H.~N.}\ \bibnamefont {Rimmer}}, \ and\
  \bibinfo {author} {\bibfnamefont {A.~J.}\ \bibnamefont {Misquitta}},\
  }\bibfield  {title} {\enquote {\bibinfo {title} {Simulation study of pressure
  and temperature dependence of the negative thermal expansion in
  {{Zn}}({{CN}}) 2},}\ }\href {\doibase 10.1103/PhysRevB.88.104306} {\bibfield
  {journal} {\bibinfo  {journal} {Physical Review B}\ }\textbf {\bibinfo
  {volume} {88}},\ \bibinfo {pages} {104306} (\bibinfo {year}
  {2013})}\BibitemShut {NoStop}%
\bibitem [{\citenamefont {Trousselet}, \citenamefont {Boutin},\ and\
  \citenamefont {Coudert}(2015)}]{Trousselet2015}%
  \BibitemOpen
  \bibfield  {author} {\bibinfo {author} {\bibfnamefont {F.}~\bibnamefont
  {Trousselet}}, \bibinfo {author} {\bibfnamefont {A.}~\bibnamefont {Boutin}},
  \ and\ \bibinfo {author} {\bibfnamefont {F.-X.}\ \bibnamefont {Coudert}},\
  }\bibfield  {title} {\enquote {\bibinfo {title} {Novel {{Porous Polymorphs}}
  of {{Zinc Cyanide}} with {{Rich Thermal}} and {{Mechanical Behavior}}},}\
  }\href {\doibase 10.1021/acs.chemmater.5b01366} {\bibfield  {journal}
  {\bibinfo  {journal} {Chemistry of Materials}\ }\textbf {\bibinfo {volume}
  {27}},\ \bibinfo {pages} {4422--4430} (\bibinfo {year} {2015})}\BibitemShut
  {NoStop}%
\bibitem [{\citenamefont {Tkatchenko}\ \emph {et~al.}(2012)\citenamefont
  {Tkatchenko}, \citenamefont {DiStasio}, \citenamefont {Car},\ and\
  \citenamefont {Scheffler}}]{Tkatchenko2012}%
  \BibitemOpen
  \bibfield  {author} {\bibinfo {author} {\bibfnamefont {A.}~\bibnamefont
  {Tkatchenko}}, \bibinfo {author} {\bibfnamefont {R.~A.}\ \bibnamefont
  {DiStasio}}, \bibinfo {author} {\bibfnamefont {R.}~\bibnamefont {Car}}, \
  and\ \bibinfo {author} {\bibfnamefont {M.}~\bibnamefont {Scheffler}},\
  }\bibfield  {title} {\enquote {\bibinfo {title} {Accurate and {{Efficient
  Method}} for {{Many-Body}} van der {{Waals Interactions}}},}\ }\href
  {\doibase 10.1103/PhysRevLett.108.236402} {\bibfield  {journal} {\bibinfo
  {journal} {Phys. Rev. Lett.}\ }\textbf {\bibinfo {volume} {108}},\ \bibinfo
  {pages} {236402} (\bibinfo {year} {2012})}\BibitemShut {NoStop}%
\bibitem [{\citenamefont {Willatt}, \citenamefont {Musil},\ and\ \citenamefont
  {Ceriotti}(2018)}]{Willatt2018}%
  \BibitemOpen
  \bibfield  {author} {\bibinfo {author} {\bibfnamefont {M.~J.}\ \bibnamefont
  {Willatt}}, \bibinfo {author} {\bibfnamefont {F.}~\bibnamefont {Musil}}, \
  and\ \bibinfo {author} {\bibfnamefont {M.}~\bibnamefont {Ceriotti}},\
  }\bibfield  {title} {\enquote {\bibinfo {title} {Feature optimization for
  atomistic machine learning yields a data-driven construction of the periodic
  table of the elements},}\ }\href {\doibase 10.1039/C8CP05921G} {\bibfield
  {journal} {\bibinfo  {journal} {Physical Chemistry Chemical Physics}\
  }\textbf {\bibinfo {volume} {20}},\ \bibinfo {pages} {29661--29668} (\bibinfo
  {year} {2018})}\BibitemShut {NoStop}%
\end{thebibliography}%

\end{document}